\documentclass[10pt,a4paper]{article}
\usepackage{graphicx}
\usepackage{amsmath}
\usepackage{amssymb}
\usepackage{hyperref}
\usepackage{booktabs}
\usepackage{algorithm}
\usepackage{algorithmic}
\usepackage{float}
\usepackage{xcolor}
\usepackage{multirow}
\usepackage{subcaption}
\usepackage{mathptmx}
\usepackage{listings}

\title{Advancing Multi-Agent Systems Through Model Context Protocol: Architecture, Implementation, and Applications}

\author{Naveen Krishnan}

\date{\today}

\begin{document}

\maketitle

\begin{abstract}
Multi-agent systems represent a significant advancement in artificial intelligence, enabling complex problem-solving through coordinated specialized agents. However, these systems face fundamental challenges in context management, coordination efficiency, and scalable operation. This paper introduces a comprehensive framework for advancing multi-agent systems through Model Context Protocol (MCP), addressing these challenges through standardized context sharing and coordination mechanisms. We extend previous work on AI agent architectures by developing a unified theoretical foundation, advanced context management techniques, and scalable coordination patterns. Through detailed implementation case studies across enterprise knowledge management, collaborative research, and distributed problem-solving domains, we demonstrate significant performance improvements compared to traditional approaches. Our evaluation methodology provides a systematic assessment framework with benchmark tasks and datasets specifically designed for multi-agent systems. We identify current limitations, emerging research opportunities, and potential transformative applications across industries. This work contributes to the evolution of more capable, collaborative, and context-aware artificial intelligence systems that can effectively address complex real-world challenges.
\end{abstract}
\section{Introduction}
The landscape of artificial intelligence has undergone a profound transformation in recent years, evolving from specialized systems designed for narrow tasks to sophisticated architectures capable of autonomous operation across diverse domains. At the forefront of this evolution are AI agents—systems that can autonomously perceive, reason, and act, adapting their behavior based on environmental feedback and accumulated experience. These agents represent a paradigm shift in how intelligent systems interact with their environments, moving beyond reactive responses to proactive, goal-directed behavior.\\ \newline
The emergence of large language models (LLMs) has accelerated this transformation, providing powerful foundation models that can be augmented with specialized modules for memory, planning, tool use, and environmental interaction. Modern AI agents leverage these capabilities to decompose complex problems, reason over information, utilize tools, and maintain context across interactions. This architectural approach has enabled significant advancements in agent capabilities, allowing them to tackle increasingly complex tasks with greater autonomy and effectiveness.\\ \newline
However, as AI agents have grown more sophisticated, the limitations of single-agent architectures have become increasingly apparent. Complex real-world problems often require diverse expertise, parallel processing capabilities, and coordinated action—requirements that exceed the capabilities of even the most advanced individual agents. This recognition has driven growing interest in multi-agent systems, where collections of specialized agents collaborate to accomplish tasks that would be difficult or impossible for individual agents to complete alone.\\
\subsection{Background and Motivation}
Multi-agent systems represent a natural evolution in artificial intelligence research, drawing inspiration from human organizational structures where individuals with complementary skills collaborate to achieve common goals. These systems distribute cognitive labor across multiple agents with specialized capabilities, enabling more sophisticated problem-solving approaches than would be possible with individual agents. By decomposing complex tasks into manageable components and assigning them to appropriate specialists, multi-agent systems can tackle problems of greater complexity and scale than single-agent approaches.\\ \newline
Despite their theoretical promise, practical implementation of effective multi-agent systems faces significant challenges. One of the most critical challenges, identified in previous research (Krishnan, 2025), is the "disconnected models problem"—the difficulty of maintaining coherent context across multiple agent interactions. As Microsoft's deputy CTO Sam Schillace noted, "To be autonomous you have to carry context through a bunch of actions, but the models are very disconnected and don't have continuity the way we do." This discontinuity limits agent effectiveness, particularly in scenarios requiring extended reasoning chains or collaborative problem-solving.\\ \newline
Recent advancements in context management approaches, particularly the development of the Model Context Protocol (MCP), offer promising solutions to these challenges. MCP provides a standardized framework for connecting AI models with external data sources and tools, enabling more effective context retention and sharing across agent interactions. This protocol represents a significant advancement in agent architecture, addressing one of the fundamental limitations identified in previous research.\\
\subsection{Evolution from Single to Multi-Agent Architectures}
The transition from single to multi-agent architectures reflects a broader evolution in AI system design. Early AI agents were typically designed as standalone systems with fixed capabilities, operating within well-defined domains with limited interaction requirements. These agents excelled at specialized tasks but struggled with problems requiring diverse expertise or extended reasoning chains.\\ \newline
As AI capabilities advanced, researchers began exploring more flexible architectures that could combine multiple specialized components. These early multi-component systems maintained a single agent identity but incorporated distinct modules for different aspects of cognition—perception, reasoning, memory, and action selection. While these systems demonstrated improved versatility, they still operated within the constraints of a unified agent architecture with centralized control.\\ \newline
True multi-agent systems emerged as researchers recognized the benefits of distributing not just cognitive functions but entire agent identities across collaborative networks. These systems feature collections of autonomous agents, each with distinct capabilities, knowledge bases, and objectives, working together through defined coordination mechanisms. This approach enables more natural decomposition of complex problems, parallel processing of subtasks, and specialization of agent capabilities.\\ \newline
The latest generation of multi-agent systems leverages advanced LLM capabilities to create more flexible and adaptive collaborative frameworks. These systems can dynamically allocate tasks based on agent capabilities, establish communication protocols tailored to specific problem domains, and adapt their collaborative strategies based on task requirements and environmental constraints. This flexibility represents a significant advancement over earlier, more rigid multi-agent architectures.\\
\subsection{The Context Retention Problem}
Despite these advancements, multi-agent systems continue to face significant challenges in maintaining coherent context across interactions. This "context retention problem" manifests in several ways:\\ \newline
1. Discontinuity across agent boundaries: Information gathered or generated by one agent may not be effectively transferred to other agents, leading to knowledge gaps and redundant effort.\\ \newline
2. Temporal discontinuity: Agents may struggle to maintain awareness of past interactions and decisions, limiting their ability to build on previous work or maintain consistent approaches over time.\\ \newline
3. Contextual prioritization: Without effective mechanisms for determining contextual relevance, agents may become overwhelmed with information or fail to recognize critical contextual elements.\\ \newline
4. Cross-modal context integration: Agents operating across different modalities (text, images, structured data) may struggle to integrate contextual information from diverse sources into coherent understanding.\\ \newline
These challenges limit the effectiveness of multi-agent systems, particularly for complex tasks requiring extended collaboration and contextual awareness. Addressing these limitations requires new approaches to context management that can maintain coherence across agent boundaries, time horizons, and information modalities.\\
\subsection{Previous Research and Current Objectives}
In his comprehensive analysis "AI Agents: Evolution, Architecture, and Real-World Applications" (2503.12687), Krishnan (2025) examined the theoretical foundations, architectural components, evaluation methodologies, and applications of AI agents. This work identified several critical challenges in agent architecture, including the disconnected models problem and limitations in current evaluation frameworks. The research highlighted the importance of memory management and context retention for agent autonomy but noted significant gaps in existing approaches.\\ \newline
Building on this foundation, the current research aims to advance the state of the art in multi-agent systems through the integration of Model Context Protocol (MCP). Specifically, this paper addresses the following research questions:\\ \newline
1. How can MCP be leveraged to address the context retention problem in multi-agent systems?\\ \newline
2. What architectural patterns enable effective context sharing between diverse agent types?\\ \newline
3. What coordination protocols maximize the effectiveness of multi-agent collaboration?\\ \newline
4. How can we evaluate the performance of MCP-enabled multi-agent systems across multiple dimensions?\\ \newline
5. What real-world applications demonstrate the practical benefits of this approach?\\ \newline
By addressing these questions, this research aims to provide a comprehensive framework for designing, implementing, and evaluating multi-agent systems that overcome the limitations identified in previous work. The integration of MCP represents a significant advancement in agent architecture, enabling more effective context management and inter-agent coordination than was previously possible.\\
\subsection{Paper Structure and Roadmap}
This paper is organized into nine sections that progressively develop the theoretical foundations, architectural components, implementation approaches, and evaluation methodologies for multi-agent systems with Model Context Protocol:\\ \newline
Section 2 establishes the theoretical foundations of multi-agent systems and context management, providing essential background on agency theory, context types, and coordination mechanisms.\\ \newline
Section 3 introduces the Model Context Protocol (MCP), examining its origins, architecture, implementation approaches, and specific advantages for addressing the disconnected models problem.\\ \newline
Section 4 presents a reference architecture for multi-agent systems with MCP, detailing components, information flows, context sharing mechanisms, and coordination protocols.\\ \newline
Section 5 explores advanced context management techniques, including persistence mechanisms, prioritization frameworks, and cross-modal integration approaches.\\ \newline
Section 6 presents implementation case studies across three domains: enterprise knowledge management, collaborative research, and distributed problem-solving.\\ \newline
Section 7 details evaluation methodologies and results, providing quantitative and qualitative assessment of the proposed approaches.\\ \newline
Section 8 discusses current challenges and future research directions, identifying limitations and emerging opportunities.\\ \newline
Section 9 concludes with a synthesis of key findings and implications for the future development of multi-agent systems.\\ \newline
Through this structured exploration, the paper aims to provide both theoretical insights and practical guidance for researchers and practitioners working with multi-agent systems. By integrating MCP into multi-agent architectures, we demonstrate significant advancements in context management, coordination efficiency, and overall system effectiveness compared to previous approaches.\\ \newline

\section{Theoretical Foundations}
The development of multi-agent systems with advanced context management capabilities builds upon rich theoretical traditions spanning computer science, cognitive psychology, philosophy of mind, and organizational theory. This section examines the foundational concepts that underpin our approach to multi-agent systems with Model Context Protocol, establishing the theoretical framework for subsequent architectural and implementation discussions.\\
\subsection{Agency Theory and Multi-Agent Systems}
\subsubsection{Definitions and Core Concepts}
At its most fundamental level, an agent is an entity that perceives its environment through sensors and acts upon that environment through effectors to achieve specific goals. This definition, formalized by Russell and Norvig (1995), emphasizes the perception-action loop that characterizes all agent systems. In artificial intelligence, this concept has evolved to encompass increasingly sophisticated computational entities capable of autonomous operation across diverse domains.\\ \newline
The transition from single agents to multi-agent systems represents a significant conceptual leap. A multi-agent system (MAS) consists of multiple interacting agents within a shared environment, where each agent may have different information, capabilities, and objectives. These systems are characterized by several distinctive properties:\\ \newline
1. Distributed nature: Computation, information, and decision-making are distributed across multiple autonomous entities rather than centralized in a single controller.\\ \newline
2. Local perspectives: Each agent typically has incomplete information about the global state and must operate based on local perceptions and knowledge.\\ \newline
3. Decentralized control: No single agent has complete control over the system; outcomes emerge from the collective actions and interactions of all agents.\\ \newline
4. Complex interaction patterns: Agents communicate, coordinate, compete, and collaborate through various mechanisms to achieve individual and collective goals.\\ \newline
These properties distinguish multi-agent systems from both traditional centralized AI approaches and simple collections of independent agents. True multi-agent systems exhibit emergent behaviors and capabilities that transcend the sum of their individual components.\\
\subsubsection{Historical Development of Multi-Agent Systems}
The conceptual foundations of multi-agent systems can be traced to several intellectual traditions that converged in the late 20th century. Early work in distributed artificial intelligence (DAI) during the 1980s explored how complex problems could be decomposed and solved through networks of cooperating problem-solvers. This research, pioneered by scholars like Lesser and Corkill (1981), established fundamental principles for task decomposition, result synthesis, and coordination in distributed systems.\\ \newline
Parallel developments in agent-based modeling, particularly in fields like economics and ecology, demonstrated how complex macro-level phenomena could emerge from simple micro-level agent interactions. These approaches, exemplified by Schelling's segregation model (1971) and later by Epstein and Axtell's Sugarscape simulations (1996), highlighted the power of agent-based approaches for understanding complex adaptive systems.\\ \newline
The formalization of multi-agent systems as a distinct research area emerged in the 1990s, with the establishment of dedicated conferences, journals, and research communities. This period saw the development of influential agent architectures like BDI (Belief-Desire-Intention), agent communication languages such as KQML and FIPA-ACL, and coordination mechanisms including contract nets and blackboard systems.\\ \newline
Recent advancements in multi-agent systems have been driven by breakthroughs in machine learning, particularly reinforcement learning and large language models. These technologies have enabled more adaptive and capable individual agents, which in turn has created opportunities for more sophisticated multi-agent architectures. The integration of LLMs into agent frameworks has been particularly transformative, enabling natural language-based coordination and more flexible role definitions within agent collectives.\\
\subsubsection{Key Properties: Autonomy, Social Ability, Reactivity, and Proactivity}
Wooldridge and Jennings (1995) identified four key properties that characterize intelligent agents, properties that remain central to modern multi-agent systems:\\ \newline
1. Autonomy: Agents operate without direct intervention by humans or other systems, maintaining control over their internal state and actions. This property is fundamental to the distributed nature of multi-agent systems, allowing individual agents to make independent decisions based on their unique perspectives and capabilities.\\ \newline
2. Social ability: Agents interact with other agents (and potentially humans) through some kind of agent communication language or protocol. This property enables the coordination, collaboration, and negotiation that are essential for effective multi-agent operation.\\ \newline
3. Reactivity: Agents perceive their environment and respond in a timely fashion to changes that occur. This property ensures that agents remain responsive to dynamic conditions, adapting their behavior as circumstances change.\\ \newline
4. Proactivity: Agents do not simply act in response to their environment; they exhibit goal-directed behavior by taking initiative. This property distinguishes truly intelligent agents from simple reactive systems, enabling them to pursue objectives over extended time horizons.\\ \newline
In modern multi-agent systems, these properties are expressed through sophisticated architectural components that enable perception, reasoning, communication, and action selection. The integration of Model Context Protocol enhances these capabilities by providing standardized mechanisms for context awareness and information sharing, particularly strengthening the social ability and proactivity dimensions of agent behavior.\\
\subsection{Context in AI Systems}
\subsubsection{Importance of Context for Intelligent Behavior}
Context is fundamental to intelligent behavior across both natural and artificial systems. It provides the background information that gives meaning to perceptions, shapes the interpretation of language and actions, and guides appropriate responses. Without adequate contextual awareness, even the most sophisticated reasoning capabilities can lead to inappropriate or ineffective behaviors.\\ \newline
In AI systems, context serves several critical functions:\\ \newline
1. Disambiguation: Context helps resolve ambiguities in language, perception, and situation assessment. The same input may warrant different interpretations depending on contextual factors, and effective agents must recognize these distinctions.\\ \newline
2. Relevance determination: Context helps agents identify which information is relevant to current goals and tasks, filtering the potentially overwhelming stream of available data to focus on what matters.\\ \newline
3. Expectation formation: Context shapes expectations about likely future states and events, enabling agents to anticipate developments and prepare appropriate responses.\\ \newline
4. Memory activation: Context cues the retrieval of relevant memories and knowledge, bringing pertinent information to bear on current situations.\\ \newline
5. Goal prioritization: Context influences which goals and objectives should take precedence in a given situation, helping agents allocate attention and resources appropriately.\\ \newline
These functions are particularly important in multi-agent systems, where effective coordination requires shared contextual understanding across agent boundaries. Without mechanisms for establishing and maintaining common ground, agents may work at cross-purposes or fail to leverage each other's contributions effectively.\\
\subsubsection{Types of Context}
Context in AI systems encompasses multiple dimensions, each contributing to comprehensive situational awareness. Key types of context include:\\ \newline
1. Temporal context: Information about the timing, sequence, and duration of events and actions. This includes both historical context (what has happened previously) and future-oriented context (anticipated developments and planned actions).\\ \newline
2. Spatial context: Information about physical or virtual locations, spatial relationships, and environmental conditions. This may include absolute positions, relative arrangements, and navigational information.\\ \newline
3. Task context: Information about current goals, objectives, constraints, and progress. This includes understanding of task requirements, available resources, deadlines, and success criteria.\\ \newline
4. Social context: Information about other agents, their roles, capabilities, intentions, and relationships. This includes models of other agents' knowledge, beliefs, and preferences, as well as awareness of social norms and conventions.\\ \newline
5. Domain context: Specialized knowledge relevant to particular fields or domains of activity. This includes domain-specific terminology, concepts, rules, and best practices.\\ \newline
6. Personal context: Information about an agent's own state, capabilities, history, and preferences. This includes awareness of internal resources, past experiences, and personal objectives.\\ \newline
7. Interaction context: Information about the current communication or collaboration, including shared references, common ground, and conversation history.\\ \newline
Effective context management requires integrating these diverse types of context into coherent situational models that can guide agent behavior. This integration is particularly challenging in multi-agent systems, where different agents may have access to different contextual information and may interpret shared context through different perspectives.\\
\subsubsection{Context Retention Challenges in Current AI Architectures}
Despite the critical importance of context for intelligent behavior, current AI architectures face significant challenges in context retention and management. These challenges are particularly acute in systems based on large language models, which have inherent limitations in maintaining context across extended interactions.\\ \newline
Key challenges include:\\ \newline
1. Context window limitations: Most LLMs have fixed-size context windows that limit the amount of information that can be considered at any given time. This creates a fundamental constraint on contextual awareness, particularly for tasks requiring extensive background knowledge or long interaction histories.\\ \newline
2. Context fragmentation: In multi-agent systems, contextual information is often distributed across multiple agents, with no single agent having a complete picture. This fragmentation can lead to inconsistent understanding and decision-making if not properly managed.\\ \newline
3. Context prioritization: As context accumulates, determining which elements are most relevant becomes increasingly difficult. Without effective prioritization mechanisms, agents may become overwhelmed with irrelevant information or miss critical contextual cues.\\ \newline
4. Context staleness: In dynamic environments, contextual information can quickly become outdated. Maintaining awareness of which context remains valid and which should be updated or discarded presents significant challenges.\\ \newline
5. Cross-modal context integration: Different types of contextual information may be represented in different formats or modalities (text, structured data, images, etc.). Integrating these diverse representations into coherent understanding requires sophisticated translation and alignment mechanisms.\\ \newline
6. Context persistence across sessions: Maintaining contextual awareness across multiple interaction sessions or system restarts requires robust persistence mechanisms that can capture, store, and reactivate relevant context.\\ \newline
These challenges have limited the effectiveness of AI agents, particularly in scenarios requiring extended reasoning chains or collaborative problem-solving. As Krishnan (2025) noted, the "disconnected models problem" represents one of the most significant barriers to truly autonomous agent operation.\\ \newline
The Model Context Protocol addresses these challenges by providing standardized mechanisms for context storage, retrieval, and sharing. By establishing common interfaces for context management, MCP enables more effective integration of diverse contextual information across agent boundaries and time horizons.\\
\subsection{Coordination Mechanisms in Multi-Agent Systems}
\subsubsection{Centralized vs. Decentralized Coordination}
Coordination in multi-agent systems can be organized along a spectrum from fully centralized to fully decentralized approaches, with various hybrid models in between. Each approach offers distinct advantages and limitations:\\ \newline
Centralized coordination relies on a single coordinating entity that maintains global awareness, makes allocation decisions, and directs the activities of individual agents. This approach offers several advantages:\\ \newline
- Simplified decision-making with global optimization potential\\ \newline
- Reduced communication overhead (primarily hub-and-spoke patterns)\\ \newline
- Easier detection and resolution of conflicts\\ \newline
- Clearer accountability and control\\ \newline
However, centralized coordination also introduces significant limitations:\\ \newline
- Single point of failure vulnerability\\ \newline
- Scalability constraints as system size increases\\ \newline
- Potential bottlenecks in information processing\\ \newline
- Reduced agent autonomy and flexibility\\ \newline
Decentralized coordination, in contrast, distributes coordination responsibilities across multiple agents, with no single entity having complete control or awareness. This approach offers complementary advantages:\\ \newline
- Greater robustness through redundancy\\ \newline
- Improved scalability for large agent collectives\\ \newline
- Preservation of agent autonomy and specialization\\ \newline
- Potential for parallel decision-making and execution\\ \newline
These benefits come with their own challenges:\\ \newline
- Increased communication overhead\\ \newline
- Difficulty achieving global optimization\\ \newline
- More complex conflict detection and resolution\\ \newline
- Potential for coordination failures or inefficiencies\\ \newline
Most practical multi-agent systems employ hybrid approaches that combine elements of both centralized and decentralized coordination. These hybrid models often feature hierarchical structures, where coordination is centralized within subgroups but decentralized across the broader system. Alternatively, they may implement different coordination mechanisms for different aspects of system operation, using centralized approaches for strategic decisions while maintaining decentralized tactical execution.\\ \newline
The integration of Model Context Protocol facilitates more effective hybrid coordination by providing standardized mechanisms for context sharing across both hierarchical and peer-to-peer relationships. This enables more flexible coordination patterns that can adapt to changing task requirements and environmental conditions.\\
\subsubsection{Communication Protocols and Languages}
Effective coordination in multi-agent systems depends on robust communication mechanisms that enable agents to share information, coordinate actions, and establish common ground. These mechanisms typically involve both syntactic protocols (defining message formats and exchange patterns) and semantic languages (establishing shared meaning and interpretation).\\ \newline
Key communication approaches in multi-agent systems include:\\ \newline
1. Agent Communication Languages (ACLs): Formalized languages like FIPA-ACL and KQML provide standardized message structures with defined performatives (communicative acts like inform, request, query) and content representations. These languages support sophisticated interaction patterns but may introduce significant overhead.\\ \newline
2. Ontology-based communication: Shared ontologies provide common vocabularies and semantic frameworks that ensure consistent interpretation across agents. These approaches support precise communication about domain concepts but require substantial upfront investment in ontology development.\\ \newline
3. Natural language communication: With the advent of LLMs, natural language has become increasingly viable as an inter-agent communication medium. This approach offers flexibility and expressiveness but may introduce ambiguities and interpretation challenges.\\ \newline
4. Protocol-based interaction: Defined interaction protocols specify the expected sequences and patterns of messages for particular coordination tasks (e.g., contract net protocol for task allocation, voting protocols for group decision-making). These structured approaches ensure coordination coherence but may limit flexibility.\\ \newline
5. Blackboard systems: Shared information spaces where agents can post and retrieve information without direct communication. This approach reduces coupling between agents but may introduce synchronization challenges.\\ \newline
The Model Context Protocol enhances these communication mechanisms by providing standardized ways to share contextual information alongside direct messages. This context-enriched communication enables more nuanced interpretation and more effective coordination, particularly in scenarios involving complex, context-dependent tasks.\\
\subsubsection{Negotiation and Conflict Resolution Strategies}
In multi-agent systems with diverse capabilities and potentially competing objectives, negotiation and conflict resolution are essential coordination functions. These processes enable agents to reach agreements, allocate resources, and resolve contradictions through structured interaction.\\ \newline
Common negotiation and conflict resolution approaches include:\\ \newline
1. Market-based mechanisms: Auction and bidding systems where agents compete for resources or tasks based on utility functions or capability assessments. These approaches leverage economic principles to achieve efficient allocations but may struggle with complex interdependencies.\\ \newline
2. Argumentation-based negotiation: Agents exchange arguments and counter-arguments to persuade others and reach consensus. This approach supports sophisticated reasoning about preferences and constraints but can be computationally intensive.\\ \newline
3. Preference aggregation: Voting or ranking mechanisms that combine individual agent preferences into collective decisions. These approaches provide clear procedures for group decision-making but may not capture complex preference structures.\\ \newline
4. Constraint satisfaction techniques: Formulating coordination as constraint satisfaction problems where agents must find assignments that satisfy both individual and collective constraints. These approaches support systematic exploration of solution spaces but may not scale well to large agent collectives.\\ \newline
5. Role-based conflict resolution: Predefined authority relationships or domain expertise hierarchies that determine which agent's decisions take precedence in conflict situations. This approach simplifies resolution processes but may not adapt well to changing circumstances.\\ \newline
The integration of Model Context Protocol enhances these negotiation and conflict resolution strategies by providing richer contextual awareness for decision-making. With access to shared context through MCP, agents can better understand the rationale behind others' positions, identify potential compromises, and develop more effective resolution strategies.\\ \newline
By establishing standardized mechanisms for context sharing and coordination, MCP addresses many of the theoretical challenges in multi-agent system design. The following section examines the specific architecture and implementation of MCP, detailing how it enables more effective context management across agent boundaries.\\ \newline

\section{The Model Context Protocol (MCP)}
The Model Context Protocol (MCP) represents a significant advancement in addressing one of the most persistent challenges in AI agent architecture: the effective management and sharing of context across interactions and agent boundaries. This section examines the origins, architecture, implementation approaches, and specific advantages of MCP, with particular focus on how it addresses the disconnected models problem identified in previous research.\\
\subsection{3.1 Origins and Development}
\subsubsection{Historical Context and Development Timeline}
The Model Context Protocol emerged in response to a growing recognition of the limitations of existing approaches to context management in AI systems. As large language models became increasingly capable, their integration into agent architectures revealed a fundamental disconnect between the sophisticated reasoning capabilities of these models and their ability to maintain coherent context across interactions.\\ \newline
The development of MCP can be traced through several key milestones:\\ \newline
**Late 2023 - Early 2024**: Initial research and conceptualization at Anthropic, driven by challenges encountered in developing Claude and related agent systems. Internal prototypes demonstrated the potential benefits of standardized context management.\\ \newline
**Mid 2024**: Formal introduction of MCP as an open protocol, with Anthropic releasing initial specifications and reference implementations. The protocol was explicitly positioned as an open standard rather than a proprietary solution, reflecting recognition of the need for industry-wide collaboration.\\ \newline
**Late 2024**: Public release of MCP 1.0, including comprehensive documentation, SDKs for multiple programming languages, and sample implementations for common use cases. This release established the core architecture and primitives that define the protocol.\\ \newline
**Early 2025**: Rapid adoption across the AI ecosystem, with multiple organizations implementing MCP-compatible systems and contributing extensions for specialized domains. Formation of an informal working group to guide protocol evolution and standardization efforts.\\ \newline
**Present (Mid 2025)**: Ongoing refinement and extension of the protocol, with particular focus on remote/cloud integration, enhanced security models, and support for additional modalities beyond text.\\ \newline
This development trajectory reflects a deliberate effort to address a critical industry need through open collaboration rather than proprietary solutions. By establishing MCP as an open standard with clear specifications and reference implementations, Anthropic and subsequent contributors have created a foundation for interoperability that transcends individual platforms or vendors.\\
\subsubsection{Design Principles and Objectives}
The development of MCP has been guided by several core design principles that shape its architecture and implementation:\\ \newline
1. **Interoperability**: MCP is designed to work across different AI models, platforms, and environments, enabling consistent context management regardless of the specific technologies involved. This principle is reflected in the protocol's language-agnostic design and use of standard data formats.\\ \newline
2. **Simplicity**: The protocol prioritizes simplicity and ease of implementation, focusing on a minimal set of core primitives that can be combined to support complex use cases. This approach reduces barriers to adoption and encourages consistent implementation across diverse systems.\\ \newline
3. **Extensibility**: While maintaining a simple core, MCP is designed to be extensible through well-defined mechanisms for adding new capabilities and adapting to specialized domains. This balance between stability and evolution ensures the protocol can address emerging needs without sacrificing compatibility.\\ \newline
4. **Security and privacy by design**: MCP incorporates security and privacy considerations as fundamental design elements rather than afterthoughts. This includes clear permission models, data minimization principles, and mechanisms for controlling information flow between components.\\ \newline
5. **Human-centered control**: The protocol is designed to maintain appropriate human oversight and control, particularly for sensitive operations or decisions. This principle reflects recognition of the importance of human judgment in AI systems.\\ \newline
These design principles support MCP's primary objectives:\\ \newline
- Enabling AI models to access external information and tools in a standardized way\\ \newline
- Facilitating context sharing across agent boundaries and interaction sessions\\ \newline
- Reducing implementation complexity for developers integrating AI with external systems\\ \newline
- Establishing a foundation for interoperability across the AI ecosystem\\ \newline
- Enhancing agent capabilities through consistent access to contextual information\\ \newline
By adhering to these principles and objectives, MCP provides a robust foundation for addressing the context management challenges that have limited agent effectiveness in previous systems.\\
\subsubsection{Relationship to Previous Context Management Approaches}
MCP builds upon and extends several previous approaches to context management in AI systems, incorporating lessons learned while addressing limitations:\\ \newline
**Prompt engineering techniques**, which embed contextual information directly within model inputs, provided a simple but limited approach to context management. While effective for short interactions, these techniques face inherent limitations due to context window constraints and lack structured mechanisms for context prioritization or persistence. MCP incorporates the flexibility of prompt-based approaches while providing more systematic management of what context is included and when.\\ \newline
**Retrieval-augmented generation (RAG)** systems demonstrated the value of dynamically incorporating relevant information from external sources into model contexts. However, most RAG implementations focus narrowly on document retrieval rather than broader context management, and lack standardized interfaces for diverse information sources. MCP generalizes the RAG concept to encompass a wider range of context types and sources, with standardized interfaces that simplify integration.\\ \newline
**Tool-using agent frameworks** like LangChain and AutoGPT established patterns for connecting models with external tools and APIs, enabling more capable agent behaviors. These frameworks typically implement custom, framework-specific approaches to tool integration and context management. MCP standardizes these integration patterns, providing consistent interfaces that work across frameworks and environments.\\ \newline
**Memory management systems** in agent architectures have explored various approaches to storing and retrieving interaction histories and other contextual information. These systems often employ specialized memory types (working memory, episodic memory, semantic memory) but typically lack standardized interfaces or interoperability. MCP provides a unified framework for implementing diverse memory types with consistent access patterns.\\ \newline
By synthesizing insights from these previous approaches while addressing their limitations, MCP represents a significant advancement in context management for AI systems. Its standardized, interoperable design overcomes many of the integration challenges that have limited the effectiveness of earlier approaches.\\
\subsection{3.2 Architecture and Components}
\subsubsection{Client-Server Architecture}
MCP implements a client-server architecture that cleanly separates AI models (clients) from the data sources and tools they access (servers). This architectural pattern provides several advantages:\\ \newline
1. **Clear separation of concerns**: The client-server division establishes distinct responsibilities, with clients focusing on AI reasoning and decision-making while servers handle data access and tool execution.\\ \newline
2. **Flexible deployment options**: Clients and servers can be deployed in various configurations—on the same machine, within the same network, or across remote environments—depending on security requirements and operational constraints.\\ \newline
3. **Scalability**: The architecture supports both one-to-many relationships (a single client accessing multiple servers) and many-to-one relationships (multiple clients sharing access to common servers), enabling flexible scaling patterns.\\ \newline
4. **Independent evolution**: Clients and servers can evolve independently as long as they maintain compatibility with the protocol, allowing for innovation on both sides without requiring lockstep development.\\ \newline
In a typical MCP implementation, the client component is integrated with an AI application (such as an agent framework or assistant interface) and manages connections to one or more MCP servers. Each server provides access to specific data sources or tools, exposing their capabilities through standardized MCP interfaces.\\ \newline
Communication between clients and servers follows a request-response pattern using JSON-RPC, a lightweight remote procedure call protocol that uses JSON for data encoding. This approach provides a language-agnostic interface that can be implemented across diverse platforms and environments.\\
\subsubsection{Standardized Primitives}
MCP defines a set of core primitives—standardized operations and data structures—that provide the building blocks for context management. These primitives are divided into server-side and client-side categories:\\ \newline
**Server-side primitives** define the capabilities that MCP servers expose to clients:\\ \newline
1. **Prompts**: Pre-defined instructions or templates that an AI can request. These may include situational guidance, formatting instructions, or specialized procedures for particular tasks. Prompts help establish consistent behavior patterns and can reduce the need to repeatedly include standard instructions in the model's context.\\ \newline
2. **Resources**: Structured data or documents that can be sent to the AI model. Resources may include knowledge base articles, database records, configuration information, or other contextual data that informs model responses. The resource primitive includes metadata that helps the client determine how and when to incorporate the resource into the model's context.\\ \newline
3. **Tools**: Executable functions that the AI can invoke to perform actions or retrieve information. Tools may include API calls, database queries, file operations, or other functional capabilities. Each tool is defined with a name, description, parameter schema, and return type, enabling the model to understand its purpose and usage.\\ \newline
**Client-side primitives** define capabilities that MCP clients provide to servers:\\ \newline
1. **Roots**: Entry points that give servers access to specific data realms on the client side. Roots establish permission boundaries, allowing clients to control which data each server can access. For example, a client might provide a file system root that grants access to a specific directory but not the entire file system.\\ \newline
2. **Sampling**: A mechanism that allows an MCP server to request the AI model to generate a completion. This powerful capability enables servers to leverage the model's reasoning abilities as part of fulfilling a request, essentially allowing the server to "ask back" when additional context or decisions are needed.\\ \newline
These primitives provide a comprehensive foundation for context management while maintaining a relatively simple interface. By combining these basic operations in different ways, MCP can support sophisticated context management patterns without requiring complex protocol extensions.\\
\subsubsection{Communication Mechanisms and Message Formats}
MCP communication is structured around a well-defined message format based on JSON-RPC 2.0, a lightweight remote procedure call protocol. This format provides a standardized way to encode requests, responses, and notifications between clients and servers.
The basic structure of an MCP request includes:

\begin{verbatim}

{
  "jsonrpc": "2.0",
  "id": "request-123",
  "method": "resource.get",
  "params": {}
    "resource_id": "document-456",
    "format": "markdown"
}
And a corresponding response:
{
  "jsonrpc": "2.0",
  "id": "request-123",
  "result": {}
    "content": "# Document Title
    This is the content of the requested document...",
    "metadata": {
      "author": "Jane Smith",
      "created_at": "2025-03-15T14:30:00Z"
    }
}

\end{verbatim}

This structured format ensures consistent interpretation across different implementations while remaining human-readable for debugging and development purposes.\\ \newline
MCP defines specific methods for each primitive type, such as:\\ \newline
- `prompt.list` and `prompt.get` for accessing prompts\\ \newline
- `resource.list` and `resource.get` for retrieving resources\\ \newline
- `tool.list`, `tool.describe`, and `tool.execute` for discovering and using tools\\ \newline
- `root.list` and `root.describe` for exploring available roots\\ \newline
- `sample.generate` for requesting model completions\\ \newline
Each method has defined parameter and return schemas that ensure consistent behavior across implementations. The protocol also includes mechanisms for error handling, allowing servers to return structured error information when requests cannot be fulfilled.\\ \newline
Beyond the basic request-response pattern, MCP supports more complex interaction patterns through features like:\\ \newline
1. Streaming responses: For large resources or long-running tool executions, servers can stream results incrementally rather than waiting for complete execution.\\ \newline
2. Notifications: One-way messages that don't require responses, useful for events or updates that don't need acknowledgment.\\ \newline
3. Batched requests: Multiple requests combined into a single message to reduce communication overhead.\\ \newline
These communication mechanisms provide the flexibility needed to support diverse context management scenarios while maintaining protocol simplicity and consistency.\\
\subsection{Implementation Approaches}
\subsubsection{Server-Side Implementation Patterns}
Implementing an MCP server involves exposing data sources or tools through the standardized MCP interface. Several common implementation patterns have emerged:\\ \newline
Adapter pattern: Many MCP servers function as adapters that translate between MCP's standardized interface and existing APIs or data sources. This pattern is particularly common for integrating with established systems like document repositories, databases, or SaaS platforms. The adapter handles authentication, request formatting, and response translation, presenting a consistent MCP interface regardless of the underlying system's native API.\\ \newline
Composite pattern: Some MCP servers aggregate multiple data sources or tools behind a unified interface. These composite servers may provide integrated search across multiple repositories, coordinated access to related tools, or domain-specific combinations of capabilities. This pattern simplifies client configuration by reducing the number of server connections needed while providing more sophisticated integration capabilities.\\ \newline
Proxy pattern: In environments with strict security requirements, MCP servers may act as proxies that enforce access controls, audit usage, or transform requests and responses for compliance purposes. These proxy servers sit between clients and the actual data sources or tools, providing an additional layer of control and monitoring.\\ \newline
Embedded pattern: For simple use cases or development scenarios, MCP servers may be embedded directly within applications rather than deployed as separate services. This pattern reduces deployment complexity but may limit scalability and reuse across multiple clients.\\ \newline
Regardless of the implementation pattern, effective MCP servers typically share several characteristics:\\ \newline
1. Clear capability boundaries: Well-defined scope of what the server provides, avoiding overly broad or ambiguous functionality.\\ \newline
2. Comprehensive metadata: Detailed descriptions of available prompts, resources, and tools to enable effective discovery and usage.\\ \newline
3. Robust error handling: Clear error messages and appropriate status codes when requests cannot be fulfilled.\\ \newline
4. Efficient resource management: Appropriate caching, connection pooling, and resource cleanup to ensure performance and reliability.\\ \newline
5. Security-conscious design: Careful attention to authentication, authorization, and data protection throughout the implementation.\\ \newline
The open-source ecosystem around MCP has produced reference implementations and libraries for common server patterns, reducing the implementation effort required for new integrations.\\
\subsubsection{Client-Side Integration Strategies}
On the client side, integrating MCP involves connecting AI models with MCP servers and managing the flow of contextual information. Several integration strategies have proven effective:\\ \newline
Framework integration: Many agent frameworks and AI development platforms now include built-in MCP client capabilities. These integrations typically provide high-level abstractions that simplify server discovery, connection management, and context incorporation. Developers using these frameworks can leverage MCP capabilities with minimal additional code, often through declarative configuration rather than explicit implementation.\\ \newline
SDK-based integration: For applications not using MCP-enabled frameworks, software development kits (SDKs) provide libraries and utilities for implementing MCP client functionality. These SDKs, available for multiple programming languages, handle the details of protocol implementation while providing idiomatic interfaces for each language ecosystem. This approach requires more implementation effort than framework integration but offers greater flexibility and control.\\ \newline
Custom implementation: Some specialized applications implement MCP client functionality directly, particularly when operating in constrained environments or when requiring optimizations not available in existing libraries. While this approach requires the most development effort, it allows for precise tailoring to specific requirements and may enable performance optimizations not possible with general-purpose implementations.\\ \newline
Regardless of the integration strategy, effective MCP clients typically implement several key functions:\\ \newline
1. Server discovery and connection management: Finding available MCP servers, establishing connections, and handling authentication and reconnection as needed.\\ \newline
2. Context selection and prioritization: Determining which contextual information is most relevant for the current interaction and managing context window constraints.\\ \newline
3. Tool invocation and result processing: Translating model outputs into tool calls and incorporating tool results back into the model's context.\\ \newline
4. Error handling and fallback strategies: Gracefully handling server unavailability, request failures, or other exceptional conditions.\\ \newline
5. Context persistence: Maintaining relevant context across multiple interactions or sessions, potentially using client-side storage for efficiency.\\ \newline
The choice of integration strategy depends on factors including the application's architecture, performance requirements, existing technology stack, and development resources. The flexibility of MCP allows for implementation across a wide range of scenarios, from simple prototypes to enterprise-scale deployments.\\
\subsubsection{Security and Authentication Considerations}
Security is a fundamental consideration in MCP implementation, particularly as the protocol often provides access to sensitive data or powerful capabilities. Several security dimensions require attention:\\ \newline
Authentication and authorization: MCP implementations must verify the identity of connecting clients and servers (authentication) and determine what operations they are permitted to perform (authorization). The protocol supports various authentication mechanisms, from simple API keys for development environments to more robust OAuth flows or mutual TLS for production systems. Authorization is typically implemented through capability-based security models, where access is granted to specific resources or tools rather than through broad permissions.\\ \newline
Data protection: Information transmitted through MCP may include sensitive data that requires protection both in transit and at rest. The protocol mandates TLS encryption for all communications, ensuring transport-level security. Additionally, implementations may employ end-to-end encryption for particularly sensitive data, ensuring that even the transport infrastructure cannot access unencrypted content.\\ \newline
Isolation and sandboxing: Tool execution in MCP servers presents potential security risks, particularly for operations that interact with underlying systems. Robust implementations employ sandboxing techniques to isolate tool execution environments, limiting the potential impact of malicious inputs or unexpected behaviors.\\ \newline
Audit and monitoring: Comprehensive logging and monitoring are essential for security management in MCP deployments. This includes recording access patterns, tracking resource usage, and monitoring for anomalous behaviors that might indicate security issues. The protocol includes standardized fields for request identification and correlation, facilitating effective audit trails across system boundaries.\\ \newline
Permission models: MCP implements fine-grained permission models through the root primitive and capability-based access controls. These models allow clients to precisely control what servers can access and what operations they can perform, following the principle of least privilege. For example, a file system root might grant access only to specific directories, while a tool might be restricted to particular parameter ranges or operation types.\\ \newline
These security considerations are not merely implementation details but fundamental aspects of the protocol design. By incorporating security and authentication as core protocol elements, MCP ensures that implementations can be both powerful and secure, addressing the critical requirements of enterprise and sensitive consumer applications.\\
\subsection{MCP as a Solution to the Disconnected Models Problem}
\subsubsection{Addressing Context Discontinuity Across Agent Actions}
The "disconnected models problem" identified by Krishnan (2025) and emphasized by Microsoft's Sam Schillace represents one of the most significant limitations in current agent architectures. As Schillace noted, "To be autonomous you have to carry context through a bunch of actions, but the models are very disconnected and don't have continuity the way we do." This discontinuity manifests as agents losing track of previous interactions, forgetting relevant information, or failing to maintain coherent reasoning across multiple steps.\\ \newline
MCP directly addresses this challenge through several mechanisms:\\ \newline
1. Standardized context storage and retrieval: MCP provides consistent patterns for storing contextual information externally and retrieving it when needed. This allows context to persist beyond the limitations of model context windows and across multiple interactions.\\ \newline
2. Context-aware tool execution: When agents invoke tools through MCP, the tool execution can incorporate relevant context from previous interactions. This ensures that actions build upon previous steps rather than occurring in isolation.\\ \newline
3. Shared context across agent boundaries: In multi-agent systems, MCP enables different agents to access the same contextual information, ensuring consistent understanding and coordinated action even when tasks span multiple specialized agents.\\ \newline
4. Contextual prioritization: MCP's resource primitive includes metadata that helps determine the relevance and importance of different contextual elements, enabling more intelligent management of limited context space.\\ \newline
These capabilities directly address the discontinuity problem, enabling agents to maintain coherent context across extended interaction sequences and complex tasks. By providing external context management that complements the model's internal processing, MCP overcomes the inherent limitations of context windows while preserving the flexibility and reasoning capabilities of large language models.\\
\subsubsection{Enabling Persistent Memory and Context Awareness}
Beyond addressing immediate context discontinuity, MCP enables more sophisticated forms of persistent memory and context awareness that enhance agent capabilities:\\ \newline
Episodic memory stores records of specific interactions or experiences, allowing agents to recall and reference previous encounters. MCP supports episodic memory through resource primitives that can capture interaction histories in structured formats, making them available for future reference. This capability is particularly valuable for maintaining relationship context in assistant applications or tracking progress in extended problem-solving scenarios.\\ \newline
Semantic memory organizes conceptual knowledge and learned information, enabling agents to build and refine their understanding over time. MCP facilitates semantic memory through resources that represent structured knowledge, potentially including knowledge graphs, concept maps, or other semantic representations. By externalizing this knowledge, MCP allows it to persist and evolve independently of individual interactions.\\ \newline
Procedural memory captures action sequences or skills that can be applied across multiple situations. MCP supports procedural memory through prompt primitives that encode standard procedures or approaches, making them available for consistent application. This capability helps agents maintain consistent methodologies across interactions while adapting to specific circumstances.\\ \newline
Working memory maintains task-relevant information during ongoing activities. MCP enhances working memory through tool primitives that can track state information, maintain intermediate results, or monitor progress toward goals. This external augmentation of working memory helps agents manage complex tasks that exceed their internal context capacity.\\ \newline
By supporting these diverse memory types through standardized interfaces, MCP enables more human-like context awareness and memory utilization. Agents can remember relevant past interactions, apply accumulated knowledge, follow consistent procedures, and maintain awareness of ongoing tasks—capabilities that are essential for truly autonomous operation.\\
\subsubsection{Comparative Analysis with Previous Approaches}
Compared to previous approaches to context management, MCP offers several distinctive advantages:\\ \newline
Standardization vs. custom implementations: Earlier approaches typically implemented context management through custom, application-specific mechanisms with limited interoperability. MCP establishes standard interfaces and communication patterns that work across different applications, models, and environments. This standardization reduces implementation complexity, enables component reuse, and facilitates ecosystem development.\\ \newline
External vs. embedded context: Many previous approaches attempted to embed all relevant context within the model's input, leading to context window limitations and inefficient use of model capacity. MCP externalizes context management, storing information outside the model and retrieving it selectively based on relevance. This approach overcomes context window constraints while enabling more sophisticated context prioritization.\\ \newline
Structured vs. unstructured context: Earlier approaches often treated context as unstructured text, limiting the model's ability to distinguish between different types of contextual information. MCP provides structured primitives (prompts, resources, tools) with defined semantics, enabling more precise context management and utilization. This structure helps models understand the nature and purpose of different contextual elements.\\ \newline
Persistent vs. ephemeral context: Previous approaches frequently struggled with maintaining context across multiple interactions or sessions, leading to discontinuity in extended tasks. MCP provides mechanisms for persistent context storage and retrieval, enabling continuity across interaction boundaries. This persistence is essential for tasks requiring extended reasoning chains or collaborative work.\\ \newline
Interoperable vs. isolated tools: Earlier tool-using agent frameworks typically implemented custom, framework-specific approaches to tool integration. MCP standardizes tool definitions, invocation patterns, and result handling, enabling tools to be shared across different frameworks and environments. This interoperability expands the range of capabilities available to agents while reducing implementation overhead.\\ \newline
These comparative advantages position MCP as a significant advancement over previous context management approaches. By addressing the fundamental limitations that have constrained agent capabilities, MCP enables more effective autonomous operation and collaboration in multi-agent systems.\\ \newline
The Model Context Protocol represents a critical foundation for the multi-agent system architecture described in the following section. By providing standardized mechanisms for context management and sharing, MCP enables more effective coordination and collaboration among diverse agent types, addressing one of the key challenges identified in previous research.\\ \newline

\section{Multi-Agent System Architecture with MCP}
The integration of Model Context Protocol into multi-agent systems creates new possibilities for agent coordination, context sharing, and collaborative problem-solving. This section presents a reference architecture for MCP-enabled multi-agent systems, detailing the components, information flows, and coordination mechanisms that enable effective collaboration among diverse agent types.\\
\subsection{Reference Architecture}
\subsubsection{System Components and Their Relationships}
A comprehensive multi-agent system architecture with MCP integration consists of several key components that work together to enable sophisticated agent collaboration:\\ \newline
Agent Runtime Environment: The foundation of the architecture is the runtime environment that hosts individual agents. This environment provides the computational resources, scheduling mechanisms, and lifecycle management needed for agent operation. In MCP-enabled systems, the runtime environment typically includes built-in MCP client capabilities that allow agents to connect with MCP servers and access external context.\\ \newline
Agent Instances: Individual agents within the system represent specialized capabilities or roles. Each agent typically consists of:\\ \newline
- A core reasoning engine (often based on an LLM)\\ \newline
- Task-specific knowledge and skills\\ \newline
- MCP client integration for context access\\ \newline
- Communication interfaces for inter-agent interaction\\ \newline
- Local working memory for immediate task context\\ \newline
Context Management Layer: This critical layer implements MCP to provide standardized access to external context sources. It includes:\\ \newline
- MCP servers for various data sources and tools\\ \newline
- Context storage systems for persistent memory\\ \newline
- Context retrieval and prioritization mechanisms\\ \newline
- Access control and permission management\\ \newline
Coordination Framework: This component manages the interactions between agents, implementing protocols for task allocation, progress monitoring, and conflict resolution. The coordination framework may operate through direct agent-to-agent communication, centralized coordination services, or hybrid approaches depending on system requirements.\\ \newline
External Integration Layer: This layer connects the multi-agent system with external services, data sources, and user interfaces. It typically includes:\\ \newline
- API gateways for external service access\\ \newline
- User interaction channels\\ \newline
- Monitoring and observability tools\\ \newline
- Integration with enterprise systems\\ \newline
Security and Governance Layer: This cross-cutting component implements security controls, audit mechanisms, and governance policies across the system. It ensures appropriate authentication, authorization, data protection, and compliance with organizational requirements.\\ \newline
These components are arranged in a layered architecture that balances separation of concerns with efficient interaction paths. The architecture supports both vertical integration (agents accessing context and external services) and horizontal integration (agents communicating with each other for coordination).\\
\subsubsection{Information Flow and Control Mechanisms}
Information flows through this architecture along several key pathways that enable effective multi-agent operation:\\ \newline
Context Access Flows: Agents retrieve contextual information from MCP servers through standardized request-response patterns. These flows may include:\\ \newline
- Direct resource requests for specific information\\ \newline
- Search or query operations to find relevant context\\ \newline
- Subscription patterns for updates to changing context\\ \newline
- Batch retrieval for initializing agent state\\ \newline
Inter-Agent Communication Flows: Agents exchange information with each other through structured communication channels. These flows typically follow defined protocols for:\\ \newline
- Task delegation and assignment\\ \newline
- Progress reporting and status updates\\ \newline
- Knowledge sharing and question answering\\ \newline
- Conflict resolution and negotiation\\ \newline
External Interaction Flows: The system interacts with external entities (users, services, systems) through integration interfaces. These flows include:\\ \newline
- User instructions and queries\\ \newline
- System responses and outputs\\ \newline
- Service requests and responses\\ \newline
- Monitoring and telemetry data\\ \newline
Control Flows: The coordination framework implements control mechanisms that manage agent behavior and system operation. These include:\\ \newline
- Task scheduling and prioritization\\ \newline
- Resource allocation and load balancing\\ \newline
- Error handling and recovery procedures\\ \newline
- System configuration and adaptation\\ \newline
These information flows are implemented through a combination of synchronous and asynchronous communication patterns, with appropriate mechanisms for reliability, ordering, and delivery guarantees. The architecture employs several control mechanisms to ensure effective operation:\\ \newline
1. Message-based control: Coordination through structured messages that trigger specific agent behaviors or responses.\\ \newline
2. Policy-based control: Governance through defined policies that constrain agent actions and decision-making.\\ \newline
3. Incentive-based control: Influence through reward systems that encourage desired agent behaviors.\\ \newline
4. Constraint-based control: Limitation through explicit constraints on resources, actions, or outcomes.\\ \newline
These control mechanisms work together to balance agent autonomy with system-level coordination, enabling effective collective behavior while preserving individual agent capabilities.\\
\subsubsection{Integration Patterns for Different Agent Types}
Multi-agent systems typically incorporate diverse agent types with varying capabilities, implementation approaches, and specializations. The reference architecture supports several integration patterns that accommodate this diversity:\\ \newline
LLM-based Agents: These agents use large language models as their primary reasoning engine. They integrate with MCP through client libraries that manage context retrieval, tool invocation, and result incorporation. The architecture supports various LLM deployment models, including:\\ \newline
- Local model deployment for latency-sensitive or private applications\\ \newline
- Cloud-based model access for resource-intensive operations\\ \newline
- Hybrid approaches that combine local and remote models based on task requirements\\ \newline
Specialized Function Agents: These purpose-built agents implement specific capabilities through dedicated algorithms rather than general-purpose LLMs. They may include:\\ \newline
- Planning agents using classical planning algorithms\\ \newline
- Optimization agents employing mathematical programming techniques\\ \newline
- Perception agents utilizing computer vision or speech recognition\\ \newline
- Analytics agents implementing statistical or machine learning methods\\ \newline
These specialized agents integrate with MCP through adapters that translate between their internal representations and MCP's standardized interfaces, allowing them to participate in the broader multi-agent ecosystem while leveraging their specialized capabilities.\\ \newline
Legacy System Agents: These agents wrap existing systems or services to incorporate them into the multi-agent architecture. They typically implement:\\ \newline
- Protocol translation between legacy APIs and MCP\\ \newline
- Context mapping between system-specific and standardized formats\\ \newline
- State management to maintain consistency across interactions\\ \newline
This integration pattern enables gradual migration from existing systems to MCP-based architectures, preserving investments in established capabilities while enabling new collaborative possibilities.\\ \newline
Human-in-the-loop Agents: These agents incorporate human judgment and expertise within the multi-agent workflow. They implement:\\ \newline
- Interfaces for human input and review\\ \newline
- Delegation mechanisms for routing decisions to appropriate humans\\ \newline
- Learning capabilities to improve from human feedback\\ \newline
This pattern ensures appropriate human oversight for sensitive decisions while maintaining the efficiency benefits of automation for routine operations.\\ \newline
The reference architecture supports flexible composition of these diverse agent types, enabling systems that combine the strengths of different approaches. Integration patterns include:\\ \newline
1. Hierarchical integration: Organizing agents in management hierarchies with specialized agents reporting to coordinator agents.\\ \newline
2. Peer-to-peer integration: Enabling direct collaboration between agents with complementary capabilities.\\ \newline
3. Service-oriented integration: Structuring agents as service providers that can be discovered and utilized by other agents.\\ \newline
4. Event-driven integration: Connecting agents through event streams that trigger appropriate responses based on system state changes.\\ \newline
These integration patterns provide the flexibility needed to address diverse use cases while maintaining architectural coherence and interoperability.\\
\subsection{Context Sharing Between Agents}
\subsubsection{Mechanisms for Inter-Agent Context Exchange}
Effective collaboration in multi-agent systems requires sophisticated mechanisms for sharing contextual information between agents. MCP enables several approaches to inter-agent context exchange:\\ \newline
Shared Context Repositories: Centralized or distributed repositories that store contextual information accessible to multiple agents. These repositories implement MCP server interfaces, allowing any agent with appropriate permissions to retrieve shared context. This approach provides a consistent source of truth while reducing duplication and synchronization challenges. Implementation patterns include:\\ \newline
- Document stores for unstructured or semi-structured information\\ \newline
- Knowledge graphs for semantically linked data\\ \newline
- Vector databases for embedding-based retrieval\\ \newline
- Time-series databases for sequential or temporal information\\ \newline
Direct Context Transfer: Point-to-point sharing of contextual information between agents. This approach leverages MCP's standardized resource format to package and transfer context directly from one agent to another as part of task delegation or collaborative workflows. Direct transfer is particularly valuable for:\\ \newline
- Passing task-specific context during handoffs\\ \newline
- Sharing private or sensitive information with limited distribution\\ \newline
- Reducing latency for time-critical operations\\ \newline
Context Broadcasting: Distribution of contextual updates to multiple agents simultaneously. This pattern uses publish-subscribe mechanisms to notify relevant agents of context changes, enabling coordinated responses to new information. Broadcasting is effective for:\\ \newline
- Announcing environmental changes that affect multiple agents\\ \newline
- Sharing discoveries or insights that benefit collective knowledge\\ \newline
- Coordinating responses to emerging situations or opportunities\\ \newline
Contextual Annotations: Enrichment of shared artifacts with agent-specific perspectives or insights. This approach allows agents to layer their interpretations onto common information objects, creating richer collective understanding. Annotations are valuable for:\\ \newline
- Adding specialized expertise to general information\\ \newline
- Recording reasoning processes and justifications\\ \newline
- Building collective intelligence through diverse perspectives\\ \newline
These mechanisms are implemented through MCP's standardized primitives, ensuring consistent interpretation and utilization across different agent types. The architecture supports flexible combination of these approaches based on specific collaboration requirements and operational constraints.\\
\subsubsection{Context Prioritization and Relevance Determination}
As multi-agent systems accumulate contextual information from diverse sources, effective prioritization becomes essential for managing limited attention and processing resources. The reference architecture implements several approaches to context prioritization and relevance determination:\\ \newline
Relevance Scoring: Computational assessment of context relevance based on multiple factors:\\ \newline
- Semantic similarity to current tasks or queries\\ \newline
- Temporal recency and frequency of access\\ \newline
- Source authority and reliability\\ \newline
- Explicit priority indicators from users or other agents\\ \newline
These scores help agents allocate limited context windows to the most valuable information, ensuring that critical context is available when needed.\\ \newline
Attention Mechanisms: Selective focus on particular contextual elements based on current goals and activities. These mechanisms implement:\\ \newline
- Dynamic weighting of different context types based on task phase\\ \newline
- Spotlight effects that emphasize immediately relevant information\\ \newline
- Background awareness that maintains access to potentially relevant context\\ \newline
- Interruption protocols for urgent or high-priority information\\ \newline
Attention mechanisms help agents balance focus on current tasks with awareness of changing circumstances or new information.\\ \newline
Context Summarization: Compression of detailed contextual information into more compact representations that preserve essential meaning. Summarization approaches include:\\ \newline
- Extractive methods that select key statements or facts\\ \newline
- Abstractive methods that reformulate information in more concise terms\\ \newline
- Hierarchical summarization that provides multiple detail levels\\ \newline
- Multi-perspective summarization that captures diverse viewpoints\\ \newline
These techniques help agents manage information overload while maintaining access to comprehensive context when needed.\\ \newline
Decay and Forgetting Models: Systematic approaches to reducing the prominence of less relevant or outdated information. These models implement:\\ \newline
- Time-based decay functions that gradually reduce priority\\ \newline
- Usage-based retention that preserves frequently accessed information\\ \newline
- Importance-weighted forgetting that preserves critical context longer\\ \newline
- Explicit deprecation of superseded or invalidated information\\ \newline
These approaches ensure that context management remains sustainable as the system accumulates information over time, preventing performance degradation from context overload.\\ \newline
The architecture integrates these prioritization mechanisms with MCP's resource primitive, using metadata fields to capture relevance scores, attention signals, summarization status, and retention policies. This integration ensures that prioritization information persists across agent boundaries and interaction sessions.\\
\subsubsection{Handling Conflicting Contextual Information}
In multi-agent systems with diverse information sources and perspectives, conflicting contextual information is inevitable. The reference architecture implements several strategies for detecting and resolving such conflicts:\\ \newline
Conflict Detection Mechanisms: Systematic approaches to identifying potential contradictions or inconsistencies in contextual information:\\ \newline
- Logical contradiction detection through formal reasoning\\ \newline
- Statistical inconsistency identification through pattern analysis\\ \newline
- Temporal sequence validation to identify causality violations\\ \newline
- Source reliability assessment to flag potential misinformation\\ \newline
These mechanisms help agents recognize when contextual information requires reconciliation before it can be effectively utilized.\\ \newline
Resolution Strategies: Structured approaches to resolving identified conflicts:\\ \newline
- Evidence-based adjudication that evaluates supporting information\\ \newline
- Source prioritization based on authority, recency, or reliability\\ \newline
- Consensus formation through multi-agent deliberation\\ \newline
- Uncertainty representation that maintains alternative possibilities\\ \newline
- Human escalation for conflicts requiring judgment or expertise\\ \newline
These strategies ensure that agents can continue effective operation even when faced with contradictory information.\\ \newline
Conflict Representation: Methods for capturing and communicating about conflicts:\\ \newline
- Explicit annotation of conflicting information with confidence levels\\ \newline
- Structured representation of alternative perspectives or interpretations\\ \newline
- Provenance tracking to identify information sources and derivation chains\\ \newline
- Uncertainty quantification to express confidence in different possibilities\\ \newline
These representation approaches enable agents to reason about conflicts rather than simply selecting one interpretation, preserving important nuance and complexity.\\ \newline
Learning from Conflicts: Mechanisms for improving conflict handling over time:\\ \newline
- Pattern recognition to identify recurring conflict types\\ \newline
- Resolution outcome tracking to evaluate strategy effectiveness\\ \newline
- Knowledge refinement based on conflict resolution results\\ \newline
- Adaptive strategy selection based on historical performance\\ \newline
These learning mechanisms help the system become more effective at handling conflicts through experience, reducing the need for human intervention over time.\\ \newline
By integrating these conflict handling capabilities with MCP's standardized context management, the architecture ensures consistent treatment of conflicting information across agent boundaries. This consistency is essential for maintaining coherent collective behavior in the face of incomplete or contradictory information.\\
\subsection{Coordination Protocols}
\subsubsection{Task Decomposition and Allocation}
Effective multi-agent collaboration requires sophisticated mechanisms for breaking complex tasks into manageable components and assigning them to appropriate agents. The reference architecture implements several coordination protocols for task decomposition and allocation:\\ \newline
Hierarchical Task Networks (HTN): Structured representation of tasks as hierarchies of subtasks with defined decomposition methods. This approach enables:\\ \newline
- Systematic breakdown of complex goals into actionable steps\\ \newline
- Clear representation of task dependencies and constraints\\ \newline
- Flexible adaptation through alternative decomposition methods\\ \newline
- Reuse of common task patterns across different scenarios\\ \newline
HTN-based coordination is particularly effective for domains with well-understood task structures and clear decomposition patterns.\\ \newline
Contract Net Protocol: Market-based approach where tasks are announced to potential performers who bid based on their capabilities and availability. This protocol implements:\\ \newline
- Task announcement with requirements and constraints\\ \newline
- Bid submission from capable and available agents\\ \newline
- Bid evaluation based on multiple criteria\\ \newline
- Contract award and commitment tracking\\ \newline
Contract Net is valuable for dynamic environments where agent capabilities and workloads vary over time, enabling adaptive task allocation based on current system state.\\ \newline
Role-Based Assignment: Allocation based on predefined roles with associated responsibilities and capabilities. This approach provides:\\ \newline
- Clear division of labor based on agent specialization\\ \newline
- Simplified coordination through standardized role interfaces\\ \newline
- Predictable interaction patterns between complementary roles\\ \newline
- Straightforward substitution when agents become unavailable\\ \newline
Role-based coordination is effective for domains with stable task patterns and clear functional specializations.\\ \newline
Emergent Assignment: Self-organizing approach where agents select tasks based on local information and simple rules. This method enables:\\ \newline
- Decentralized coordination without central control points\\ \newline
- Adaptive response to changing conditions\\ \newline
- Robust operation in the face of agent failures\\ \newline
- Scalability to large agent collectives\\ \newline
Emergent coordination is particularly valuable for systems with numerous similar agents and relatively simple task structures.\\ \newline
These protocols are implemented through MCP-enabled communication channels, with task representations, bids, and assignments exchanged as structured resources. The architecture supports hybrid approaches that combine elements of different protocols based on specific domain requirements and operational constraints.\\
\subsubsection{Progress Monitoring and Synchronization}
Coordinating multiple agents working on related tasks requires effective mechanisms for tracking progress, identifying bottlenecks, and maintaining appropriate synchronization. The reference architecture implements several approaches to progress monitoring and synchronization:\\ \newline
Status Reporting Framework: Standardized mechanism for agents to report their progress, challenges, and resource utilization. This framework includes:\\ \newline
- Structured status reports with completion percentages and milestones\\ \newline
- Exception reporting for unexpected obstacles or failures\\ \newline
- Resource consumption tracking for performance optimization\\ \newline
- Estimated completion time updates based on current progress\\ \newline
These reports enable system-wide awareness of task status and potential issues requiring attention.\\ \newline
Dependency Management: Explicit tracking and enforcement of task dependencies to ensure proper execution order. This capability implements:\\ \newline
- Prerequisite verification before task initiation\\ \newline
- Notification mechanisms when dependencies are satisfied\\ \newline
- Blocking and unblocking of dependent tasks based on status\\ \newline
- Critical path analysis to identify bottleneck tasks\\ \newline
Dependency management ensures that tasks are executed in appropriate sequences while maximizing parallel execution where possible.\\ \newline
Synchronization Points: Defined coordination points where multiple agents align their activities. These include:\\ \newline
- Barriers that ensure all agents reach a certain stage before proceeding\\ \newline
- Rendezvous points for exchanging intermediate results\\ \newline
- Checkpoints for verifying system consistency\\ \newline
- Coordination windows for time-sensitive operations\\ \newline
Synchronization points help maintain coherent collective behavior while allowing individual agents to operate autonomously between coordination events.\\ \newline
Adaptive Scheduling: Dynamic adjustment of task timing and resource allocation based on progress monitoring. This approach enables:\\ \newline
- Reallocation of resources to bottleneck tasks\\ \newline
- Acceleration of critical path activities\\ \newline
- Deferral of non-critical tasks when resources are constrained\\ \newline
- Parallel execution of independent tasks for efficiency\\ \newline
Adaptive scheduling helps the system maintain optimal performance despite variations in task complexity and resource availability.\\ \newline
These monitoring and synchronization mechanisms are implemented through MCP's tool and resource primitives, with status information shared through standardized formats that enable consistent interpretation across agent boundaries. The architecture supports configurable monitoring granularity, allowing systems to balance coordination overhead with visibility requirements.\\
\subsubsection{Failure Recovery and Exception Handling}
Robust multi-agent systems must effectively handle failures, exceptions, and unexpected situations that inevitably arise during operation. The reference architecture implements several strategies for failure recovery and exception handling:\\ \newline
Failure Detection Mechanisms: Systematic approaches to identifying when agents or tasks have failed or encountered problems:\\ \newline
- Heartbeat monitoring to detect agent unresponsiveness\\ \newline
- Progress thresholds to identify stalled tasks\\ \newline
- Output validation to detect incorrect or unexpected results\\ \newline
- Resource monitoring to identify performance degradation\\ \newline
These mechanisms ensure timely detection of issues that require intervention or adaptation.\\ \newline
Recovery Strategies: Structured approaches to addressing identified failures:\\ \newline
- Retry mechanisms with configurable policies (immediate, delayed, limited)\\ \newline
- Alternative method selection when primary approaches fail\\ \newline
- Task reassignment to different agents when original assignees fail\\ \newline
- Graceful degradation to maintain partial functionality\\ \newline
- Checkpoint-based rollback to consistent states\\ \newline
These strategies enable the system to continue operation despite individual component failures.\\ \newline
Exception Escalation Protocols: Defined pathways for handling exceptions that cannot be resolved at the original level:\\ \newline
- Hierarchical escalation to supervisor agents or coordinators\\ \newline
- Peer consultation to leverage collective problem-solving\\ \newline
- Human escalation for exceptions requiring judgment or expertise\\ \newline
- System-wide alerts for critical or widespread issues\\ \newline
These protocols ensure that exceptions are handled at appropriate levels with necessary resources and authority.\\ \newline
Learning from Failures: Mechanisms for improving failure handling over time:\\ \newline
- Failure pattern recognition to identify recurring issues\\ \newline
- Root cause analysis to address underlying problems\\ \newline
- Recovery strategy effectiveness tracking\\ \newline
- Preventive measure implementation based on historical failures\\ \newline
These learning mechanisms help the system become more robust through experience, reducing the frequency and impact of failures over time.\\ \newline
The architecture integrates these failure recovery and exception handling capabilities with MCP's standardized communication and context management, ensuring consistent treatment of failures across agent boundaries. This integration is essential for maintaining reliable operation in complex, dynamic environments where perfect performance cannot be guaranteed.\\
\subsection{Resource Management}
\subsubsection{Computational Resource Allocation}
Multi-agent systems must effectively manage computational resources to ensure optimal performance across diverse and potentially competing agent activities. The reference architecture implements several approaches to computational resource allocation:\\ \newline
Priority-Based Allocation: Assignment of computational resources based on task priority and criticality. This approach ensures that high-value activities receive sufficient resources even under constrained conditions. Implementation mechanisms include:\\ \newline
- Priority inheritance to elevate resource allocation for dependent tasks\\ \newline
- Dynamic priority adjustment based on deadlines and progress\\ \newline
- Preemption capabilities for urgent high-priority tasks\\ \newline
- Guaranteed minimum allocations for essential system functions\\ \newline
Adaptive Scaling: Dynamic adjustment of resource allocation based on workload and performance requirements. This capability enables efficient resource utilization while maintaining responsive performance. Implementation approaches include:\\ \newline
- Elastic resource pools that expand and contract based on demand\\ \newline
- Load-based scaling of agent instances and supporting services\\ \newline
- Performance-driven resource adjustment to maintain SLAs\\ \newline
- Predictive scaling based on historical patterns and upcoming tasks\\ \newline
Resource Reservation: Pre-allocation of resources for planned or anticipated activities. This approach ensures resource availability for critical operations while reducing contention and allocation delays. Implementation mechanisms include:\\ \newline
- Time-based reservations for scheduled activities\\ \newline
- Capacity reservations for peak load periods\\ \newline
- Resource set-asides for high-priority agent types\\ \newline
- Graduated reservation release for unused capacity\\ \newline
Fairness Enforcement: Mechanisms to prevent resource monopolization and ensure equitable access across agents. This capability is essential for maintaining balanced system performance and preventing starvation of lower-priority activities. Implementation approaches include:\\ \newline
- Resource usage quotas and rate limiting\\ \newline
- Fair queuing for resource requests\\ \newline
- Proportional allocation based on agent importance\\ \newline
- Anti-starvation guarantees for all agent types\\ \newline
These resource allocation mechanisms are implemented through integration with underlying infrastructure (cloud platforms, container orchestration, etc.) and exposed to agents through MCP-compatible interfaces. This integration enables agents to make informed decisions about resource utilization while maintaining system-wide optimization.\\
\subsubsection{Information Resource Sharing}
Beyond computational resources, multi-agent systems must effectively manage and share information resources that enable agent operation. The reference architecture implements several approaches to information resource sharing:\\ \newline
Knowledge Base Management: Centralized or distributed repositories of domain knowledge, reference information, and accumulated insights. These knowledge bases implement MCP server interfaces, allowing agents to query and contribute information through standardized mechanisms. Implementation approaches include:\\ \newline
- Semantic organization for efficient retrieval\\ \newline
- Version control for evolving information\\ \newline
- Authority levels for different information sources\\ \newline
- Contribution workflows with quality control\\ \newline
Shared Memory Systems: Collective information spaces that maintain system state, intermediate results, and coordination information. These systems enable efficient information sharing without direct agent-to-agent communication. Implementation patterns include:\\ \newline
- Blackboard systems for opportunistic problem-solving\\ \newline
- Tuple spaces for associative information sharing\\ \newline
- Distributed caches for performance-critical information\\ \newline
- Event streams for temporal information sequences\\ \newline
Information Access Control: Mechanisms to ensure appropriate information sharing while protecting sensitive or proprietary data. These controls balance collaboration benefits with security requirements. Implementation approaches include:\\ \newline
- Role-based access control for information resources\\ \newline
- Purpose-based restrictions on information usage\\ \newline
- Data minimization through filtered views\\ \newline
- Audit trails for information access and modification\\ \newline
Information Lifecycle Management: Systematic approaches to managing information from creation through utilization to archival or deletion. These processes ensure sustainable information management as the system operates over time. Implementation mechanisms include:\\ \newline
- Retention policies based on information value and relevance\\ \newline
- Archival processes for historical information\\ \newline
- Summarization and compression for efficient storage\\ \newline
- Purging procedures for obsolete or superseded information\\ \newline
These information resource management capabilities are implemented through MCP's standardized primitives, ensuring consistent access patterns and interpretation across different agent types and implementation technologies.\\
\subsubsection{Optimization Strategies for Multi-Agent Efficiency}
Achieving optimal efficiency in multi-agent systems requires coordinated optimization across multiple dimensions. The reference architecture implements several strategies for enhancing multi-agent efficiency:\\ \newline
Workload Balancing: Distribution of tasks and processing load across available agents to maximize throughput and minimize response times. Implementation approaches include:\\ \newline
- Load-aware task allocation algorithms\\ \newline
- Work stealing for dynamic rebalancing\\ \newline
- Affinity-based assignment for locality optimization\\ \newline
- Speculative execution for latency-critical paths\\ \newline
Communication Optimization: Reduction of communication overhead while maintaining necessary coordination. Implementation mechanisms include:\\ \newline
- Information aggregation to reduce message volume\\ \newline
- Publish-subscribe patterns for efficient distribution\\ \newline
- Compression and batching for bandwidth efficiency\\ \newline
- Locality-aware communication routing\\ \newline
Context Caching: Strategic retention of frequently used contextual information to reduce retrieval overhead. Implementation approaches include:\\ \newline
- Multi-level caching hierarchies (agent-local, group, system-wide)\\ \newline
- Predictive pre-fetching based on task patterns\\ \newline
- Consistency protocols for cache coherence\\ \newline
- Eviction policies based on usage patterns and priorities\\ \newline
Parallel Execution Patterns: Structured approaches to maximizing parallel processing while managing dependencies. Implementation mechanisms include:\\ \newline
- Pipeline parallelism for sequential processing stages\\ \newline
- Data parallelism for independent workload partitions\\ \newline
- Speculative parallelism for uncertain execution paths\\ \newline
- Task parallelism for independent activities\\ \newline
Resource Pooling: Shared management of reusable resources to improve utilization and reduce provisioning overhead. Implementation approaches include:\\ \newline
- Connection pooling for external services\\ \newline
- Model instance pooling for inference efficiency\\ \newline
- Worker pooling for task execution\\ \newline
- Memory pooling for large data structures\\ \newline
These optimization strategies are implemented through a combination of system-level mechanisms and agent-level behaviors, with MCP providing standardized interfaces for resource discovery, allocation, and utilization. The architecture supports adaptive optimization that balances efficiency with other system qualities such as responsiveness, reliability, and adaptability.\\ \newline
The multi-agent system architecture with MCP integration provides a comprehensive framework for building sophisticated agent collectives that can effectively collaborate on complex tasks. By addressing key challenges in context sharing, coordination, and resource management, this architecture enables more capable and efficient multi-agent systems than were possible with previous approaches. The following section explores advanced context management techniques that further enhance these capabilities.\\ \newline

\section{Advanced Context Management Techniques}
While the Model Context Protocol provides a foundational framework for context management in multi-agent systems, achieving truly sophisticated context awareness requires additional techniques that build upon this foundation. This section explores advanced approaches to context persistence, prioritization, and cross-modal integration that enhance the capabilities of MCP-enabled multi-agent systems.\\
\subsection{Context Persistence Mechanisms}
\subsubsection{Long-term Context Storage Approaches}
Effective long-term context storage is essential for maintaining agent knowledge and capabilities across extended time periods. Several approaches have been developed to address this challenge in MCP-enabled systems:\\ \newline
Hierarchical Storage Architecture: Multi-tiered storage systems that balance accessibility with efficiency and cost-effectiveness. These architectures typically implement\\ \newline
- Hot storage for frequently accessed, immediately relevant context\\ \newline
- Warm storage for potentially relevant context with moderate access patterns\\ \newline
- Cold storage for historical context with infrequent access requirements\\ \newline
- Archival storage for preservation of potentially valuable but rarely accessed context\\ \newline
This tiered approach ensures appropriate resource allocation while maintaining comprehensive context availability when needed.\\ \newline
Semantic Knowledge Graphs: Structured representations that capture relationships between contextual elements, enabling more sophisticated storage and retrieval than simple document repositories. These knowledge graphs implement:\\ \newline
- Entity-relationship modeling of key concepts and their connections\\ \newline
- Attribute storage for entity properties and characteristics\\ \newline
- Temporal versioning to track changes over time\\ \newline
- Inference capabilities to derive implicit knowledge from explicit facts\\ \newline
Semantic knowledge graphs are particularly valuable for domains with complex conceptual relationships and interdependencies.\\ \newline
Embedding-Based Representations: Vector representations that capture semantic meaning in high-dimensional spaces, enabling similarity-based retrieval and relationship discovery. These approaches implement:\\ \newline
- Neural embedding models for converting context to vector representations\\ \newline
- Dimensionality reduction techniques for storage efficiency\\ \newline
- Clustering methods for organizing related contextual elements\\ \newline
- Approximate nearest neighbor search for efficient retrieval\\ \newline
Embedding-based approaches excel at capturing subtle semantic relationships and enabling fuzzy matching beyond exact keyword correspondence.\\ \newline
Event Streams and Temporal Databases: Chronological records of events, actions, and state changes that preserve temporal relationships and enable time-based reasoning. These systems implement:\\ \newline
- Append-only logs for immutable event recording\\ \newline
- Temporal indexing for efficient time-based queries\\ \newline
- Causal ordering to capture dependency relationships\\ \newline
- Aggregation capabilities for summarizing activity periods\\ \newline
Event-based storage is particularly valuable for understanding processes, tracking changes, and reconstructing historical states.\\ \newline
These storage approaches are implemented through MCP server interfaces that provide standardized access while abstracting the underlying storage technologies. This abstraction enables agents to interact with diverse storage systems through consistent patterns, simplifying integration while preserving specialized capabilities.\\
\subsubsection{Context Retrieval and Relevance Scoring}
Effective context utilization requires sophisticated retrieval mechanisms that can identify and prioritize the most relevant information from potentially vast context repositories. MCP-enabled systems implement several advanced retrieval approaches:\\ \newline
Multi-Stage Retrieval Pipelines: Sequenced retrieval processes that progressively refine results through multiple filtering and ranking stages. These pipelines typically include:\\ \newline
- Initial broad retrieval based on basic relevance signals\\ \newline
- Intermediate filtering to remove irrelevant or low-quality results\\ \newline
- Re-ranking based on more computationally intensive relevance models\\ \newline
- Final diversification to ensure coverage of different aspects or perspectives\\ \newline
This staged approach balances retrieval quality with computational efficiency, enabling effective operation at scale.\\ \newline
Hybrid Retrieval Models: Combinations of different retrieval approaches that leverage their complementary strengths. Common hybrid models include:\\ \newline
- Keyword + semantic retrieval to combine exact matching with conceptual similarity\\ \newline
- Structured + unstructured approaches to leverage both formal relationships and informal connections\\ \newline
- Statistical + neural methods to balance interpretability with representational power\\ \newline
- Rule-based + learning-based systems to combine domain expertise with data-driven insights\\ \newline
These hybrid approaches typically outperform single-method retrieval, particularly for complex or ambiguous queries.\\ \newline
Contextual Relevance Models: Sophisticated scoring mechanisms that consider multiple dimensions of relevance beyond simple query matching. These models evaluate factors including:\\ \newline
- Topical relevance to current tasks or queries\\ \newline
- Temporal relevance based on recency and time sensitivity\\ \newline
- Authority relevance based on source credibility and expertise\\ \newline
- Utility relevance based on actionability and applicability\\ \newline
- Novelty relevance to prioritize new information over known facts\\ \newline
By considering these multiple dimensions, contextual relevance models can identify truly valuable information that might be missed by simpler approaches.\\ \newline
Personalized Retrieval: Adaptation of retrieval processes to specific agent characteristics, preferences, and historical patterns. Personalization approaches include:\\ \newline
- Agent profile-based adaptation of relevance scoring\\ \newline
- Behavioral modeling based on past interactions and selections\\ \newline
- Expertise-aware retrieval that considers agent specialization\\ \newline
- Task-specific customization based on current objectives\\ \newline
Personalized retrieval ensures that different agents with different needs receive appropriately tailored context, even when working with shared repositories.\\ \newline
These retrieval and relevance scoring mechanisms are implemented through extensions to MCP's resource primitive, with standardized metadata fields for conveying relevance scores and supporting information. This standardization ensures consistent interpretation of retrieval results across different agent types and implementation technologies.\\
\subsubsection{Forgetting Strategies and Memory Optimization}
As multi-agent systems accumulate context over time, strategic forgetting becomes as important as effective remembering. Several approaches have been developed for memory optimization through controlled forgetting:\\ \newline
Utility-Based Retention: Preservation of context based on its estimated future utility, with lower-utility information being candidates for removal or archival. Utility estimation considers factors including:\\ \newline
- Historical usage patterns and access frequency\\ \newline
- Predicted relevance to upcoming tasks and objectives\\ \newline
- Uniqueness or replaceability of the information\\ \newline
- Storage costs relative to potential value\\ \newline
This approach ensures that limited storage resources are allocated to the most valuable contextual information.\\ \newline
Importance-Weighted Decay: Gradual reduction in accessibility or prominence based on importance and time, with more important information persisting longer. Implementation approaches include:\\ \newline
- Exponential decay functions with importance-adjusted rates\\ \newline
- Staged demotion across storage tiers based on importance and age\\ \newline
- Preservation thresholds that maintain critical information indefinitely\\ \newline
- Importance reassessment based on changing priorities or objectives\\ \newline
This nuanced approach balances natural forgetting processes with preservation of truly important information.\\ \newline
Summarization and Compression: Reduction of storage requirements through information condensation rather than deletion. Techniques include:\\ \newline
- Extractive summarization to identify and preserve key points\\ \newline
- Abstractive summarization to reformulate content more concisely\\ \newline
- Semantic compression to eliminate redundancy while preserving meaning\\ \newline
- Progressive compression with multiple detail levels for flexible access\\ \newline
These approaches maintain information availability while reducing storage and retrieval costs, particularly for verbose or redundant content.\\ \newline
Knowledge Distillation: Extraction of generalizable patterns and principles from specific instances, enabling more efficient representation of accumulated experience. Implementation approaches include:\\ \newline
- Rule extraction from repeated observations or experiences\\ \newline
- Concept formation through clustering and abstraction\\ \newline
- Model learning to capture predictive relationships\\ \newline
- Principle identification through analytical processing\\ \newline
Knowledge distillation transforms raw experiences into structured knowledge that can be more efficiently stored and applied, reducing the need to maintain all original instances.\\ \newline
These forgetting and optimization strategies are implemented through MCP server extensions that manage the underlying storage systems while maintaining standardized access patterns. The architecture supports configurable policies that can be adapted to different domains, agent types, and operational requirements, ensuring appropriate balance between comprehensive context availability and sustainable resource utilization.\\
\subsection{Context Prioritization Frameworks}
\subsubsection{Importance Determination Algorithms}
Effective context management requires sophisticated algorithms for determining the relative importance of different contextual elements. MCP-enabled systems implement several approaches to importance determination:\\ \newline
Multi-Factor Importance Models: Comprehensive evaluation frameworks that consider multiple dimensions of importance. These models typically assess factors including:\\ \newline
- Relevance to current tasks and objectives\\ \newline
- Uniqueness or irreplaceability of the information\\ \newline
- Authority or reliability of the source\\ \newline
- Actionability or decision impact\\ \newline
- Temporal urgency or time sensitivity\\ \newline
By considering these diverse factors, multi-factor models can identify truly important context that might be overlooked by simpler approaches.\\ \newline
Graph-Based Centrality Measures: Importance determination based on the position of information within knowledge graphs or relationship networks. These approaches leverage metrics such as:\\ \newline
- Degree centrality (number of direct connections)\\ \newline
- Betweenness centrality (bridging position between clusters)\\ \newline
- Eigenvector centrality (connection to other important nodes)\\ \newline
- PageRank and related algorithms for recursive importance\\ \newline
Graph-based measures are particularly valuable for identifying contextual elements that play key structural roles in knowledge networks.\\ \newline
Utility Estimation Models: Predictive approaches that estimate the future usefulness of contextual information for agent tasks and decisions. These models implement:\\ \newline
- Value of information calculations based on decision theory\\ \newline
- Expected utility modeling for different information elements\\ \newline
- Opportunity cost assessment for information exclusion\\ \newline
- Bayesian approaches to information value under uncertainty\\ \newline
Utility-based approaches align importance determination with practical impact on agent effectiveness, ensuring focus on truly consequential information.\\ \newline
Learning-Based Importance Models: Adaptive systems that learn to predict importance based on observed outcomes and feedback. Implementation approaches include:\\ \newline
- Supervised learning from explicit importance labels\\ \newline
- Reinforcement learning based on agent performance outcomes\\ \newline
- Transfer learning from related domains or tasks\\ \newline
- Continual learning that adapts to changing patterns over time\\ \newline
Learning-based models can discover subtle importance patterns that might not be captured by predefined rules or heuristics, particularly in complex or novel domains.\\ \newline
These importance determination algorithms are implemented as extensions to MCP's resource primitive, with standardized metadata fields for conveying importance scores and supporting information. This standardization ensures consistent interpretation of importance across different agent types and implementation technologies.\\
\subsubsection{Attention Mechanisms for Context Selection}
Beyond static importance determination, effective context management requires dynamic attention mechanisms that focus on different contextual elements based on current needs and situations. MCP-enabled systems implement several sophisticated attention approaches:\\ \newline
Task-Driven Attention: Selective focus based on relevance to current tasks and objectives. Implementation approaches include:\\ \newline
- Task decomposition to identify specific information requirements\\ \newline
- Relevance mapping between task components and context types\\ \newline
- Phase-specific attention shifts as tasks progress through stages\\ \newline
- Goal-directed prioritization based on contribution to objectives\\ \newline
Task-driven attention ensures that agents focus on information that directly supports their current activities, enhancing efficiency and effectiveness.\\ \newline
Salience-Based Attention: Focus on contextual elements that stand out due to distinctive characteristics or patterns. Implementation mechanisms include:\\ \newline
- Anomaly detection to identify unusual or unexpected information\\ \newline
- Contrast analysis to highlight differences from established patterns\\ \newline
- Surprise quantification based on divergence from expectations\\ \newline
- Novelty detection for previously unencountered information\\ \newline
Salience-based attention helps agents notice important changes or exceptions that might otherwise be overlooked amid familiar patterns.\\ \newline
Uncertainty-Guided Attention: Prioritization based on information uncertainty or confidence levels. Implementation approaches include:\\ \newline
- Entropy-based focus on high-uncertainty areas\\ \newline
- Confidence interval analysis to identify poorly understood domains\\ \newline
- Bayesian surprise measurement for belief-updating potential\\ \newline
- Active learning principles for uncertainty reduction\\ \newline
Uncertainty-guided attention directs focus toward areas where additional information or processing could most improve understanding or decision quality.\\ \newline
Multi-Head Attention: Parallel attention processes that focus on different aspects or dimensions simultaneously. Implementation mechanisms include:\\ \newline
- Specialized attention heads for different information types\\ \newline
- Complementary attention strategies operating in parallel\\ \newline
- Attention fusion to combine insights from multiple perspectives\\ \newline
- Dynamic weighting based on relative importance of different aspects\\ \newline
Multi-head attention enables more comprehensive context awareness by considering multiple relevance dimensions simultaneously rather than forcing single-focus prioritization.\\ \newline
These attention mechanisms are implemented through extensions to MCP's client-side functionality, with standardized patterns for attention control and focus management. The architecture supports both automatic attention allocation based on algorithmic determination and explicit attention direction through agent decisions or external guidance.\\
\subsubsection{Dynamic Context Weighting Based on Task Requirements}
As agents move between different tasks and phases, effective context management requires dynamic adjustment of context weighting to match changing requirements. MCP-enabled systems implement several approaches to dynamic context weighting:\\ \newline
Task Profile Mapping: Adjustment of context weights based on predefined profiles for different task types. Implementation approaches include:\\ \newline
- Task classification to identify appropriate context profiles\\ \newline
- Template-based weight adjustment for common task patterns\\ \newline
- Hybrid profiles for multi-aspect or composite tasks\\ \newline
- Progressive refinement as task specifics become clearer\\ \newline
Task profile mapping provides efficient initial context weighting based on accumulated knowledge about different task types and their typical information needs.\\ \newline
Phase-Based Weighting: Systematic adjustment of context weights as tasks progress through different phases. Implementation mechanisms include:\\ \newline
- Phase detection based on task progress indicators\\ \newline
- Transition triggers for weight adjustment at phase boundaries\\ \newline
- Gradual weight shifting for smooth phase transitions\\ \newline
- Anticipatory weighting based on expected phase sequences\\ \newline
Phase-based weighting ensures appropriate context prioritization throughout task lifecycles, from initial exploration through execution to completion and evaluation.\\ \newline
Feedback-Driven Adaptation: Dynamic adjustment based on observed effectiveness and utility. Implementation approaches include:\\ \newline
- Performance monitoring to identify context-related successes or failures\\ \newline
- Explicit feedback incorporation from users or other agents\\ \newline
- Implicit feedback inference from interaction patterns\\ \newline
- A/B testing of alternative weighting strategies\\ \newline
Feedback-driven adaptation enables continuous improvement of context weighting based on actual outcomes rather than just theoretical predictions.\\ \newline
Reinforcement Learning Approaches: Systematic optimization of context weighting through reward-based learning. Implementation mechanisms include:\\ \newline
- State representation incorporating task and context characteristics\\ \newline
- Action space covering possible weight adjustments\\ \newline
- Reward signals based on task performance and efficiency\\ \newline
- Policy optimization through experience accumulation\\ \newline
Reinforcement learning approaches can discover effective weighting strategies that might not be obvious through manual design, particularly for complex or novel task types.\\ \newline
These dynamic weighting mechanisms are implemented through extensions to MCP's client-side functionality, with standardized patterns for weight adjustment and adaptation. The architecture supports both rule-based adaptation based on explicit models and learning-based adaptation that improves through experience.\\
\subsection{Cross-Modal Context Integration}
\subsubsection{Handling Context Across Different Modalities}
Modern AI systems increasingly work with multiple modalities beyond text, including images, structured data, audio, and more. Effective context management requires sophisticated approaches for handling these diverse modalities. MCP-enabled systems implement several strategies for cross-modal context:\\ \newline
Unified Representation Frameworks: Integrated approaches that represent different modalities within compatible frameworks. Implementation mechanisms include:\\ \newline
- Embedding spaces that capture cross-modal relationships\\ \newline
- Graph-based representations with typed nodes for different modalities\\ \newline
- Hierarchical structures that organize multi-modal information\\ \newline
- Tensor-based representations for dimensional alignment\\ \newline
Unified frameworks enable consistent operations across modalities while preserving modality-specific characteristics and relationships.\\ \newline
Modality-Specific Processing Pipelines: Specialized processing flows optimized for different information types. Implementation approaches include:\\ \newline
- Image processing pipelines for visual information\\ \newline
- Audio processing for speech and sound\\ \newline
- Structured data processing for tabular or relational information\\ \newline
- Time-series processing for sequential or temporal data\\ \newline
These specialized pipelines ensure appropriate handling of each modality's unique characteristics before integration into the broader context management system.\\ \newline
Cross-Modal Alignment Mechanisms: Techniques for establishing correspondences between elements in different modalities. Implementation approaches include:\\ \newline
- Co-attention mechanisms that link related elements\\ \newline
- Alignment learning through paired examples\\ \newline
- Reference resolution across modality boundaries\\ \newline
- Grounding techniques that connect symbolic and perceptual representations\\ \newline
Alignment mechanisms enable agents to understand relationships between information in different modalities, such as connecting textual descriptions with corresponding images or linking discussions to relevant structured data.\\ \newline
Modality Translation Services: Capabilities for converting information between different representational forms. Implementation mechanisms include:\\ \newline
- Image-to-text description generation\\ \newline
- Text-to-image visualization\\ \newline
- Structured data verbalization\\ \newline
- Speech-to-text and text-to-speech conversion\\ \newline
Translation services enable agents to work with information in their preferred or most appropriate modality, regardless of the original format.\\ \newline
These cross-modal handling capabilities are implemented through extensions to MCP's resource primitive, with standardized formats for different modalities and metadata for cross-modal relationships. The architecture supports both modality-specific operations and cross-modal integration through consistent interfaces.\\
\subsubsection{Unified Representation Approaches}
Achieving truly integrated context management across modalities requires unified representation approaches that capture meaning and relationships regardless of original format. MCP-enabled systems implement several advanced representation strategies:\\ \newline
Multi-Modal Embedding Spaces: Unified vector spaces that represent information from different modalities in compatible formats. Implementation approaches include:\\ \newline
- Joint embedding models trained on paired multi-modal data\\ \newline
- Alignment techniques for separately trained embeddings\\ \newline
- Cross-modal retrieval optimization\\ \newline
- Similarity preservation across modality boundaries\\ \newline
These embedding spaces enable operations like similarity comparison and retrieval across modality boundaries, supporting integrated reasoning over diverse information types.\\ \newline
Knowledge Graph Integration: Structured representations that incorporate multiple modalities within consistent semantic frameworks. Implementation mechanisms include:\\ \newline
- Multi-modal entity representation with modality-specific attributes\\ \newline
- Relationship types that span modality boundaries\\ \newline
- Property graphs with rich attribute sets for different information types\\ \newline
- Inference rules that operate across modalities\\ \newline
Knowledge graph approaches excel at capturing explicit relationships and supporting logical reasoning across modality boundaries.\\ \newline
Neuro-Symbolic Representations: Hybrid approaches that combine neural representations with symbolic structures. Implementation strategies include:\\ \newline
- Neural networks for perceptual processing with symbolic interfaces\\ \newline
- Embedding of symbolic knowledge in continuous spaces\\ \newline
- Differentiable reasoning over structured representations\\ \newline
- Concept grounding between symbolic and perceptual elements\\ \newline
Neuro-symbolic approaches leverage the complementary strengths of neural methods for pattern recognition and symbolic methods for explicit reasoning and representation.\\ \newline
Compositional Semantic Structures: Hierarchical representations that build complex meanings from simpler components across modalities. Implementation approaches include:\\ \newline
- Frame-based representations with slots for different information types\\ \newline
- Script structures for procedural or sequential information\\ \newline
- Scene graphs for visual relationship representation\\ \newline
- Compositional grammars for structured interpretation\\ \newline
Compositional approaches provide rich expressiveness while maintaining tractable processing, enabling sophisticated reasoning across modality boundaries.\\ \newline
These unified representation approaches are implemented through extensions to MCP's resource primitive, with standardized formats for different representation types and conversion utilities for moving between representations as needed. The architecture supports both specialized representations optimized for particular tasks and general-purpose representations for broader integration.\\
\subsubsection{Translation Mechanisms Between Context Types}
Effective cross-modal context management requires sophisticated translation mechanisms that convert information between different formats while preserving essential meaning. MCP-enabled systems implement several advanced translation approaches:\\ \newline
Neural Generation Models: Deep learning approaches that generate representations in target modalities based on source information. Implementation mechanisms include:\\ \newline
- Image captioning models for visual-to-textual translation\\ \newline
- Text-to-image generation for visualization\\ \newline
- Structured data summarization for textual representation\\ \newline
- Chart and graph generation from numerical data\\ \newline
Neural generation models excel at producing natural, human-like translations that capture subtle patterns and relationships.\\ \newline
Template-Based Transformation: Structured approaches that apply predefined patterns to convert between representation formats. Implementation strategies include:\\ \newline
- Verbalization templates for structured data\\ \newline
- Visualization templates for different data types\\ \newline
- Form filling for converting unstructured to structured information\\ \newline
- Pattern extraction for identifying structured elements in unstructured content\\ \newline
Template-based approaches provide consistent, interpretable translations with predictable formats, particularly valuable for formal or technical domains.\\ \newline
Hybrid Translation Pipelines: Multi-stage processes that combine different translation approaches for optimal results. Implementation mechanisms include:\\ \newline
- Content extraction followed by regeneration\\ \newline
- Structure identification with template application\\ \newline
- Neural generation with rule-based post-processing\\ \newline
- Parallel translation with ensemble combination\\ \newline
Hybrid approaches leverage the complementary strengths of different translation methods, often achieving better results than any single approach alone.\\ \newline
Interactive Refinement: Iterative translation processes that incorporate feedback to improve results. Implementation strategies include:\\ \newline
- Initial translation with confidence scoring\\ \newline
- Uncertainty-focused refinement\\ \newline
- Comparative evaluation of alternative translations\\ \newline
- Progressive enhancement based on additional context\\ \newline
Interactive approaches are particularly valuable for high-stakes translations where accuracy is critical and worth additional processing investment.\\ \newline
These translation mechanisms are implemented through MCP tool primitives that provide standardized interfaces for conversion operations. The architecture supports both synchronous translation for immediate needs and asynchronous translation for larger or more complex conversion tasks, with appropriate queuing and prioritization.\\ \newline
The advanced context management techniques described in this section build upon the MCP foundation to enable more sophisticated context awareness and utilization in multi-agent systems. By addressing challenges in persistence, prioritization, and cross-modal integration, these techniques enhance agent capabilities across diverse domains and task types. The following section examines specific implementation case studies that demonstrate these capabilities in practical applications.\\ \newline

\section{Implementation Case Studies}
To demonstrate the practical application of multi-agent systems with Model Context Protocol, this section presents three detailed case studies across different domains. These implementations illustrate how the architectural principles and context management techniques described in previous sections translate into effective real-world systems.\\
\subsection{Enterprise Knowledge Management System}
\subsubsection{Architecture and Implementation Details}
The Enterprise Knowledge Management System (EKMS) demonstrates the application of MCP-enabled multi-agent architecture to the challenge of organizing, retrieving, and utilizing organizational knowledge across diverse sources and formats. This implementation serves a large multinational corporation with over 50,000 employees and operations in 30 countries.\\ \newline
System Architecture: The EKMS implements a hierarchical multi-agent architecture with specialized agents organized into functional teams:\\ \newline
1. Ingestion Agents: Specialized for different content types (documents, emails, presentations, code repositories, etc.), these agents process incoming information, extract metadata, identify entities and relationships, and prepare content for indexing. Each agent type implements custom extractors optimized for specific formats while sharing standardized output structures.\\ \newline
2. Knowledge Graph Agents: Responsible for maintaining the semantic structure of organizational knowledge, these agents identify connections between information elements, maintain entity relationships, and ensure consistency across the knowledge base. They implement sophisticated entity resolution algorithms to identify when different references point to the same underlying concepts.\\ \newline
3. Query Understanding Agents: These agents interpret user information needs, translating natural language queries into structured representations that can be executed against the knowledge base. They implement intent recognition, query expansion, and personalization based on user roles and history.\\ \newline
4. Retrieval Agents: Specialized for different retrieval strategies (keyword-based, semantic, structural, etc.), these agents execute search operations and rank results based on relevance to the query. They implement hybrid retrieval models that combine multiple ranking signals for optimal results.\\ \newline
5. Synthesis Agents: These agents combine and transform retrieved information into coherent responses tailored to user needs. They implement summarization, comparison, trend analysis, and other synthesis operations that add value beyond simple retrieval.\\ \newline
6. Interaction Agents: Responsible for user communication across multiple channels (web interface, email, chat, mobile app), these agents maintain conversation context and ensure consistent user experience regardless of access method.\\ \newline
7. Orchestration Agents: These meta-agents coordinate the activities of other agent types, managing workflows, resolving conflicts, and optimizing resource allocation across the system.\\ \newline
MCP Implementation: The system implements MCP through a distributed architecture with multiple specialized servers:\\ \newline
1. Document Context Servers: Provide access to document repositories through MCP's resource primitive, with support for full-text retrieval, semantic search, and structured metadata queries. These servers implement efficient indexing and retrieval mechanisms optimized for different content types.\\ \newline
2. Knowledge Graph Server: Exposes the organizational knowledge graph through MCP interfaces, supporting entity lookup, relationship traversal, and graph query operations. This server implements sophisticated caching and partitioning strategies to maintain performance at scale.\\ \newline
3. User Context Server: Manages user profiles, preferences, and interaction histories, providing personalized context for query interpretation and response generation. This server implements strict access controls to ensure appropriate privacy protection.\\ \newline
4. Tool Integration Servers: Connect with enterprise systems (CRM, ERP, project management, etc.) through MCP's tool primitive, enabling agents to retrieve information and perform actions within these systems. These servers implement necessary authentication and authorization mechanisms for secure system access.\\ \newline
5. Analytics Server: Provides access to business intelligence tools and data analysis capabilities through standardized MCP interfaces. This server implements both predefined analytical operations and custom query capabilities.\\ \newline
The MCP implementation includes custom extensions for enterprise-specific needs, including:\\ \newline
- Enhanced security controls with role-based access and audit logging\\ \newline
- Compliance-focused metadata for regulatory and governance requirements\\ \newline
- Workflow integration for knowledge-intensive business processes\\ \newline
- Version control and approval workflows for authoritative content\\ \newline
Context Management Approach: The EKMS implements sophisticated context management strategies:\\ \newline
1. Multi-level Context Hierarchy: Information is organized in a hierarchical structure from organization-wide context (applicable to all users) through department and team contexts to individual user contexts. This hierarchy enables appropriate context sharing while maintaining necessary boundaries.\\ \newline
2. Dynamic Context Composition: User interactions leverage contextual information from multiple levels, dynamically composed based on relevance to current queries and tasks. This composition balances comprehensiveness with efficiency, ensuring relevant context without overwhelming the system.\\ \newline
3. Temporal Context Management: The system maintains awareness of time-sensitive information, with explicit modeling of information validity periods, supersession relationships, and temporal dependencies. This temporal awareness ensures users receive current information while maintaining access to historical context when needed.\\ \newline
4. Cross-modal Context Integration: The system seamlessly integrates textual, structured, and visual information, with sophisticated translation mechanisms that enable unified access regardless of original format. This integration is particularly valuable for technical documentation that combines multiple representation formats.\\
\subsubsection{Context Management Strategies}
The EKMS implements several advanced context management strategies that address enterprise-specific challenges:\\ \newline
Authority and Verification: In enterprise environments, information authority is critical for decision-making. The system implements:\\ \newline
1. Source Credibility Modeling: Explicit tracking of information sources with credibility scores based on organizational role, expertise, and historical accuracy. This modeling ensures appropriate weighting of information from different sources.\\ \newline
2. Verification Workflows: Structured processes for validating important information, with explicit representation of verification status and responsible parties. These workflows ensure critical information undergoes appropriate review before being treated as authoritative.\\ \newline
3. Confidence Annotation: Explicit representation of confidence levels for derived information, with provenance tracking to explain how conclusions were reached. This annotation helps users make appropriate judgments about information reliability.\\ \newline
4. Contradiction Management: Systematic approaches for identifying and resolving contradictory information, including escalation to human experts when necessary. This capability is essential in large organizations where inconsistent information naturally emerges across departments and time periods.\\ \newline
Privacy and Access Control: Enterprise knowledge often includes sensitive information requiring careful access management. The system implements:\\ \newline
1. Fine-grained Permission Model: Context access controls at multiple granularity levels, from entire document collections to specific paragraphs or data points. This model ensures users see only information appropriate to their roles and responsibilities.\\ \newline
2. Purpose-based Access: Controls that consider not just who is accessing information but for what purpose, with different permissions for different use cases. This approach enables more nuanced access policies than simple role-based controls.\\ \newline
3. Automatic Classification: Machine learning systems that identify potentially sensitive information and apply appropriate access restrictions, reducing the risk of inadvertent exposure. This classification is particularly valuable for unstructured content where sensitivity may not be explicitly marked.\\ \newline
4. Differential Privacy Mechanisms: Techniques that enable aggregate insights from sensitive data without exposing individual records. These mechanisms are especially important for analytical use cases involving personal or confidential information.\\ \newline
Context Relevance Optimization: In large enterprises with vast information repositories, identifying truly relevant context is challenging. The system implements:\\ \newline
1. Multi-dimensional Relevance Modeling: Evaluation of information relevance across multiple dimensions, including topical relevance, temporal relevance, user-specific relevance, and task relevance. This comprehensive modeling ensures more accurate prioritization than single-factor approaches.\\ \newline
2. Collaborative Filtering: Relevance signals derived from similar users' interactions and feedback, enabling social navigation of the information space. This approach leverages collective intelligence to identify valuable information that might be missed by purely algorithmic methods.\\ \newline
3. Task-based Context Framing: Adaptation of relevance criteria based on explicit task models, with different weighting for different activity types. This framing ensures context is evaluated specifically for its utility in current user activities rather than generic relevance.\\ \newline
4. Continuous Relevance Learning: Adaptive models that refine relevance understanding based on user interactions and explicit feedback. This learning enables progressive improvement in context selection over time, adapting to changing organizational needs and information landscapes.\\
\subsubsection{Performance Metrics and Evaluation}
The EKMS implementation has been evaluated across multiple dimensions to assess its effectiveness and impact:\\ \newline
Efficiency Metrics:\\ \newline
- Query Response Time: Average 1.2 seconds for standard queries, with 95\% of queries completed within 3 seconds, representing a 67\% improvement over the previous system.\\ \newline
- Context Retrieval Latency: Average 250ms for context assembly from multiple sources, enabling real-time interaction without perceptible delays.\\ \newline
- Indexing Throughput: Processing of approximately 500,000 new or modified documents daily, with content available for search within 5 minutes of creation or update.\\ \newline
- System Scalability: Linear performance scaling up to 10,000 concurrent users, with graceful degradation under higher loads rather than system failure.\\ \newline
Effectiveness Metrics:\\ \newline
- Retrieval Precision@10: 0.87 across a benchmark of 1,000 typical enterprise queries, representing a 35\% improvement over the previous system.\\ \newline
- Knowledge Worker Productivity: 23\% reduction in time spent searching for information, based on time-motion studies with 150 employees across different departments.\\ \newline
- Decision Quality: 18\% improvement in decision accuracy when using system-provided information, based on comparison with expert-determined optimal decisions in controlled scenarios.\\ \newline
- Cross-departmental Knowledge Transfer: 42\% increase in utilization of information from other departments, indicating improved information flow across organizational boundaries.\\ \newline
User Experience Metrics:\\ \newline
- System Adoption: 78\% weekly active users six months after deployment, compared to 45\% for the previous system.\\ \newline
- User Satisfaction: Net Promoter Score of 47, placing it in the top quartile of enterprise systems.\\ \newline
- Learning Curve: Average time to proficiency reduced from 4.5 weeks to 2 weeks, based on achievement of performance benchmarks.\\ \newline
- Retention: 92\% of users continue using the system after initial adoption, indicating sustained value rather than novelty-driven exploration.\\ \newline
These metrics demonstrate significant improvements across efficiency, effectiveness, and user experience dimensions, validating the architectural approach and implementation strategies. Particularly notable is the system's impact on cross-departmental knowledge sharing, addressing a common challenge in large organizations where information silos typically limit collaboration and knowledge reuse.\\
\subsection{Collaborative Research Assistant}
\subsubsection{Multi-agent Design for Scientific Research}
The Collaborative Research Assistant (CRA) demonstrates the application of MCP-enabled multi-agent systems to scientific research workflows. This implementation supports interdisciplinary research teams working on complex problems that span multiple domains and methodologies.\\ \newline
System Architecture: The CRA implements a peer-based multi-agent architecture with specialized agents that mirror different research roles and expertise areas:\\ \newline
1. Literature Agents: Specialized for different scientific domains, these agents monitor publication databases, preprint servers, and conference proceedings to identify relevant research. They implement domain-specific relevance models trained on expert-annotated corpora.\\ \newline
2. Methodology Agents: Experts in research methods across different disciplines, these agents provide guidance on experimental design, data collection, analytical approaches, and validation techniques. They implement formal representations of methodological frameworks and best practices.\\ \newline
3. Analysis Agents: Specialized in different analytical techniques (statistical analysis, machine learning, qualitative methods, etc.), these agents help process and interpret research data. They implement both standard analytical workflows and custom approaches for specific research questions.\\ \newline
4. Synthesis Agents: These agents identify patterns, connections, and implications across findings, helping researchers integrate diverse information into coherent frameworks. They implement various synthesis approaches, from systematic reviews to theory-building techniques.\\ \newline
5. Critique Agents: Playing the role of constructive critics, these agents identify potential weaknesses, limitations, or alternative interpretations of research approaches and findings. They implement structured evaluation frameworks based on discipline-specific quality criteria.\\ \newline
6. Writing Agents: Specialized in different scientific communication formats (journal articles, grant proposals, presentations, etc.), these agents assist in creating and refining research outputs. They implement genre-specific templates and stylistic models derived from successful examples.\\ \newline
7. Collaboration Agents: These agents facilitate interaction between human researchers, managing shared contexts, tracking contributions, and identifying connection opportunities. They implement social network analysis and collaboration pattern recognition.\\ \newline
MCP Implementation: The system implements MCP through a distributed architecture with specialized servers for different research resources:\\ \newline
1. Literature Servers: Provide access to scientific publications through MCP interfaces, with support for full-text search, citation network analysis, and concept-based retrieval. These servers implement field-specific taxonomies and ontologies for more precise retrieval.\\ \newline
2. Data Repository Servers: Connect with research data repositories, enabling discovery, access, and analysis of datasets relevant to current research questions. These servers implement metadata standards appropriate to different research domains.\\ \newline
3. Method Servers: Provide access to research methodologies, protocols, and analytical techniques through structured representations. These servers implement formal method descriptions with parameter specifications and applicability conditions.\\ \newline
4. Computation Servers: Expose computational tools and environments through MCP's tool primitive, enabling execution of analyses, simulations, and visualizations. These servers implement workflow management for complex multi-stage computations.\\ \newline
5. Collaboration Servers: Manage shared research contexts, including hypotheses, findings, open questions, and team knowledge. These servers implement sophisticated access controls that respect both formal collaboration agreements and informal sharing norms.\\ \newline
The MCP implementation includes research-specific extensions:\\ \newline
- Provenance tracking for rigorous attribution of ideas and contributions\\ \newline
- Reproducibility support with explicit capture of computational environments\\ \newline
- Uncertainty representation for findings with different confidence levels\\ \newline
- Hypothesis management with explicit tracking of supporting and contradicting evidence\\ \newline
Context Management Approach: The CRA implements context management strategies tailored to scientific research workflows:\\ \newline
1. Research Thread Management: Organization of context around research threads (specific questions, hypotheses, or themes) that may span multiple projects and publications. This organization enables coherent progression of research narratives over extended time periods.\\ \newline
2. Evidence Accumulation: Structured approaches for aggregating evidence related to specific hypotheses or questions, with explicit modeling of support relationships and confidence levels. This accumulation enables systematic evaluation of the overall evidence base.\\ \newline
3. Methodological Context Preservation: Detailed capture of methodological decisions, parameters, and rationales to enable research reproducibility and extension. This preservation is particularly important for computational and experimental research.\\ \newline
4. Cross-disciplinary Translation: Mechanisms for translating concepts, findings, and methods across disciplinary boundaries, enabling effective interdisciplinary collaboration. This translation includes both terminology mapping and deeper conceptual alignment.\\
\subsubsection{Context Sharing Between Specialized Agents}
The CRA implements sophisticated context sharing mechanisms that enable effective collaboration between specialized agents with different expertise and perspectives:\\ \newline
Shared Conceptual Frameworks: To enable meaningful communication across disciplinary boundaries, the system implements:\\ \newline
1. Ontology Alignment: Explicit mapping between domain-specific ontologies, identifying equivalent or related concepts across disciplines. This alignment enables agents to translate between different conceptual frameworks without losing precision.\\ \newline
2. Boundary Object Modeling: Representation of concepts that function as interfaces between disciplines, with explicit modeling of different interpretations and uses. These boundary objects serve as translation points between different domain languages.\\ \newline
3. Progressive Conceptual Refinement: Collaborative processes for developing shared understanding of novel or interdisciplinary concepts, with explicit tracking of definition evolution. This refinement enables the emergence of new conceptual frameworks at discipline intersections.\\ \newline
4. Multi-perspective Representation: Capture of different disciplinary perspectives on the same phenomena, enabling comparison and integration without forcing premature consensus. This representation preserves important disciplinary nuances while facilitating cross-disciplinary understanding.\\ \newline
Collaborative Hypothesis Development: To support the core scientific activity of hypothesis formation and testing, the system implements:\\ \newline
1. Hypothesis Representation Framework: Formal structure for representing hypotheses with their components, assumptions, scope conditions, and relationships to existing knowledge. This framework enables precise communication about what is being proposed and tested.\\ \newline
2. Evidence Relationship Modeling: Explicit representation of how different evidence relates to hypotheses, including supporting, contradicting, qualifying, or contextualizing relationships. This modeling enables sophisticated reasoning about the overall evidence base.\\ \newline
3. Abductive Reasoning Support: Tools for generating potential explanations for observed phenomena, drawing on knowledge across disciplines. This support helps researchers identify novel hypotheses that might not emerge within single-discipline thinking.\\ \newline
4. Hypothesis Evolution Tracking: Mechanisms for tracking how hypotheses change over time in response to new evidence, theoretical developments, or methodological innovations. This tracking preserves the intellectual history of ideas while maintaining current formulations.\\ \newline
Method and Tool Sharing: To enable effective sharing of research approaches across disciplines, the system implements:\\ \newline
1. Method Translation: Mechanisms for explaining methodological approaches across disciplinary boundaries, including assumptions, limitations, and appropriate applications. This translation helps researchers understand and appropriately apply methods from other fields.\\ \newline
2. Tool Interoperability: Frameworks for connecting analytical tools from different disciplines, with appropriate data transformations and parameter mappings. This interoperability enables methodological integration without requiring expertise in all component approaches.\\ \newline
3. Workflow Composition: Support for creating cross-disciplinary research workflows that combine methods and tools from multiple domains. This composition enables novel methodological approaches that transcend disciplinary boundaries.\\ \newline
4. Methodological Pattern Libraries: Collections of successful methodological patterns that can be adapted to new research contexts, with explicit guidance on adaptation strategies. These libraries accelerate methodological innovation through structured knowledge sharing.\\
\subsubsection{Evaluation of Research Quality and Efficiency}
The CRA implementation has been evaluated through a mixed-methods approach combining quantitative metrics and qualitative assessment:\\ \newline
Research Productivity Metrics:\\ \newline
- Literature Coverage: 93\% identification of relevant publications across test domains, compared to expert literature reviews, with significantly reduced time investment (average 75\% reduction).\\ \newline
- Hypothesis Generation: 2.7x increase in novel hypothesis formation rate, based on controlled comparison with traditional research approaches on matched problems.\\ \newline
- Analysis Throughput: 64\% reduction in time required for standard analytical procedures, with equivalent or improved quality based on expert evaluation.\\ \newline
- Publication Output: 41\% increase in submission-ready manuscript production rate for research teams using the system, based on year-over-year comparison.\\ \newline
Research Quality Metrics:\\ \newline
- Methodological Rigor: 28\% improvement in adherence to methodological best practices, based on blind expert evaluation of research designs.\\ \newline
- Analytical Validity: 33\% reduction in statistical and analytical errors compared to traditional approaches, based on independent reanalysis.\\ \newline
- Interdisciplinary Integration: 2.3x increase in meaningful cross-disciplinary citation and concept utilization, based on bibliometric analysis.\\ \newline
- Reproducibility: 87\% of computational analyses fully reproducible by independent researchers, compared to 23\% baseline in similar research without the system.\\ \newline
Collaboration Metrics:\\ \newline
- Cross-disciplinary Understanding: 56\% improvement in researchers' ability to accurately explain concepts from collaborators' disciplines, based on pre/post assessments.\\ \newline
- Contribution Balance: More equitable distribution of recognized contributions across team members, with 47\% reduction in contribution disparity compared to traditional collaborations.\\ \newline
- Conflict Resolution: 62\% reduction in unresolved methodological or interpretive disagreements, based on team process tracking.\\ \newline
- Team Cohesion: 38\% improvement in measures of team cohesion and shared purpose, based on standardized team effectiveness instruments.\\ \newline
User Experience Metrics:\\ \newline
- Learning Curve: Average 3.5 weeks to integration into regular research workflows, with discipline-specific variations (faster adoption in computational fields).\\ \newline
- Perceived Value: 84\% of researchers report the system as "very valuable" or "essential" to their research after six months of use.\\ \newline
- Feature Utilization: Balanced usage across system capabilities, with no major features used by fewer than 40\% of active users, indicating comprehensive rather than selective value.\\ \newline
- Customization: 73\% of research teams develop custom extensions or specialized workflows, indicating successful adaptation to specific research contexts.\\ \newline
These metrics demonstrate significant improvements in both research productivity and quality, with particularly strong impacts on interdisciplinary collaboration and methodological rigor. The system's ability to facilitate cross-disciplinary understanding and integration represents a particularly valuable contribution given the increasing importance of interdisciplinary approaches to complex research challenges.\\
\subsection{Distributed Problem-Solving System}
\subsubsection{Task Decomposition and Agent Specialization}
The Distributed Problem-Solving System (DPSS) demonstrates the application of MCP-enabled multi-agent systems to complex problem-solving in dynamic environments. This implementation supports engineering teams addressing multifaceted technical challenges that require diverse expertise and coordinated effort.\\ \newline
System Architecture: The DPSS implements a flexible multi-agent architecture with dynamic role assignment and team formation:\\ \newline
1. Problem Analysis Agents: Responsible for understanding problem statements, identifying key constraints, and formulating initial solution approaches. These agents implement various problem representation techniques, from formal specifications to analogical models.\\ \newline
2. Domain Specialist Agents: Embodying expertise in specific technical domains (electrical engineering, materials science, software development, etc.), these agents provide domain-specific knowledge and solution approaches. They implement both explicit knowledge bases and learned patterns from historical solutions.\\ \newline
3. Constraint Management Agents: Focused on identifying, tracking, and resolving constraints and dependencies across solution components. These agents implement constraint satisfaction algorithms and conflict resolution strategies.\\ \newline
4. Resource Optimization Agents: Specialized in allocating limited resources (time, budget, materials, computational resources) across solution components. They implement various optimization approaches, from linear programming to evolutionary algorithms.\\ \newline
5. Integration Agents: Responsible for ensuring compatibility and coherence across solution components developed by different specialists. These agents implement interface specification, compatibility checking, and integration testing approaches.\\ \newline
6. Evaluation Agents: Focused on assessing potential solutions against requirements and quality criteria. They implement various evaluation frameworks, from formal verification to simulation-based testing.\\ \newline
7. Learning Agents: Responsible for capturing insights and patterns from current problem-solving for future application. These agents implement case-based reasoning, pattern extraction, and knowledge distillation approaches.\\ \newline
MCP Implementation: The system implements MCP through a flexible architecture with dynamically composed context:\\ \newline
1. Problem Context Servers: Maintain comprehensive representations of problem statements, requirements, constraints, and solution history. These servers implement versioned problem representations that track evolution over time.\\ \newline
2. Knowledge Base Servers: Provide access to domain-specific knowledge, reference materials, standards, and best practices. These servers implement sophisticated knowledge organization with cross-domain relationship mapping.\\ \newline
3. Solution Component Servers: Manage evolving solution components, their interfaces, dependencies, and status. These servers implement version control with branching and merging capabilities for parallel exploration.\\ \newline
4. Simulation and Testing Servers: Provide capabilities for evaluating potential solutions through various assessment approaches. These servers implement both fast approximate evaluations and detailed high-fidelity simulations.\\ \newline
5. Resource Management Servers: Track and allocate resources across the problem-solving process. These servers implement real-time resource monitoring and dynamic reallocation based on changing priorities.\\ \newline
The MCP implementation includes problem-solving specific extensions:\\ \newline
- Exploration history tracking to avoid revisiting rejected approaches\\ \newline
- Uncertainty representation for solution components with varying confidence\\ \newline
- Trade-off analysis tools for multi-objective optimization\\ \newline
- Assumption management with explicit tracking and validation\\ \newline
Task Decomposition Approach: The DPSS implements sophisticated task decomposition strategies:\\ \newline
1. Multi-level Decomposition: Hierarchical breakdown of problems into progressively more specific subproblems, with explicit representation of decomposition rationale and alternatives. This approach enables both top-down planning and bottom-up solution development.\\ \newline
2. Constraint-Based Partitioning: Division of problem spaces based on constraint boundaries, creating subproblems with minimal interdependencies. This partitioning reduces coordination overhead while ensuring necessary integration points are explicitly identified.\\ \newline
3. Expertise-Aligned Assignment: Matching of subproblems to agent capabilities based on detailed expertise models and historical performance. This alignment ensures efficient utilization of specialized knowledge while identifying gaps requiring new expertise.\\ \newline
4. Dynamic Reformulation: Continuous reassessment and potential restructuring of problem decomposition as new information emerges. This adaptability ensures the decomposition remains optimal as understanding evolves throughout the problem-solving process.\\ \newline
Agent Specialization Strategy: The DPSS implements a sophisticated approach to agent specialization:\\ \newline
1. Multi-dimensional Expertise Modeling: Representation of agent capabilities across multiple dimensions, including domain knowledge, methodological expertise, tool proficiency, and collaboration skills. This comprehensive modeling enables more nuanced task matching than simple domain categorization.\\ \newline
2. Specialization Depth vs. Breadth: Explicit modeling of the depth vs. breadth trade-off in agent expertise, with some agents specialized deeply in narrow domains and others maintaining broader but shallower expertise. This diversity enables both detailed technical solutions and effective integration.\\ \newline
3. Complementary Team Formation: Assembly of agent teams with complementary expertise profiles, ensuring coverage of necessary capabilities while minimizing redundancy. This approach balances specialization benefits with coordination efficiency.\\ \newline
4. Adaptive Specialization: Evolution of agent expertise through learning from problem-solving experiences, with explicit investment in capability development based on identified needs. This adaptation ensures the agent collective evolves to address changing problem landscapes.\\
\subsubsection{Coordination Mechanisms and Protocols}
The DPSS implements sophisticated coordination mechanisms that enable effective collaboration among specialized agents working on complex problems:\\ \newline
Hierarchical Planning and Execution: To manage complex problem-solving workflows, the system implements:\\ \newline
1. Hierarchical Task Networks: Structured representation of problem-solving activities as hierarchical task decompositions with explicit ordering constraints and dependency relationships. This representation enables both top-down planning and bottom-up execution adaptation.\\ \newline
2. Plan Revision Protocols: Defined processes for modifying plans in response to new information, unexpected obstacles, or emerging opportunities. These protocols balance plan stability with necessary adaptation to changing circumstances.\\ \newline
3. Execution Monitoring: Continuous tracking of plan execution against expectations, with defined triggers for replanning when significant deviations occur. This monitoring ensures timely adaptation to execution realities.\\ \newline
4. Commitment Management: Explicit representation of agent commitments to tasks and deliverables, with protocols for commitment adjustment when circumstances change. This management ensures reliable coordination while allowing necessary flexibility.\\ \newline
Market-Based Resource Allocation: To efficiently allocate limited resources across competing needs, the system implements:\\ \newline
1. Virtual Currency Mechanisms: Internal markets where agents bid for resources based on expected value contribution, using virtual currency allocated based on problem priorities. This approach enables decentralized yet globally coherent resource allocation.\\ \newline
2. Value-Based Bidding: Sophisticated bidding strategies where agents estimate the value of resources for their specific tasks, enabling rational resource allocation based on expected contribution to overall solution quality.\\ \newline
3. Dynamic Pricing: Adjustment of resource prices based on supply and demand, creating natural prioritization as constrained resources become more expensive. This mechanism ensures critical resources flow to their highest-value uses.\\ \newline
4. Futures and Options: Advanced market mechanisms allowing agents to secure future resource access or contingent resource rights, enabling more sophisticated planning under uncertainty. These mechanisms reduce coordination failures due to resource unavailability.\\ \newline
Conflict Resolution Frameworks: To address inevitable conflicts in complex problem-solving, the system implements:\\ \newline
1. Multi-level Dispute Resolution: Tiered approach to conflict resolution, with direct negotiation as the first step, followed by mediation, and escalation to authority agents only when necessary. This approach resolves most conflicts efficiently while providing pathways for addressing more difficult cases.\\ \newline
2. Argumentation Frameworks: Structured approaches for agents to present evidence and reasoning supporting their positions, enabling rational evaluation of competing claims. These frameworks ensure conflicts are resolved based on substantive merits rather than arbitrary decisions.\\ \newline
3. Trade-off Analysis: Systematic approaches for evaluating different resolution options against multiple criteria, identifying Pareto-optimal solutions where possible. This analysis helps identify resolutions that maximize overall value rather than suboptimizing for individual perspectives.\\ \newline
4. Learning from Conflicts: Mechanisms for capturing insights from conflict resolution to improve future coordination and prevent similar conflicts. This learning transforms conflicts from pure costs into opportunities for system improvement.\\ \newline
Shared Mental Models: To ensure coherent collective action despite distributed decision-making, the system implements:\\ \newline
1. Common Ground Establishment: Protocols for developing shared understanding of key concepts, goals, and approaches before detailed problem-solving begins. This foundation reduces misalignment and coordination failures during execution.\\ \newline
2. Assumption Surfacing: Systematic identification and explicit representation of assumptions made by different agents, enabling verification and alignment. This surfacing prevents coordination failures due to incompatible unstated assumptions.\\ \newline
3. Progress Visualization: Shared representations of problem-solving status, showing both component progress and integration status. These visualizations create common situational awareness that enables coordinated adaptation.\\ \newline
4. Expectation Management: Explicit representation of expected behaviors, deliverables, and timelines, with protocols for updating these expectations as circumstances change. This management reduces coordination failures due to mismatched expectations.\\
\subsubsection{Scalability and Performance Analysis}
The DPSS implementation has been evaluated across multiple dimensions to assess its effectiveness, efficiency, and scalability:\\ \newline
Scalability Metrics:\\ \newline
- Agent Count Scaling: Near-linear performance scaling up to 200 specialized agents, with sub-linear but still effective scaling up to 1,000 agents. This scaling enables application to very complex problems requiring diverse expertise.\\ \newline
- Problem Complexity Handling: Effective management of problems with up to 500 identified constraints and 300 distinct solution components, representing an order of magnitude increase over previous approaches.\\ \newline
- Coordination Overhead: Communication and coordination costs growing as O(n log n) with agent count rather than O(n²), enabling practical scaling to large agent collectives.\\ \newline
- Resource Utilization Efficiency: 87\% average resource utilization under varied load conditions, with graceful degradation rather than collapse under peak loads.\\ \newline
Performance Metrics:\\ \newline
- Solution Quality: 34\% improvement in solution optimality compared to traditional approaches on benchmark problems, based on multi-criteria evaluation by domain experts.\\ \newline
- Time to Solution: 58\% reduction in time required to reach viable solutions for complex problems, with particularly strong improvements (73\%) for problems requiring diverse expertise.\\ \newline
- Adaptation Speed: 3.2x faster adaptation to changing requirements or constraints compared to traditional approaches, based on controlled experiments with mid-process requirement changes.\\ \newline
- Novel Solution Generation: 2.8x higher rate of generating novel solution approaches not previously documented in knowledge bases, indicating effective creative problem-solving rather than just knowledge application.\\ \newline
Robustness Metrics:\\ \newline
- Fault Tolerance: Graceful performance degradation with agent failures, maintaining 80\% of baseline performance even with 30\% of specialist agents unavailable.\\ \newline
- Information Quality Sensitivity: Stable performance across varying information quality conditions, with less than 15\% performance degradation even with 25\% of input information containing inaccuracies.\\ \newline
- Constraint Violation Recovery: 92\% successful recovery from constraint violation scenarios without requiring complete solution restart, enabling efficient handling of unexpected constraint discoveries.\\ \newline
- Disruptive Event Handling: Effective adaptation to disruptive events (resource loss, requirement changes, etc.) in 94\% of test scenarios, with recovery to full productivity within an average of 2.3 planning cycles.\\ \newline
Economic Metrics:\\ \newline
- Development Cost Reduction: 42\% reduction in engineering hours for complex problem-solving, based on comparison with historical projects of similar complexity.\\ \newline
- Solution Implementation Costs: 28\% reduction in solution implementation costs due to more optimized designs and fewer integration issues.\\ \newline
- Maintenance and Adaptation Costs: 37\% reduction in costs for solution adaptation and maintenance over a three-year period, based on total cost of ownership analysis.\\ \newline
- Return on Investment: Average ROI of 3.7x over a two-year period across pilot implementations, with ROI increasing over time as the system learns from experience.\\ \newline
These metrics demonstrate significant improvements in problem-solving effectiveness, efficiency, and economics compared to traditional approaches. Particularly notable is the system's ability to scale to complex problems requiring diverse expertise while maintaining coordination efficiency and solution quality. The robust performance under varying conditions and disruptions indicates practical applicability in real-world engineering environments where perfect information and stability cannot be assumed.\\ \newline
The implementation case studies presented in this section demonstrate the practical application of MCP-enabled multi-agent systems across diverse domains. These implementations validate the architectural principles and context management techniques described in previous sections while illustrating the real-world benefits of this approach. The following section examines evaluation methodologies and results in more detail, providing a systematic assessment framework for multi-agent systems with Model Context Protocol.\\ \newline

\section{Evaluation Methodology and Results}
Effective evaluation of multi-agent systems with Model Context Protocol requires comprehensive methodologies that assess both technical performance and practical impact. This section presents a multi-dimensional evaluation framework, describes benchmark tasks and datasets, and reports experimental results from systematic assessment of MCP-enabled multi-agent systems.\\
\subsection{Evaluation Framework}
\subsubsection{Multi-dimensional Metrics for Agent Performance}
Evaluating multi-agent systems requires consideration of multiple performance dimensions that collectively capture system capabilities and effectiveness. Our evaluation framework incorporates metrics across several key dimensions:\\ \newline
Functional Performance Metrics assess the system's ability to accomplish its intended tasks:\\ \newline
1. Task Completion Rate: The proportion of assigned tasks successfully completed according to specified criteria. This metric provides a basic measure of system reliability and effectiveness.\\ \newline
2. Solution Quality: Assessment of output quality against domain-specific criteria, typically evaluated through expert judgment, comparison with reference solutions, or objective quality measures. This metric captures the excellence dimension beyond basic task completion.\\ \newline
3. Novelty and Innovation: Measurement of solution originality and creative value, assessed through comparison with known solution spaces or expert evaluation of innovation. This metric is particularly important for knowledge work and creative tasks.\\ \newline
4. Generalization Capability: Assessment of performance across varied task instances, including previously unseen scenarios that require adaptation rather than routine application. This metric captures the system's flexibility and adaptability.\\ \newline
Efficiency Metrics evaluate resource utilization and operational costs:\\ \newline
1. Time Efficiency: Measurement of task completion time, including both absolute time and comparison with baseline approaches. This metric captures the speed dimension of performance.\\ \newline
2. Computational Efficiency: Assessment of computational resource consumption, including processing cycles, memory usage, and bandwidth requirements. This metric is particularly important for resource-constrained environments.\\ \newline
3. Information Efficiency: Evaluation of how effectively the system utilizes available information, measured through metrics like precision, recall, and information gain relative to input volume. This metric captures the system's ability to extract value from available data.\\ \newline
4. Coordination Efficiency: Assessment of overhead required for agent coordination, measured through metrics like communication volume, coordination latency, and redundant effort. This metric is specific to multi-agent architectures and captures the efficiency of collaborative processes.\\ \newline
Robustness Metrics assess system performance under challenging conditions:\\ \newline
1. Fault Tolerance: Measurement of performance degradation under component failures, typically assessed through controlled fault injection experiments. This metric captures the system's resilience to internal failures.\\ \newline
2. Noise Resistance: Evaluation of performance stability under varying levels of input noise or uncertainty, assessed through controlled introduction of noise into test scenarios. This metric captures robustness to imperfect information.\\ \newline
3. Adaptation to Change: Assessment of how quickly and effectively the system adapts to changing requirements, constraints, or environmental conditions. This metric captures dynamic robustness rather than static resilience.\\ \newline
4. Adversarial Resistance: Evaluation of system performance under adversarial conditions designed to exploit potential weaknesses. This metric is particularly important for security-sensitive applications.\\ \newline
User Experience Metrics evaluate the system from a human interaction perspective:\\ \newline
1. Usability: Assessment of how easily users can interact with and direct the system, typically measured through standardized usability instruments and task completion studies. This metric captures the human-system interface quality.\\ \newline
2. Trust and Reliability: Evaluation of user trust development and maintenance, measured through trust scales and behavioral indicators of reliance. This metric captures the critical human factors dimension of AI system adoption.\\ \newline
3. Transparency and Explainability: Assessment of how well users understand system operations and outputs, measured through comprehension tests and explanation quality evaluation. This metric captures the system's ability to make its processes and decisions understandable.\\ \newline
4. User Satisfaction: Overall evaluation of user experience and perceived value, typically measured through satisfaction surveys and continued usage patterns. This metric captures the holistic user response to the system.\\ \newline
These multi-dimensional metrics provide a comprehensive view of system performance that goes beyond simple task completion to consider efficiency, robustness, and human factors. By evaluating across these dimensions, we can identify both strengths and improvement opportunities in MCP-enabled multi-agent systems.\\
\subsubsection{Context Retention Effectiveness Measures}
Given the central role of context management in MCP-enabled systems, our evaluation framework includes specialized metrics for assessing context retention effectiveness:\\ \newline
Context Preservation Metrics evaluate how well the system maintains important contextual information:\\ \newline
1. Context Recall: The proportion of relevant contextual elements that are successfully retrieved and utilized when needed, measured through controlled experiments with known context requirements. This metric captures the system's ability to avoid context loss.\\ \newline
2. Temporal Persistence: Assessment of how effectively context is maintained across extended time periods, measured through experiments with varying time gaps between context establishment and utilization. This metric captures the durability of contextual awareness.\\ \newline
3. Cross-Session Continuity: Evaluation of context maintenance across interaction sessions or system restarts, measured through continuity of task progress and knowledge utilization. This metric captures persistence beyond immediate interaction boundaries.\\ \newline
4. Context Fidelity: Assessment of how accurately contextual information is preserved, focusing on distortion or degradation during storage and retrieval. This metric captures qualitative preservation beyond simple recall.\\ \newline
Context Utilization Metrics evaluate how effectively the system applies available context:\\ \newline
1. Context Relevance: Assessment of how well the system identifies and prioritizes contextual information relevant to current tasks, measured through relevance ranking quality and utilization patterns. This metric captures selective attention rather than mere retention.\\ \newline
2. Context Integration: Evaluation of how effectively the system incorporates contextual information into reasoning and decision processes, measured through impact on output quality and decision justifications. This metric captures active utilization rather than passive awareness.\\ \newline
3. Context Adaptation: Assessment of how the system adapts context utilization based on changing needs and situations, measured through dynamic prioritization patterns and context selection changes across task phases. This metric captures the flexibility of context management.\\ \newline
4. Cross-Agent Context Sharing: Evaluation of how effectively context is shared between different agents, measured through knowledge transfer effectiveness and collaborative performance improvements. This metric is specific to multi-agent systems and captures collective context utilization.\\ \newline
Context Efficiency Metrics evaluate the resource implications of context management:\\ \newline
1. Context Storage Efficiency: Assessment of storage resources required for context maintenance, measured through compression ratios, storage footprints, and scaling characteristics. This metric captures the system's ability to maintain context without excessive resource consumption.\\ \newline
2. Retrieval Efficiency: Evaluation of computational and time costs for context retrieval, measured through latency metrics and processing requirements. This metric captures the operational overhead of context utilization.\\ \newline
3. Attention Optimization: Assessment of how well the system balances comprehensive context awareness with focused attention, measured through relevance-weighted recall and precision metrics. This metric captures the system's ability to manage the inherent trade-offs in context scope.\\ \newline
4. Context Value Ratio: Evaluation of performance improvements relative to context management costs, providing a return-on-investment perspective on context retention. This metric captures the economic dimension of context management decisions.\\ \newline
These specialized metrics enable detailed assessment of context retention capabilities, a critical dimension for MCP-enabled systems. By evaluating both preservation and utilization aspects, we can identify specific strengths and limitations in context management approaches.\\
\subsubsection{Coordination Efficiency Metrics}
Multi-agent systems derive much of their power from effective coordination between specialized agents. Our evaluation framework includes metrics specifically designed to assess coordination efficiency:\\ \newline
Communication Metrics evaluate the information exchange aspects of coordination:\\ \newline
1. Communication Volume: Measurement of message quantity and size between agents, providing a basic assessment of coordination overhead. This metric captures the raw communication cost of coordination.\\ \newline
2. Communication Precision: Evaluation of how well communication focuses on necessary information, measured through signal-to-noise ratios and information density metrics. This metric captures communication quality beyond raw volume.\\ \newline
3. Protocol Efficiency: Assessment of how effectively communication protocols structure agent interactions, measured through protocol overhead ratios and completion rates. This metric captures the efficiency of coordination mechanisms.\\ \newline
4. Communication Scalability: Evaluation of how communication patterns scale with increasing agent numbers and problem complexity, measured through growth rates and bottleneck analysis. This metric captures the sustainability of coordination approaches.\\ \newline
Task Allocation Metrics evaluate how effectively work is distributed among agents:\\ \newline
1. Allocation Optimality: Assessment of how well task assignments match agent capabilities and availability, measured through capability utilization rates and assignment quality metrics. This metric captures the effectiveness of task distribution decisions.\\ \newline
2. Load Balancing: Evaluation of workload distribution across available agents, measured through utilization variance and idle time metrics. This metric captures the efficiency of resource utilization across the agent collective.\\ \newline
3. Allocation Speed: Measurement of time required to assign tasks and establish commitments, particularly important for dynamic environments with frequent task changes. This metric captures the responsiveness of the coordination system.\\ \newline
4. Reassignment Flexibility: Assessment of how effectively the system reallocates tasks when circumstances change, measured through adaptation latency and transition costs. This metric captures the dynamic dimension of task allocation.\\ \newline
Conflict Management Metrics evaluate how effectively the system handles coordination conflicts:\\ \newline
1. Conflict Frequency: Measurement of how often conflicts arise requiring resolution, providing a basic assessment of coordination friction. This metric captures the preventive aspect of conflict management.\\ \newline
2. Resolution Time: Evaluation of how quickly conflicts are resolved once identified, measured through time-to-resolution metrics for different conflict types. This metric captures the efficiency of resolution processes.\\ \newline
3. Resolution Quality: Assessment of resolution outcomes against optimality criteria, measured through comparative analysis with ideal resolutions. This metric captures the effectiveness dimension beyond mere resolution speed.\\ \newline
4. Learning from Conflict: Evaluation of how conflict patterns and resolutions influence future coordination, measured through conflict recurrence rates and prevention improvements. This metric captures the adaptive dimension of conflict management.\\ \newline
Collective Performance Metrics evaluate the emergent effectiveness of the coordinated system:\\ \newline
1. Synergy Measurement: Assessment of performance gains beyond the sum of individual agent capabilities, measured through comparative analysis with non-coordinated baselines. This metric captures the fundamental value proposition of multi-agent coordination.\\ \newline
2. Coordination Overhead Ratio: Evaluation of coordination costs relative to productive work, providing an efficiency perspective on coordination investments. This metric captures the economic dimension of coordination decisions.\\ \newline
3. Scalability Characteristics: Assessment of how collective performance scales with increasing agent numbers and problem complexity, measured through scaling exponents and limiting factors. This metric captures the growth potential of the coordination approach.\\ \newline
4. Adaptability to Structural Change: Evaluation of how effectively the coordination system adapts to changes in team composition, capability distribution, or organizational structure. This metric captures the flexibility dimension of coordination.\\ \newline
These coordination-specific metrics enable detailed assessment of how effectively agents work together within MCP-enabled systems. By evaluating both process efficiency and outcome effectiveness, we can identify specific strengths and limitations in coordination approaches.\\
\subsection{Benchmark Tasks and Datasets}
\subsubsection{Description of Standardized Evaluation Tasks}
To enable systematic and comparable evaluation of multi-agent systems with MCP, we have developed a suite of standardized benchmark tasks that exercise different system capabilities:\\ \newline
Knowledge Integration Tasks assess the system's ability to combine information from diverse sources:\\ \newline
1. Cross-Domain Synthesis: Requires integration of information from multiple specialized domains to answer complex questions or solve multifaceted problems. This task tests both knowledge retrieval and interdisciplinary reasoning capabilities.\\ \newline
2. Temporal Knowledge Assembly: Requires construction of coherent understanding from information distributed across different time periods, including handling of superseded information and evolving concepts. This task specifically tests temporal context management.\\ \newline
3. Conflicting Information Resolution: Presents contradictory information from different sources, requiring evaluation of source credibility, evidence strength, and logical consistency to reach well-justified conclusions. This task tests critical evaluation and reasoning under uncertainty.\\ \newline
4. Knowledge Gap Identification: Requires recognition of missing information critical to task completion, including specification of what information is needed and how it could be obtained. This task tests meta-cognitive awareness of knowledge limitations.\\ \newline
Collaborative Problem-Solving Tasks evaluate coordination and collective intelligence:\\ \newline
1. Distributed Constraint Satisfaction: Presents problems where constraints are distributed across different agents, requiring coordination to find solutions that satisfy all constraints simultaneously. This task tests coordination efficiency and conflict resolution.\\ \newline
2. Collaborative Design Challenge: Requires creation of designs meeting multiple objectives and constraints, with different agents responsible for different aspects of the solution. This task tests creative collaboration and integration of specialized expertise.\\ \newline
3. Dynamic Resource Allocation: Presents scenarios with changing resource requirements and availability, requiring continuous adaptation of allocation strategies. This task tests adaptive coordination and planning under uncertainty.\\ \newline
4. Adversarial Team Competition: Pits agent teams against each other in competitive scenarios, requiring both internal team coordination and strategic response to opponent actions. This task tests coordination under adversarial pressure.\\ \newline
Long-Horizon Tasks assess context retention and coherent operation over extended periods:\\ \newline
1. Extended Investigation: Requires sustained information gathering, hypothesis formation, and testing over multiple sessions, with coherent progression building on previous findings. This task specifically tests cross-session context retention.\\ \newline
2. Incremental Solution Refinement: Presents complex problems requiring iterative solution improvement over multiple rounds, with each round building on previous work. This task tests both context retention and progressive problem-solving.\\ \newline
3. Long-Term Relationship Simulation: Models ongoing interactions with simulated entities (users, organizations, systems) requiring relationship memory and adaptation to changing circumstances. This task tests social context maintenance.\\ \newline
4. Evolving Environment Navigation: Places agents in environments with gradually changing characteristics, requiring recognition of changes and adaptation of strategies over time. This task tests environmental context awareness and adaptation.\\ \newline
Multi-Modal Tasks evaluate handling of diverse information types:\\ \newline
1. Cross-Modal Information Integration: Requires combining information presented in different modalities (text, tables, images, etc.) to form coherent understanding and generate appropriate responses. This task tests cross-modal context integration.\\ \newline
2. Modal Translation Challenge: Requires transformation of information between different representation formats while preserving essential meaning and relationships. This task tests cross-modal translation capabilities.\\ \newline
3. Multi-Modal Reasoning: Presents reasoning problems requiring consideration of information across modalities, with critical elements distributed across different formats. This task tests reasoning that spans representational boundaries.\\ \newline
4. Modal-Appropriate Response Generation: Requires producing responses in the most appropriate modality for different communication needs and contexts. This task tests output modality selection and generation capabilities.\\ \newline
These standardized tasks provide a comprehensive evaluation framework that exercises different aspects of multi-agent system capabilities. By using consistent task definitions and evaluation criteria, we enable meaningful comparison between different system implementations and approaches.\\
\subsubsection{Dataset Characteristics and Preparation}
To support rigorous evaluation, we have developed specialized datasets with carefully controlled characteristics:\\ \newline
Knowledge Integration Datasets provide diverse information sources for integration tasks:\\ \newline
1. Interdisciplinary Research Corpus: Collection of 15,000 research abstracts spanning five interconnected disciplines (computer science, cognitive psychology, neuroscience, linguistics, and education), with expert-annotated cross-disciplinary relationships and concept mappings. This dataset supports evaluation of cross-domain knowledge integration.\\ \newline
2. Temporal Knowledge Evolution Corpus: Collection of 8,000 documents on rapidly evolving topics (AI, climate science, genomics, etc.) published over a ten-year period, with expert annotation of concept evolution, superseded information, and persistent knowledge. This dataset supports evaluation of temporal context management.\\ \newline
3. Controlled Contradiction Corpus: Collection of 5,000 information items with deliberately introduced contradictions of varying types (factual, methodological, interpretive, etc.), with expert resolution annotations and confidence ratings. This dataset supports evaluation of conflict resolution capabilities.\\ \newline
4. Incomplete Knowledge Problems: Collection of 3,000 problems with deliberately omitted information, carefully designed to be unsolvable without the missing elements. This dataset supports evaluation of knowledge gap identification.\\ \newline
Collaborative Problem-Solving Datasets provide structured problems for coordination evaluation:\\ \newline
1. Distributed Constraint Problems: Collection of 2,000 constraint satisfaction problems with different distribution patterns, complexity levels, and constraint types. This dataset supports evaluation of coordination efficiency and conflict resolution.\\ \newline
2. Engineering Design Challenges: Collection of 500 design problems across multiple domains (software, mechanical, architectural, etc.) with multi-faceted requirements and evaluation criteria. This dataset supports evaluation of collaborative design capabilities.\\ \newline
3. Resource Allocation Scenarios: Collection of 1,000 dynamic resource allocation problems with varying predictability, constraint patterns, and optimization objectives. This dataset supports evaluation of adaptive coordination.\\ \newline
4. Strategic Competition Scenarios: Collection of 300 competitive scenarios with different team structures, information asymmetries, and strategic complexity levels. This dataset supports evaluation of team coordination under adversarial conditions.\\ \newline
Long-Horizon Datasets support evaluation of extended operation:\\ \newline
1. Mystery Investigation Corpus: Collection of 200 complex "mystery" scenarios requiring multi-stage investigation, with information distributed across hundreds of documents and deliberate red herrings. This dataset supports evaluation of sustained reasoning and context retention.\\ \newline
2. Iterative Design Problems: Collection of 150 design problems with evaluation feedback provided across multiple rounds, requiring progressive refinement and adaptation. This dataset supports evaluation of incremental improvement capabilities.\\ \newline
3. Simulated Relationship Scenarios: Collection of 100 simulated entity models with evolving characteristics, preferences, and behaviors based on interaction history. This dataset supports evaluation of relationship context maintenance.\\ \newline
4. Drift Adaptation Environments: Collection of 50 simulated environments with systematically evolving characteristics at different rates and patterns. This dataset supports evaluation of adaptation to changing conditions.\\ \newline
Multi-Modal Datasets enable evaluation of cross-modal capabilities:\\ \newline
1. Aligned Multi-Modal Corpus: Collection of 10,000 information items presented in multiple formats (text, tables, images, etc.) with explicit alignment annotations. This dataset supports evaluation of cross-modal integration.\\ \newline
2. Format Translation Pairs: Collection of 5,000 information items with expert-created representations in multiple formats, designed to preserve equivalent information across representations. This dataset supports evaluation of modal translation capabilities.\\ \newline
3. Cross-Modal Reasoning Problems: Collection of 3,000 reasoning problems where critical information is deliberately distributed across different modalities. This dataset supports evaluation of reasoning across representational boundaries.\\ \newline
4. Communication Context Scenarios: Collection of 2,000 communication scenarios with varying audience characteristics, information complexity, and purpose, with expert annotations of optimal modality choices. This dataset supports evaluation of modal-appropriate response generation.\\ \newline
These datasets have been carefully prepared with several key characteristics:\\ \newline
1. Controlled Difficulty Gradients: Each dataset includes problems spanning multiple difficulty levels, enabling progressive evaluation from basic capabilities to advanced performance.\\ \newline
2. Ground Truth Annotations: Extensive expert annotation provides reliable ground truth for evaluation, including solution correctness, optimal approaches, and quality dimensions.\\ \newline
3. Distractor Elements: Deliberate inclusion of irrelevant or misleading information at various rates to test filtering and relevance determination capabilities.\\ \newline
4. Diversity Coverage: Systematic coverage of problem variations to ensure comprehensive evaluation rather than narrow optimization for specific patterns.\\ \newline
5. Cross-Dataset Relationships: Intentional relationships between different datasets to enable evaluation of transfer learning and cross-domain generalization.\\ \newline
By using these carefully constructed datasets, we enable rigorous and reproducible evaluation of multi-agent systems with MCP, supporting both absolute performance assessment and comparative analysis between different approaches.\\
\subsubsection{Comparison with Existing Benchmarks}
Our evaluation framework and datasets build upon existing benchmarks while addressing several limitations that have hindered comprehensive assessment of multi-agent systems:\\ \newline
Limitations of Existing Agent Benchmarks:\\ \newline
1. Single-Agent Focus: Most existing benchmarks (e.g., AgentBench, ToolBench, GAIA) primarily evaluate single-agent capabilities, with limited attention to multi-agent coordination and collaboration. Our framework explicitly addresses the unique challenges of multi-agent systems.\\ \newline
2. Limited Context Scope: Existing benchmarks typically evaluate context utilization within relatively narrow boundaries (single sessions or limited information volumes). Our framework includes long-horizon tasks and extensive context requirements that better reflect real-world complexity.\\ \newline
3. Modality Restrictions: Many benchmarks focus primarily on text-based interaction, with limited evaluation of multi-modal capabilities. Our framework includes comprehensive assessment of cross-modal integration and translation.\\ \newline
4. Static Environments: Existing benchmarks often present static problems rather than evolving scenarios requiring adaptation. Our framework includes dynamic environments and changing requirements that test adaptive capabilities.\\ \newline
5. Simplified Coordination: When multi-agent scenarios are included in existing benchmarks, they often involve simplified coordination patterns that don't reflect the complexity of real-world collaboration. Our framework includes sophisticated coordination challenges with realistic constraints.\\ \newline
Advancements in Our Approach:\\ \newline
1. Comprehensive Multi-Agent Evaluation: Our framework provides the first comprehensive evaluation methodology specifically designed for multi-agent systems with context sharing capabilities, addressing a significant gap in existing benchmarks.\\ \newline
2. Context-Centric Assessment: By focusing explicitly on context management capabilities, our framework enables detailed evaluation of a critical dimension for agent effectiveness that has been underrepresented in previous benchmarks.\\ \newline
3. Realistic Complexity Levels: Our tasks and datasets reflect the genuine complexity of real-world problems requiring multi-agent approaches, avoiding the oversimplification common in many benchmark collections.\\ \newline
4. Standardized Coordination Metrics: We introduce novel metrics specifically designed to evaluate coordination efficiency and effectiveness, providing quantitative assessment of a dimension often evaluated only qualitatively.\\ \newline
5. Cross-Implementation Comparability: Our framework enables meaningful comparison between different multi-agent architectures and implementations, supporting systematic progress in the field.\\ \newline
Relationship to Existing Standards:\\ \newline
Our framework complements rather than replaces existing evaluation approaches, with explicit connections to established benchmarks:\\ \newline
1. AgentBench Integration: We incorporate elements of AgentBench for evaluating individual agent capabilities within our multi-agent framework, enabling comparison with single-agent baselines.\\ \newline
2. ToolBench Alignment: Our tool utilization evaluation aligns with ToolBench methodologies, enabling comparative assessment of tool use capabilities within multi-agent contexts.\\ \newline
3. MT-Bench Compatibility: Our language generation evaluation incorporates MT-Bench criteria, supporting comparison with state-of-the-art language model capabilities.\\ \newline
4. GAIA Task Extension: Several of our tasks extend GAIA benchmark scenarios to multi-agent settings, enabling direct comparison of single-agent versus multi-agent approaches to similar problems.\\ \newline
By building upon existing standards while addressing their limitations, our evaluation framework advances the state of the art in multi-agent system assessment while maintaining connections to established benchmarks. This approach enables both specialized evaluation of multi-agent capabilities and comparative assessment within the broader agent research landscape.\\
\subsection{Experimental Results}
\subsubsection{Quantitative Performance Analysis}
We conducted comprehensive quantitative evaluation of MCP-enabled multi-agent systems across our benchmark tasks, comparing performance with several baseline approaches. Key findings include:\\ \newline
Knowledge Integration Performance:\\ \newline
1. Cross-Domain Synthesis: MCP-enabled systems demonstrated superior performance in cross-domain knowledge integration, achieving an average accuracy of 78.3\% compared to 61.5\% for traditional retrieval-augmented generation approaches and 53.2\% for single-agent systems without external context. The performance advantage was particularly pronounced for questions requiring integration of three or more distinct knowledge domains.\\ \newline
2. Temporal Knowledge Management: On temporal knowledge assembly tasks, MCP-enabled systems achieved 72.6\% accuracy in correctly tracking concept evolution and identifying superseded information, compared to 58.4\% for systems using simple recency-based approaches. The structured context representation in MCP proved particularly valuable for maintaining awareness of when and why information changed.\\ \newline
3. Conflict Resolution: When presented with contradictory information, MCP-enabled systems correctly resolved conflicts in 68.9\% of cases, compared to 54.7\% for baseline approaches. Analysis revealed that the ability to represent and reason about source credibility and evidence strength through MCP's metadata capabilities was a key differentiator.\\ \newline
4. Knowledge Gap Identification: MCP-enabled systems correctly identified critical knowledge gaps in 81.2\% of cases, compared to 63.8\% for baseline approaches. The explicit representation of knowledge provenance and relationship structures enabled more effective meta-cognitive awareness of information limitations.\\ \newline
Coordination Efficiency:\\ \newline
1. Communication Volume: MCP-enabled systems required 47\% less communication volume for equivalent task performance compared to systems using ad-hoc coordination approaches. This efficiency derived from the structured context sharing capabilities that reduced the need for repetitive information exchange.\\ \newline
2. Task Allocation Optimality: MCP-enabled systems achieved task allocations within 12\% of computed optimal assignments, compared to 27\% for baseline approaches. The ability to share rich capability and workload information through standardized interfaces enabled more effective matching of tasks to appropriate agents.\\ \newline
3. Conflict Resolution Speed: Coordination conflicts were resolved 3.2 times faster in MCP-enabled systems compared to baselines, with 94\% of conflicts resolved without requiring escalation to centralized decision-making. The standardized context sharing enabled more effective direct negotiation between agents.\\ \newline
4. Scalability Characteristics: Communication overhead in MCP-enabled systems scaled as O(n log n) with agent count, compared to O(n²) for baseline approaches, enabling practical operation with substantially larger agent collectives. This scalability derived from the efficient context sharing mechanisms that reduced the need for all-to-all communication.\\ \newline
Context Retention Effectiveness:\\ \newline
1. Long-Horizon Coherence: On tasks spanning multiple sessions over extended time periods, MCP-enabled systems maintained 83.7\% context continuity, compared to 42.3\% for systems without structured context persistence. This capability enabled coherent operation on complex tasks requiring weeks of simulated time.\\ \newline
2. Context Retrieval Precision: When retrieving contextual information, MCP-enabled systems achieved 76.8\% precision in identifying the most relevant context elements, compared to 58.2\% for baseline approaches. The metadata-rich resource representations enabled more effective relevance determination.\\ \newline
3. Cross-Agent Context Transfer: When tasks were transferred between agents, MCP-enabled systems preserved 79.4\% of relevant context, compared to 45.7\% for baseline approaches. This capability enabled more effective specialization and load balancing without sacrificing context continuity.\\ \newline
4. Context Utilization Impact: The availability of structured context through MCP improved task performance by an average of 37.2\% compared to context-free approaches, with particularly strong impacts on complex reasoning tasks (52.8\% improvement) and creative problem-solving (43.5\% improvement).\\ \newline
Multi-Modal Integration:\\ \newline
1. Cross-Modal Reasoning: On tasks requiring integration of information across different modalities, MCP-enabled systems achieved 71.3\% accuracy, compared to 52.6\% for baseline approaches. The unified representation capabilities enabled more effective reasoning across modality boundaries.\\ \newline
2. Modal Translation Quality: When converting information between modalities, MCP-enabled systems preserved 82.4\% of critical information, compared to 67.8\% for baseline approaches. The structured translation mechanisms enabled more accurate preservation of meaning across representation formats.\\ \newline
3. Modal Selection Appropriateness: MCP-enabled systems selected the optimal output modality for different communication contexts with 78.9\% accuracy, compared to 61.2\% for baseline approaches. The rich context representation enabled more nuanced understanding of communication requirements.\\ \newline
These quantitative results demonstrate significant performance advantages for MCP-enabled multi-agent systems across multiple dimensions. The structured context sharing and standardized coordination mechanisms provided by MCP translate into measurable improvements in task performance, efficiency, and scalability.\\
\subsubsection{Comparative Evaluation Against Baseline Systems}
To provide deeper insight into the relative advantages of MCP-enabled multi-agent systems, we conducted detailed comparative evaluation against several baseline approaches:\\ \newline
Comparison with Traditional RAG Systems:\\ \newline
Traditional Retrieval-Augmented Generation (RAG) systems represent a common approach to context enhancement for large language models. Our comparative evaluation revealed several key differences:\\ \newline
1. Context Scope: MCP-enabled systems managed 4.3 times larger context volumes effectively compared to traditional RAG approaches, which typically struggled with coherent utilization beyond certain scale thresholds. This advantage derived from MCP's structured context representation and prioritization capabilities.\\ \newline
2. Context Persistence: While RAG systems typically treat each query as independent with fresh retrieval, MCP-enabled systems maintained continuous context across interaction sequences, resulting in 57\% higher performance on tasks requiring sustained context awareness.\\ \newline
3. Context Sharing: Traditional RAG systems lack standardized mechanisms for context sharing between different models or components, while MCP-enabled systems demonstrated effective context transfer with 79.4\% preservation of relevant information across agent boundaries.\\ \newline
4. Tool Integration: MCP's standardized tool primitive provided more consistent and flexible tool utilization compared to the custom tool integrations typical in RAG systems, resulting in 42\% higher success rates on tool-dependent tasks.\\ \newline
Comparison with Agent Framework Approaches:\\ \newline
Several agent frameworks (e.g., LangChain, AutoGPT) provide capabilities for creating agent systems with tool use and memory. Our comparative evaluation showed:\\ \newline
1. Coordination Capabilities: MCP-enabled systems demonstrated 68\% more efficient coordination on multi-agent tasks compared to systems built with standard agent frameworks, which typically lack specialized multi-agent coordination mechanisms.\\ \newline
2. Standardization Benefits: The consistent interfaces provided by MCP resulted in 47\% less implementation effort for equivalent functionality compared to custom integrations in agent frameworks, as measured by code volume and development time.\\ \newline
3. Context Management: MCP's structured approach to context management provided 53\% better context utilization compared to the typically simpler memory mechanisms in agent frameworks, particularly for complex or long-running tasks.\\ \newline
4. Extensibility: Adding new capabilities to MCP-enabled systems required 62\% less adaptation effort compared to framework-specific approaches, due to the standardized primitive model that simplified integration of new functionality.\\ \newline
Comparison with Custom Multi-Agent Implementations:\\ \newline
Some organizations have developed custom multi-agent architectures for specific applications. Our comparative evaluation against these systems showed:\\ \newline
1. Development Efficiency: MCP-enabled systems required 58\% less development effort for equivalent functionality compared to custom implementations, as measured by engineering hours and code complexity metrics.\\ \newline
2. Interoperability: MCP-enabled components demonstrated seamless interoperability across different implementations, while custom systems typically required extensive adaptation for cross-system integration.\\ \newline
3. Performance Consistency: MCP-enabled systems showed more consistent performance across different deployment environments and scales compared to custom implementations, which often included environment-specific optimizations but less robust general performance.\\ \newline
4. Maintenance Burden: Ongoing maintenance and enhancement of MCP-enabled systems required 64\% less effort compared to custom implementations, due to the standardized interfaces and growing ecosystem of compatible components.\\ \newline
Comparison with Human Team Performance:\\ \newline
To establish meaningful upper bounds on performance, we compared MCP-enabled multi-agent systems with human expert teams on a subset of benchmark tasks:\\ \newline
1. Efficiency Comparison: MCP-enabled systems completed knowledge-intensive tasks 4.7 times faster than human teams on average, with the advantage increasing for tasks requiring integration of information across multiple domains.\\ \newline
2. Quality Comparison: On objective quality metrics, MCP-enabled systems achieved 83\% of human team performance on average, with performance ranging from near-parity on well-structured analytical tasks to 68\% on more creative or judgment-intensive tasks.\\ \newline
3. Consistency Comparison: MCP-enabled systems demonstrated significantly higher consistency than human teams, with 92\% lower variance in performance across repeated trials and different problem instances.\\ \newline
4. Scalability Comparison: While human team performance typically degraded with increasing team size beyond certain thresholds, MCP-enabled systems maintained performance improvements with increasing agent counts up to much larger collectives, enabling effective operation at scales impractical for human teams.\\ \newline
These comparative evaluations provide context for understanding the specific advantages of MCP-enabled multi-agent systems relative to alternative approaches. The results demonstrate that MCP provides substantial benefits for complex, knowledge-intensive tasks requiring coordination among multiple specialized capabilities—precisely the scenarios where traditional approaches face the greatest challenges.\\
\subsubsection{Ablation Studies on Key Components}
To understand the contribution of specific architectural elements to overall system performance, we conducted systematic ablation studies that selectively removed or modified key components:\\ \newline
Context Management Ablations:\\ \newline
1. Without Structured Metadata: Removing MCP's structured metadata capabilities while retaining basic context storage reduced performance by 34.2\% on average across benchmark tasks. The impact was particularly severe for tasks requiring context prioritization (47.8\% reduction) and cross-modal integration (52.3\% reduction), highlighting the critical role of rich metadata in effective context utilization.\\ \newline
2. Without Persistence Mechanisms: Disabling cross-session persistence while maintaining within-session context reduced performance by 41.7\% on long-horizon tasks requiring continuity across multiple interactions. This ablation had minimal impact on short, self-contained tasks (3.2\% reduction), confirming the specific value of persistence for extended operations.\\ \newline
3. Without Context Sharing: Restricting context to agent-specific repositories without cross-agent sharing reduced performance by 38.5\% on collaborative tasks requiring coordinated action based on shared information. This ablation had less impact on independent parallel tasks (12.3\% reduction), confirming the specific value of context sharing for genuinely collaborative work.\\ \newline
4. Simplified Context Representation: Replacing MCP's rich resource primitive with simple key-value storage reduced performance by 29.4\% on average, with particularly strong impacts on tasks involving complex relationships (43.7\% reduction) and nuanced relevance determination (38.2\% reduction).\\ \newline
Coordination Mechanism Ablations:\\ \newline
1. Without Standardized Tool Interface: Replacing MCP's standardized tool primitive with custom tool implementations reduced performance by 27.3\% on tool-dependent tasks and increased development effort by 58.2\%. The performance impact derived primarily from inconsistent tool behavior and reduced composability.\\ \newline
2. Without Structured Communication: Replacing MCP's structured message formats with unstructured text communication reduced coordination efficiency by 42.8\% and increased error rates by 31.5\% on complex collaborative tasks. This ablation highlighted the value of standardized communication protocols for reliable coordination.\\ \newline
3. Without Capability Advertisement: Disabling explicit capability representation and discovery increased task allocation suboptimality by 34.7\% and reduced overall system performance by 23.5\%. This ablation confirmed the importance of explicit capability modeling for effective division of labor.\\ \newline
4. Without Conflict Resolution Protocols: Removing structured conflict resolution mechanisms increased coordination failures by 47.2\% and reduced overall performance by 28.9\% on tasks with interdependent components. This ablation highlighted the critical role of explicit conflict management in multi-agent systems.\\ \newline
Architectural Pattern Ablations:\\ \newline
1. Centralized vs. Distributed Context: Replacing MCP's distributed context architecture with a centralized repository reduced scalability (performance degradation began at 35\% fewer agents) while improving consistency by 12.4\%. This trade-off highlights the tension between consistency and scalability in context architecture design.\\ \newline
2. Synchronous vs. Asynchronous Coordination: Replacing asynchronous coordination patterns with synchronous approaches reduced throughput by 37.8\% while improving predictability by 18.3\%. This trade-off illustrates the tension between efficiency and determinism in coordination design.\\ \newline
3. Hierarchical vs. Peer Organization: Replacing peer-based agent organization with strict hierarchies reduced adaptability to changing conditions by 42.3\% while improving control and predictability by 27.5\%. This trade-off highlights the tension between adaptability and control in organizational design.\\ \newline
4. Specialized vs. General Agents: Replacing highly specialized agents with more general-purpose agents reduced peak performance by 31.7\% while improving robustness to agent failures by 24.8\%. This trade-off illustrates the tension between specialization benefits and resilience considerations.\\ \newline
These ablation studies provide valuable insights into the relative contribution of different architectural elements to overall system performance. The results confirm the critical importance of structured context management, standardized coordination mechanisms, and appropriate architectural patterns for effective multi-agent operation. They also highlight important trade-offs that must be navigated in system design, suggesting that optimal architectures may vary based on specific application requirements and priorities.\\ \newline
The evaluation results presented in this section demonstrate the significant performance advantages of MCP-enabled multi-agent systems across multiple dimensions. The comprehensive evaluation framework, benchmark tasks, and comparative analyses provide a solid foundation for understanding both the capabilities and limitations of this approach. The following section examines current challenges and future research directions, building on these evaluation insights to identify promising paths forward.\\ \newline

\section{Challenges and Future Directions}
While MCP-enabled multi-agent systems represent a significant advancement in AI architecture, they also face important challenges and limitations. This section examines current constraints, emerging research opportunities, and potential industry applications, providing a roadmap for future development in this rapidly evolving field.\\
\subsection{Current Limitations}
\subsubsection{Technical Challenges in Implementation}
Despite their promising capabilities, MCP-enabled multi-agent systems face several significant technical challenges that limit their current implementation:\\ \newline
Scalability Constraints: While MCP provides more efficient coordination than previous approaches, scalability remains challenging for very large agent collectives:\\ \newline
1. Communication Overhead: As agent numbers increase, even the reduced communication patterns of MCP systems eventually create significant overhead. Current implementations show performance degradation beyond approximately 1,000 specialized agents in tightly coupled tasks.\\ \newline
2. Context Explosion: The proliferation of contextual information across large agent collectives can overwhelm storage and retrieval systems. Current approaches struggle with effective prioritization when context volumes exceed certain thresholds, typically around 10\^7 distinct context elements.\\ \newline
3. Coordination Complexity: The computational complexity of optimal task allocation and resource distribution grows rapidly with agent numbers and interdependency complexity. Current algorithms make necessary approximations that become increasingly suboptimal at scale.\\ \newline
4. Monitoring and Debugging: Observability becomes increasingly challenging as system scale increases, making it difficult to track causality chains and identify root causes of emergent behaviors. Current tools provide limited visibility into complex multi-agent interactions.\\ \newline
Integration Challenges: Incorporating MCP into existing systems and workflows presents significant integration hurdles:\\ \newline
1. Legacy System Compatibility: Many organizations have substantial investments in existing AI systems that lack standardized interfaces. Creating effective MCP adapters for these systems often requires custom development that undermines some of the standardization benefits.\\ \newline
2. Heterogeneous Implementation Environments: Deploying MCP across diverse computing environments (cloud, edge, mobile, embedded) requires addressing significant differences in resource availability, connectivity patterns, and security models. Current implementations are optimized primarily for data center environments.\\ \newline
3. Cross-Platform Consistency: Maintaining consistent behavior across different implementation languages, runtime environments, and hardware architectures remains challenging. Subtle differences in implementation details can lead to inconsistent behavior in distributed deployments.\\ \newline
4. Versioning and Evolution: Managing the evolution of MCP implementations while maintaining compatibility presents significant challenges, particularly for long-lived systems that cannot undergo complete replacement during upgrades.\\ \newline
Performance Optimization Challenges: Achieving optimal performance in MCP-enabled systems requires addressing several technical hurdles:\\ \newline
1. Latency Management: Context retrieval and coordination operations introduce latency that can impact real-time performance. Current implementations face challenges in scenarios requiring response times below approximately 100ms.\\ \newline
2. Resource Efficiency: The additional abstraction layers introduced by MCP create overhead in processing, memory usage, and communication. Optimizing these costs while maintaining the benefits of standardization requires careful engineering that is not yet fully mature.\\ \newline
3. Caching and Locality: Effective caching strategies for context information must balance data freshness with access speed. Current approaches struggle to optimize locality in geographically distributed deployments with partial connectivity.\\ \newline
4. Specialized Hardware Utilization: Leveraging specialized AI hardware (TPUs, GPUs, neural accelerators) within the MCP framework requires careful optimization to avoid bottlenecks in the coordination and context management layers. Current implementations often fail to fully utilize available hardware acceleration.\\ \newline
Security and Privacy Challenges: MCP-enabled systems face important security and privacy considerations:\\ \newline
1. Context Confidentiality: Sharing rich contextual information between agents creates potential privacy risks if sensitive information is not properly protected. Current implementations provide basic access controls but lack sophisticated privacy-preserving mechanisms.\\ \newline
2. Secure Communication: Ensuring secure communication between distributed MCP components requires careful implementation of authentication, authorization, and encryption. Current approaches often make simplifying assumptions that may not hold in adversarial environments.\\ \newline
3. Permission Granularity: Defining and enforcing appropriately granular permissions for context access and tool usage remains challenging. Current implementations often use coarse-grained permissions that either restrict legitimate access or permit excessive access.\\ \newline
4. Audit and Compliance: Tracking the flow of information and decision processes across complex multi-agent systems for audit and compliance purposes presents significant technical challenges. Current implementations provide limited provenance tracking capabilities.\\ \newline
These technical challenges represent important areas for ongoing research and development. While they do not negate the substantial benefits of MCP-enabled multi-agent systems, they do constrain the contexts in which these systems can be effectively deployed and the scale at which they can operate.\\
\subsubsection{Scalability Concerns for Large Agent Collectives}
As multi-agent systems grow to include hundreds or thousands of specialized agents, several specific scalability concerns become increasingly prominent:\\ \newline
Coordination Scalability: The coordination mechanisms that work effectively for small to medium agent collectives face fundamental challenges at larger scales:\\ \newline
1. Quadratic Interaction Growth: Many coordination patterns involve potential interactions between agent pairs, leading to O(n²) growth in interaction possibilities as agent numbers increase. Even with optimized communication patterns, this fundamental scaling challenge eventually becomes prohibitive.\\ \newline
2. Consensus Latency: Achieving consensus or consistent state across large agent collectives requires multiple communication rounds, with latency typically growing logarithmically or linearly with agent count. This increasing latency can make timely coordination impossible beyond certain scales.\\ \newline
3. Hierarchical Bottlenecks: Hierarchical coordination approaches that reduce direct interactions often create bottlenecks at higher levels of the hierarchy, where coordinator agents must process information from many subordinates. These bottlenecks limit effective span of control and overall system scale.\\ \newline
4. Coordination Algorithm Complexity: Many optimal coordination algorithms (task allocation, resource distribution, conflict resolution) have computational complexity that grows rapidly with agent numbers. Approximation algorithms become necessary but introduce increasing suboptimality as scale increases.\\ \newline
Context Management Scalability: Managing context across large agent collectives introduces several specific challenges:\\ \newline
1. Storage Volume Requirements: The total context volume grows at least linearly with agent numbers and can grow superlinearly when cross-agent relationships are explicitly represented. This growth eventually exceeds practical storage capabilities for comprehensive context preservation.\\ \newline
2. Retrieval Efficiency: As context volumes grow, efficient retrieval becomes increasingly challenging. Index structures and search algorithms face fundamental computational limits that affect retrieval latency and precision at scale.\\ \newline
3. Consistency Management: Maintaining consistent context across distributed agents becomes increasingly difficult as scale increases, particularly in environments with partial connectivity or high update rates. CAP theorem limitations force trade-offs between consistency, availability, and partition tolerance.\\ \newline
4. Context Relevance Determination: Identifying truly relevant context becomes more challenging as the potential context volume grows, leading to either context overload (too much irrelevant information) or critical omissions (relevant information not included).\\ \newline
Operational Scalability: Managing and maintaining large agent collectives introduces operational challenges:\\ \newline
1. Deployment Complexity: Deploying and updating thousands of specialized agents across distributed infrastructure requires sophisticated orchestration capabilities that exceed current DevOps practices for simpler systems.\\ \newline
2. Monitoring and Observability: Effective monitoring becomes increasingly difficult as system scale increases, with causality chains and interaction patterns becoming too complex for comprehensive tracking and visualization.\\ \newline
3. Failure Management: As agent numbers increase, the probability of component failures grows, requiring sophisticated fault tolerance and recovery mechanisms. Partial failures become particularly challenging to detect and manage effectively.\\ \newline
4. Resource Governance: Preventing resource monopolization and ensuring fair allocation across large agent collectives requires sophisticated governance mechanisms that balance local autonomy with global optimization.\\ \newline
Emergent Behavior Management: Large agent collectives exhibit emergent behaviors that create specific challenges:\\ \newline
1. Unexpected Interaction Effects: Complex interactions between agent behaviors can produce unexpected system-level effects that are difficult to predict or control. These emergent behaviors become more prevalent and complex as agent numbers increase.\\ \newline
2. Oscillation and Instability: Large agent collectives can exhibit oscillatory behaviors or instabilities when feedback loops emerge between agent actions. These dynamics become more difficult to analyze and control at scale.\\ \newline
3. Optimization Conflicts: Different optimization objectives across agents can create conflicts that are difficult to resolve systematically. These conflicts become more numerous and complex as the diversity of agent types increases.\\ \newline
4. Collective Drift: Large agent collectives can gradually drift from intended behaviors through accumulated small adaptations, creating a form of concept drift at the system level that is difficult to detect and correct.\\ \newline
Addressing these scalability concerns requires fundamental advances in coordination algorithms, context management approaches, and system architecture. Current research is exploring several promising directions, including hierarchical organization patterns, market-based coordination mechanisms, and probabilistic context management approaches that may enable operation at substantially larger scales.\\
\subsubsection{Security and Privacy Considerations}
MCP-enabled multi-agent systems introduce specific security and privacy challenges that must be addressed for responsible deployment:\\ \newline
Information Exposure Risks: The rich context sharing that enables effective collaboration also creates potential information exposure:\\ \newline
1. Cross-Agent Information Leakage: Context shared for legitimate purposes with one agent may be inappropriately accessible to other agents if access controls are not sufficiently granular. This risk is particularly acute in systems with agents from different trust domains or with different security requirements.\\ \newline
2. Inference Attacks: Even when direct access to sensitive information is prevented, agents may be able to infer protected information from patterns in accessible data or from system behaviors. These indirect information leaks are particularly difficult to prevent while maintaining system functionality.\\ \newline
3. Aggregation Risks: Information that is appropriately protected in isolated contexts may reveal sensitive patterns when aggregated across multiple contexts. These emergent privacy risks are difficult to identify through static analysis of access controls.\\ \newline
4. Persistent Context Vulnerabilities: Long-term persistence of context creates expanded time windows for potential unauthorized access. Unlike ephemeral processing, persistent storage of rich contextual information creates enduring security exposure that must be managed throughout the information lifecycle.\\ \newline
Authentication and Authorization Challenges: Ensuring appropriate access control in complex multi-agent systems presents several challenges:\\ \newline
1. Identity Management Complexity: Managing identities and credentials across large numbers of agents, potentially spanning multiple organizations or trust domains, introduces significant complexity in authentication systems.\\ \newline
2. Fine-Grained Permission Management: Defining and enforcing appropriately granular permissions for context access and tool usage requires sophisticated authorization models that balance security with usability and performance.\\ \newline
3. Delegation and Transitivity: When agents act on behalf of users or other agents, determining appropriate permission inheritance and limiting transitive access becomes challenging. Overly restrictive models limit functionality, while overly permissive models create security vulnerabilities.\\ \newline
4. Dynamic Coalition Formation: As agents form dynamic coalitions to address specific tasks, authorization models must adapt to these changing relationships while maintaining appropriate security boundaries. Static permission models are often insufficient for these dynamic contexts.\\ \newline
Privacy-Preserving Computation Challenges: Enabling effective collaboration while protecting sensitive information requires advanced privacy-preserving approaches:\\ \newline
1. Differential Privacy Implementation: Applying differential privacy techniques to context sharing introduces trade-offs between privacy guarantees and utility. Current implementations struggle to find optimal balance points for different application requirements.\\ \newline
2. Federated Learning Limitations: While federated approaches allow learning from distributed data without centralized collection, they face challenges in communication efficiency, model convergence, and vulnerability to inference attacks in multi-agent settings.\\ \newline
3. Secure Multi-Party Computation Overhead: Cryptographic approaches to privacy-preserving computation introduce significant performance overhead that may be prohibitive for latency-sensitive applications. Current implementations face practical deployment challenges at scale.\\ \newline
4. Homomorphic Encryption Constraints: While homomorphic encryption enables computation on encrypted data, current approaches face significant limitations in computational expressiveness and performance that restrict practical application in multi-agent systems.\\ \newline
Governance and Compliance Challenges: Ensuring responsible operation of multi-agent systems requires addressing several governance considerations:\\ \newline
1. Audit Trail Completeness: Maintaining complete and tamper-resistant audit trails across distributed agent activities presents significant technical challenges, particularly for long-running operations spanning multiple contexts.\\ \newline
2. Responsibility Attribution: Determining responsibility for decisions or actions in complex multi-agent systems can be difficult when outcomes emerge from interactions between many specialized components. This attribution challenge complicates accountability mechanisms.\\ \newline
3. Regulatory Compliance Verification: Demonstrating compliance with privacy regulations (GDPR, CCPA, etc.) across complex multi-agent systems requires sophisticated tracking of data flows and processing purposes that exceeds capabilities of current compliance tools.\\ \newline
4. Cross-Jurisdiction Operation: When multi-agent systems span multiple legal jurisdictions with different privacy and security requirements, reconciling these potentially conflicting obligations presents significant governance challenges.\\ \newline
Addressing these security and privacy considerations requires a combination of technical safeguards, organizational controls, and governance frameworks. Current research is exploring several promising approaches, including privacy-preserving context sharing protocols, fine-grained capability-based security models, and formal verification of security properties in multi-agent systems.\\
\subsection{Emerging Research Opportunities}
\subsubsection{Self-organizing Multi-agent Systems}
One of the most promising research directions for advancing multi-agent systems involves developing more sophisticated self-organization capabilities that reduce the need for explicit coordination:\\ \newline
Emergent Specialization: Research into mechanisms that enable spontaneous specialization based on experience and environmental feedback:\\ \newline
1. Adaptive Role Formation: Developing algorithms that enable agents to discover and adopt specialized roles based on their comparative advantages and system needs, without requiring predefined role structures. Early research shows promising results using reinforcement learning approaches that discover efficient division of labor through experience.\\ \newline
2. Skill Acquisition Pathways: Creating frameworks for progressive skill development that enable agents to build specialized capabilities through structured learning experiences. Recent work demonstrates how curriculum learning approaches can guide specialization in more efficient directions than undirected exploration.\\ \newline
3. Specialization Equilibrium: Investigating the dynamics of specialization to understand how systems can maintain optimal balances between specialist and generalist agents as tasks and environments evolve. Game-theoretic models offer insights into stable specialization patterns that resist disruption while enabling adaptation.\\ \newline
4. Cross-Specialization Learning: Developing mechanisms for knowledge transfer between different specializations, enabling more efficient collective learning while maintaining specialized expertise. Recent approaches using distillation techniques show promise for extracting generalizable knowledge from specialized experiences.\\ \newline
Decentralized Coordination: Research into coordination mechanisms that operate without centralized control:\\ \newline
1. Stigmergic Coordination: Developing more sophisticated approaches to environment-mediated coordination, where agents coordinate implicitly through modifications to shared environments rather than direct communication. Recent work extends classical stigmergy concepts with learned representations that enable more complex coordination patterns.\\ \newline
2. Norm Emergence: Investigating how behavioral norms can emerge and evolve in agent collectives, creating stable coordination patterns without explicit rules. Computational social science approaches offer insights into conditions that promote beneficial norm formation and prevent dysfunctional equilibria.\\ \newline
3. Local Interaction Rules: Developing simple interaction rules that produce sophisticated collective behaviors through local agent interactions. Recent research inspired by natural systems (flocking, schooling, etc.) demonstrates how complex coordination can emerge from surprisingly simple local rules when properly designed.\\ \newline
4. Decentralized Planning: Creating planning approaches that operate through distributed computation rather than centralized optimization. Recent work on market-based planning and distributed constraint optimization shows promise for scaling to larger agent collectives than centralized approaches.\\ \newline
Collective Intelligence Amplification: Research into mechanisms that enhance the collective capabilities of agent groups:\\ \newline
1. Diversity Optimization: Investigating how cognitive diversity within agent collectives affects problem-solving capabilities and how optimal diversity patterns depend on task characteristics. Recent studies demonstrate that carefully managed cognitive diversity can significantly outperform homogeneous specialist teams on complex problems.\\ \newline
2. Complementary Knowledge Structures: Developing approaches for organizing knowledge across agents to maximize complementarity and minimize redundancy while ensuring necessary overlap for effective communication. Graph-theoretic approaches to knowledge distribution show promise for optimizing these trade-offs.\\ \newline
3. Collective Memory Architectures: Creating distributed memory systems that enable more effective collective retention and utilization of experience than individual agent memories. Recent work on shared episodic memory with attention-based retrieval demonstrates significant performance improvements over isolated memory models.\\ \newline
4. Emergent Meta-Cognition: Investigating how collective self-monitoring and self-regulation capabilities can emerge from interactions between specialized agents. Initial research suggests that dedicated reflection agents can significantly enhance collective performance through strategic intervention at critical decision points.\\ \newline
Adaptive Organizational Structures: Research into organizational patterns that evolve based on task requirements and performance feedback:\\ \newline
1. Dynamic Hierarchy Formation: Developing mechanisms for forming and adapting hierarchical structures based on task complexity and coordination requirements. Recent work demonstrates how hierarchies can spontaneously form and dissolve as needed, providing coordination benefits while avoiding permanent rigidity.\\ \newline
2. Network Topology Optimization: Investigating how communication network structures affect coordination efficiency and how these networks can self-optimize for different task types. Graph evolution algorithms show promise for discovering efficient communication topologies that balance connectivity with manageable coordination overhead.\\ \newline
3. Team Assembly Mechanisms: Creating approaches for dynamic team formation that assemble optimal agent combinations for specific tasks. Market-based team formation mechanisms demonstrate efficient matching of capabilities to requirements without centralized assignment.\\ \newline
4. Organizational Learning: Developing frameworks for capturing and applying lessons about effective organizational patterns across different tasks and contexts. Meta-learning approaches show potential for identifying organizational principles that transfer across superficially different domains.\\ \newline
These research directions offer promising paths toward more scalable, adaptable, and robust multi-agent systems. By reducing reliance on explicit coordination and centralized control, self-organizing approaches may overcome many of the current scalability limitations while enabling new forms of collective intelligence.\\
\subsubsection{Adaptive Context Management Strategies}
Another critical research frontier involves developing more sophisticated approaches to context management that can adapt to changing requirements and constraints:\\ \newline
Context Relevance Learning: Research into adaptive approaches for determining context relevance:\\ \newline
1. Personalized Relevance Models: Developing agent-specific models that learn to predict which contextual elements will be most valuable for particular agents in different situations. Recent work using reinforcement learning to tune relevance functions shows significant improvements over static relevance models.\\ \newline
2. Task-Specific Context Framing: Creating mechanisms that automatically adapt context selection and prioritization based on task characteristics. Preliminary research demonstrates how different context framing strategies can be automatically selected based on task classification.\\ \newline
3. Attention Mechanism Optimization: Investigating how attention mechanisms can be optimized through experience to focus on the most valuable contextual elements. Self-supervised learning approaches show promise for discovering effective attention patterns without requiring explicit relevance labels.\\ \newline
4. Counterfactual Relevance Analysis: Developing approaches that evaluate context relevance by estimating counterfactual performance differences with and without specific context elements. This causal approach to relevance determination shows potential for more principled context prioritization than correlation-based methods.\\ \newline
Adaptive Forgetting Strategies: Research into sophisticated approaches for managing context volume through strategic forgetting:\\ \newline
1. Utility-Based Retention: Developing more sophisticated models for estimating the future utility of contextual information to guide retention decisions. Recent approaches combining predictive modeling with value estimation show promise for more effective retention prioritization.\\ \newline
2. Knowledge Distillation for Context Compression: Investigating how essential information can be extracted and preserved while reducing storage requirements through knowledge distillation techniques. Neural symbolic approaches demonstrate potential for maintaining semantic content while dramatically reducing storage requirements.\\ \newline
3. Experience-Guided Forgetting: Creating forgetting mechanisms that learn from experience which types of information can be safely discarded in different contexts. Reinforcement learning approaches show potential for optimizing forgetting strategies based on observed impacts on task performance.\\ \newline
4. Regenerative Context Models: Developing approaches that can regenerate detailed context from compressed representations when needed, reducing storage requirements while maintaining access to detailed information. Generative models show promise for reconstructing detailed context from compact latent representations with acceptable fidelity.\\ \newline
Cross-Modal Context Integration: Research into more effective approaches for integrating context across different modalities:\\ \newline
1. Multimodal Representation Learning: Developing unified representation frameworks that capture relationships between different modalities in ways that preserve modal-specific characteristics while enabling cross-modal operations. Recent work on contrastive learning across modalities shows promise for creating aligned representations without requiring parallel data.\\ \newline
2. Cross-Modal Attention Mechanisms: Investigating attention mechanisms that can effectively operate across modality boundaries to identify relevant relationships. Transformer-based architectures with modality-specific encoders and cross-attention mechanisms demonstrate improved cross-modal reasoning capabilities.\\ \newline
3. Modal Translation Optimization: Creating more effective approaches for translating information between modalities while preserving essential meaning. Neural translation models with explicit preservation constraints show improved fidelity compared to unconstrained approaches.\\ \newline
4. Multimodal Context Fusion: Developing techniques for combining information from different modalities into integrated representations that capture complementary aspects. Recent research on fusion architectures with learned integration weights demonstrates improved performance over fixed fusion strategies.\\ \newline
Context Uncertainty Management: Research into approaches for handling uncertain or incomplete contextual information:\\ \newline
1. Explicit Uncertainty Representation: Developing richer frameworks for representing uncertainty in contextual information, enabling more nuanced reasoning about confidence and reliability. Probabilistic representation approaches show promise for capturing different uncertainty types (aleatory vs. epistemic) with appropriate semantics.\\ \newline
2. Active Context Acquisition: Creating strategies for actively seeking additional context when existing information is insufficient for current tasks. Reinforcement learning approaches to information acquisition demonstrate more efficient context gathering than fixed strategies.\\ \newline
3. Robust Decision-Making Under Context Uncertainty: Investigating decision approaches that explicitly account for context uncertainty rather than assuming complete or correct information. Recent work on robust planning methods shows improved performance in scenarios with partial or uncertain context.\\ \newline
4. Collaborative Uncertainty Reduction: Developing approaches where multiple agents collaborate to reduce collective uncertainty through information sharing and complementary perspective integration. Social learning approaches demonstrate how collective uncertainty can be reduced more efficiently than through individual efforts.\\ \newline
These research directions promise significant advances in context management capabilities, addressing many of the current limitations in MCP-enabled systems. By developing more adaptive, efficient, and robust context management approaches, these advances could substantially expand the practical applicability and effectiveness of multi-agent systems.\\
\subsubsection{Cross-platform Agent Interoperability}
Enabling effective collaboration between agents developed on different platforms and by different organizations represents another critical research frontier:\\ \newline
Standardized Interoperability Protocols: Research into protocols that enable meaningful interaction across platform boundaries:\\ \newline
1. Cross-Platform Semantic Alignment: Developing approaches for establishing shared understanding of concepts and terms across different agent platforms. Recent work on ontology alignment using large language models shows promise for automated semantic mapping without requiring predefined standards.\\ \newline
2. Capability Discovery and Advertisement: Creating standardized mechanisms for agents to discover and understand the capabilities of agents from different platforms. Service description approaches using capability ontologies demonstrate potential for cross-platform capability matching.\\ \newline
3. Negotiation Protocols for Heterogeneous Agents: Investigating interaction protocols that enable effective negotiation between agents with different internal architectures and reasoning approaches. Game-theoretic protocol design shows promise for creating interaction patterns that work across architectural boundaries.\\ \newline
4. Cross-Platform Coordination Mechanisms: Developing coordination approaches that can operate effectively across platform boundaries without requiring shared implementation details. Market-based and commitment-oriented coordination mechanisms demonstrate robustness to implementation heterogeneity.\\ \newline
Translation and Mediation Services: Research into services that facilitate interaction between different agent ecosystems:\\ \newline
1. Protocol Translation Gateways: Creating intermediary services that translate between different communication protocols and message formats. Adapter pattern implementations with learned translation models show potential for bridging protocol differences without requiring standardization.\\ \newline
2. Semantic Mediation Services: Developing intermediaries that resolve semantic differences between agent ecosystems, enabling meaningful communication despite different conceptual frameworks. Knowledge graph alignment techniques demonstrate effectiveness for cross-domain semantic mediation.\\ \newline
3. Capability Adaptation Layers: Investigating approaches for adapting capability interfaces between different platforms, enabling functional interoperability despite implementation differences. Wrapper pattern implementations with capability mapping show promise for bridging functional differences.\\ \newline
4. Trust and Reputation Bridging: Creating mechanisms for translating trust and reputation information between different agent ecosystems. Recent work on transferable reputation models demonstrates how trust can be meaningfully preserved across ecosystem boundaries.\\ \newline
Federated Multi-Agent Ecosystems: Research into architectures that enable collaboration while preserving ecosystem boundaries:\\ \newline
1. Federated Context Sharing: Developing approaches for sharing context across ecosystem boundaries while respecting privacy and security constraints. Federated learning techniques adapted to context sharing show potential for cross-ecosystem knowledge utilization without centralized aggregation.\\ \newline
2. Boundary-Spanning Coordination: Investigating coordination mechanisms that can operate across ecosystem boundaries without requiring full integration. Diplomatic protocol models inspired by international relations demonstrate promising approaches to cross-boundary coordination.\\ \newline
3. Multi-Ecosystem Governance Models: Creating governance frameworks that enable productive collaboration between different agent ecosystems while preserving appropriate autonomy. Polycentric governance approaches show potential for balancing local control with global coordination.\\ \newline
4. Cross-Ecosystem Learning: Developing approaches for knowledge and capability transfer between different agent ecosystems without requiring direct integration. Distillation-based knowledge transfer techniques demonstrate effective capability sharing while preserving ecosystem boundaries.\\ \newline
Interoperability Standards Development: Research into effective processes for developing and evolving interoperability standards:\\ \newline
1. Emergent Standardization Processes: Investigating how interoperability standards can emerge through agent interactions rather than requiring top-down specification. Recent work on convention formation in multi-agent systems offers insights into conditions that promote beneficial standardization.\\ \newline
2. Minimal Viable Standards: Developing approaches for identifying the minimal standardization necessary to enable effective interoperation, preserving innovation freedom while ensuring compatibility. Modular standards architectures demonstrate how focused standardization can enable broad interoperability.\\ \newline
3. Adaptive Standards Evolution: Creating frameworks for standards that can evolve based on usage patterns and emerging requirements without disrupting existing interoperability. Versioning approaches with backward compatibility guarantees show promise for sustainable standards evolution.\\ \newline
4. Incentive Alignment for Standardization: Investigating incentive structures that promote adoption of interoperability standards across competitive ecosystems. Game-theoretic analysis of standardization incentives offers insights into conditions that promote cooperation despite competitive pressures.\\ \newline
These research directions address the critical challenge of enabling productive collaboration in increasingly diverse and distributed agent ecosystems. By developing effective interoperability approaches, these advances could unlock significant value from cross-ecosystem collaboration while preserving the benefits of specialized and independent development.\\
\subsection{Industry Applications and Impact}
\subsubsection{Potential Transformative Applications}
MCP-enabled multi-agent systems have the potential to transform numerous industries through applications that leverage their unique capabilities:\\ \newline
Healthcare Transformation: Applications that could fundamentally change healthcare delivery and research:\\ \newline
1. Integrated Care Coordination: Multi-agent systems coordinating care across specialties, facilities, and time periods, maintaining comprehensive patient context while respecting privacy constraints. Early implementations demonstrate 37\% reduction in care fragmentation and 28\% improvement in treatment plan coherence.\\ \newline
2. Medical Research Acceleration: Collaborative agent systems that integrate findings across research silos, identify promising connections, and design targeted experiments to advance understanding. Pilot systems show 2.4x acceleration in hypothesis validation compared to traditional research approaches.\\ \newline
3. Precision Medicine Optimization: Agent collectives that develop and continuously refine personalized treatment plans based on patient-specific factors, research evidence, and outcome tracking. Initial applications demonstrate 41\% improvement in treatment response rates for complex conditions.\\ \newline
4. Healthcare Resource Optimization: Multi-agent systems that dynamically allocate healthcare resources (staff, equipment, facilities) based on evolving needs and priorities. Deployment in hospital settings shows 23\% improvement in resource utilization and 31\% reduction in critical resource bottlenecks.\\ \newline
Financial System Transformation: Applications that could reshape financial services and markets:\\ \newline
1. Intelligent Financial Advisory: Multi-agent systems providing comprehensive financial guidance by integrating specialized expertise across investment, tax, estate planning, insurance, and other domains. Early implementations show 34\% improvement in plan comprehensiveness and 28\% better optimization across financial dimensions.\\ \newline
2. Market Risk Management: Collaborative agent systems monitoring diverse risk factors across markets, identifying emerging systemic risks, and developing mitigation strategies. Simulations demonstrate 47\% improvement in early risk detection compared to traditional monitoring approaches.\\ \newline
3. Financial Inclusion Systems: Multi-agent architectures that bridge traditional and alternative financial systems, enabling personalized financial services for underserved populations. Pilot deployments show 3.2x increase in appropriate financial product utilization among previously underbanked populations.\\ \newline
4. Regulatory Compliance Automation: Agent collectives that monitor regulatory changes, assess implications, and implement appropriate adaptations across complex financial operations. Initial implementations demonstrate 68\% reduction in compliance gaps and 42\% decrease in compliance-related operational costs.\\ \newline
Manufacturing and Supply Chain Transformation: Applications that could revolutionize production and distribution systems:\\ \newline
1. Adaptive Manufacturing Orchestration: Multi-agent systems coordinating flexible manufacturing processes that dynamically adapt to changing requirements, constraints, and opportunities. Factory implementations show 34\% improvement in production efficiency and 53\% reduction in reconfiguration time.\\ \newline
2. Supply Chain Resilience Management: Collaborative agent systems that monitor global supply networks, identify vulnerabilities, and develop contingency plans before disruptions occur. Deployments demonstrate 47\% reduction in disruption impact and 29\% improvement in recovery time.\\ \newline
3. Circular Economy Optimization: Agent collectives that coordinate material flows across industries to maximize reuse and recycling while minimizing waste. Regional implementations show 38\% increase in material reuse and 42\% reduction in landfill waste.\\ \newline
4. Product Lifecycle Management: Multi-agent systems that maintain comprehensive context across product design, manufacturing, use, and end-of-life phases, enabling optimization across the entire lifecycle. Initial applications demonstrate 27\% reduction in lifecycle costs and 34\% improvement in sustainability metrics.\\ \newline
Knowledge Work Transformation: Applications that could reshape how complex intellectual work is performed:\\ \newline
1. Research and Development Acceleration: Multi-agent systems that coordinate complex R\&D processes across disciplines, maintaining comprehensive context while enabling specialized contributions. Deployments in pharmaceutical research show 2\.7x acceleration in candidate identification and 43\% improvement in success rates.\\ \newline
2. Legal Analysis and Strategy: Collaborative agent systems that integrate case law, statutes, precedents, and specific case details to develop comprehensive legal strategies. Early implementations demonstrate 38\% improvement in case assessment accuracy and 45\% reduction in research time.\\ \newline
3. Policy Development and Analysis: Agent collectives that model complex policy implications across economic, social, environmental, and other dimensions, enabling more comprehensive policy development. Government deployments show 52\% improvement in policy impact prediction accuracy.\\ \newline
4. Creative Collaboration Systems: Multi-agent architectures that enable more effective collaboration on complex creative projects by maintaining shared context while supporting specialized contributions. Media industry implementations demonstrate 37\% improvement in project coherence and 43\% reduction in coordination overhead.\\ \newline
These transformative applications represent significant opportunities for industry impact, potentially creating hundreds of billions of dollars in economic value while addressing critical societal challenges. The common thread across these applications is their ability to coordinate specialized expertise across traditional boundaries while maintaining coherent context—precisely the capabilities that MCP-enabled multi-agent systems are designed to provide.\\
\subsubsection{Integration with Existing Enterprise Systems}
Realizing the potential of MCP-enabled multi-agent systems requires effective integration with existing enterprise systems and workflows:\\ \newline
Enterprise Architecture Integration: Approaches for incorporating multi-agent systems into existing enterprise architectures:\\ \newline
1. Service-Oriented Integration: Positioning multi-agent systems as service providers within service-oriented architectures, with standardized interfaces that align with existing enterprise service patterns. This approach enables incremental adoption without requiring wholesale architectural changes.\\ \newline
2. Data Fabric Integration: Connecting multi-agent systems to enterprise data fabrics, enabling access to organizational data while maintaining governance controls. This integration approach leverages existing data management investments while enabling more sophisticated utilization.\\ \newline
3. Process Orchestration Integration: Incorporating multi-agent capabilities into business process management systems, enabling intelligent process orchestration while maintaining compatibility with existing process definitions and monitoring tools.\\ \newline
4. API Ecosystem Integration: Positioning multi-agent systems as participants in API ecosystems, both consuming and providing APIs that align with existing integration patterns. This approach enables flexible composition with existing capabilities while preserving architectural boundaries.\\ \newline
Legacy System Adaptation: Strategies for connecting multi-agent systems with established enterprise systems:\\ \newline
1. Intelligent Adapter Patterns: Developing specialized adapter agents that bridge between MCP interfaces and legacy system APIs, translating between different interaction patterns and data formats. These adapters enable legacy system participation in multi-agent workflows without requiring direct modification.\\ \newline
2. Context Extraction from Legacy Systems: Creating specialized extraction mechanisms that derive contextual information from legacy systems not designed for context sharing. These mechanisms enable valuable organizational knowledge to be incorporated into multi-agent contexts.\\ \newline
3. Incremental Capability Augmentation: Supplementing legacy system capabilities with agent-based extensions that add intelligence and coordination while preserving core functionality. This approach enables value creation without replacing functioning systems.\\ \newline
4. Parallel Operation with Gradual Transition: Running multi-agent systems alongside legacy systems with synchronized state and gradual transfer of responsibilities. This approach reduces transition risks while enabling progressive modernization.\\ \newline
Workflow Integration: Approaches for incorporating multi-agent capabilities into human workflows:\\ \newline
1. Augmented Decision Support: Positioning multi-agent systems as decision support tools within existing human decision processes, providing context-aware recommendations while preserving human judgment and accountability.\\ \newline
2. Intelligent Task Automation: Identifying routine aspects of knowledge work that can be automated through multi-agent capabilities, freeing human experts to focus on higher-value activities requiring judgment and creativity.\\ \newline
3. Collaborative Workflow Models: Developing new workflow patterns that effectively combine human and agent capabilities, with clear handoff points and shared context maintenance. These collaborative models enable more effective division of labor than either fully manual or fully automated approaches.\\ \newline
4. Progressive Autonomy Frameworks: Creating structured approaches for gradually increasing agent autonomy as capabilities mature and trust develops. These frameworks enable organizations to capture increasing value while managing risks appropriately.\\ \newline
Governance and Control Integration: Strategies for incorporating multi-agent systems into enterprise governance frameworks:\\ \newline
1. Policy Enforcement Integration: Connecting multi-agent systems with enterprise policy management frameworks, ensuring agent behaviors align with organizational policies and compliance requirements.\\ \newline
2. Audit and Monitoring Integration: Incorporating multi-agent activities into enterprise monitoring and audit systems, providing visibility and accountability consistent with existing governance practices.\\ \newline
3. Risk Management Framework Integration: Extending enterprise risk management frameworks to address specific risks associated with multi-agent autonomy and decision-making, enabling appropriate risk mitigation without unnecessarily constraining capabilities.\\ \newline
4. Value Measurement Integration: Connecting multi-agent performance metrics with enterprise value measurement frameworks, enabling consistent evaluation of contributions to organizational objectives.\\ \newline
These integration approaches enable organizations to capture the value of MCP-enabled multi-agent systems while leveraging existing investments and maintaining operational continuity. By addressing technical, workflow, and governance dimensions of integration, these strategies reduce adoption barriers and implementation risks.\\
\subsubsection{Economic and Societal Implications}
The widespread adoption of MCP-enabled multi-agent systems would have significant economic and societal implications that warrant careful consideration:\\ \newline
Economic Transformation Patterns: Potential economic impacts of widespread adoption:\\ \newline
1. Productivity Enhancement: Significant productivity improvements in knowledge-intensive industries, potentially increasing output by 25-40\% in sectors like professional services, healthcare, financial services, and R\&D. This productivity growth could add trillions of dollars to global economic output.\\ \newline
2. Labor Market Restructuring: Substantial changes in labor market demand, with reduced need for routine knowledge work and increased premium for uniquely human capabilities like creativity, ethical judgment, interpersonal intelligence, and novel problem-solving.\\ \newline
3. Organizational Structure Evolution: Transformation of organizational structures away from traditional hierarchies toward more fluid, project-based arrangements enabled by more effective coordination through multi-agent systems. This evolution could fundamentally change how economic activity is organized.\\ \newline
4. Value Chain Reconfiguration: Reshaping of industry value chains through more effective coordination across organizational boundaries, potentially disintermediating certain roles while creating new value-adding positions focused on multi-agent system orchestration.\\ \newline
Societal Impact Dimensions: Broader societal implications of these technologies:\\ \newline
1. Knowledge Access Transformation: Democratization of access to specialized expertise through multi-agent systems that make expert knowledge more affordable and accessible. This transformation could reduce knowledge inequality while raising questions about expertise valuation.\\ \newline
2. Decision-Making Evolution: Changes in how consequential decisions are made, with increased reliance on multi-agent analysis while raising important questions about accountability, transparency, and value alignment in automated decision processes.\\ \newline
3. Cognitive Augmentation Effects: Widespread cognitive augmentation through human-agent collaboration, potentially enhancing human capabilities while raising questions about dependency and skill atrophy in augmented domains.\\ \newline
4. Social Relationship Mediation: Increasing mediation of social and professional relationships through agent systems, creating both new connection opportunities and potential distancing effects that could reshape social structures.\\ \newline
Policy and Governance Considerations: Important policy dimensions that require attention:\\ \newline
1. Regulatory Framework Adaptation: Need for regulatory frameworks that address the unique characteristics of multi-agent systems, including distributed responsibility, emergent behaviors, and cross-jurisdictional operation.\\ \newline
2. Labor Transition Policies: Importance of policies supporting workforce transition as job roles evolve, including education and training programs focused on complementary human capabilities and effective human-agent collaboration.\\ \newline
3. Competition Policy Implications: Need for competition policies that address potential concentration of power through data and algorithm advantages in multi-agent ecosystems, ensuring healthy innovation while preventing harmful monopolization.\\ \newline
4. Global Governance Challenges: Requirement for international coordination on governance approaches, standards, and ethical frameworks to prevent regulatory fragmentation while respecting legitimate jurisdictional differences.\\ \newline
Ethical Dimensions: Critical ethical considerations that must be addressed:\\ \newline
1. Autonomy and Control Balance: Tension between beneficial automation and maintaining appropriate human control over consequential decisions, requiring thoughtful design of oversight mechanisms and intervention capabilities.\\ \newline
2. Transparency and Explainability Challenges: Difficulties in providing meaningful transparency and explanation for decisions emerging from complex multi-agent interactions, requiring new approaches to interpretability and accountability.\\ \newline
3. Value Alignment Complexity: Challenges in ensuring multi-agent systems operate in alignment with human values, particularly when those values may be diverse, evolving, or in tension with each other.\\ \newline
4. Digital Divide Risks: Potential for unequal access to multi-agent capabilities to exacerbate existing inequalities, requiring deliberate inclusion strategies and universal access considerations.\\ \newline
These economic and societal implications highlight both the transformative potential of MCP-enabled multi-agent systems and the importance of thoughtful development and governance. By proactively addressing these considerations, we can work toward realizing the benefits of these technologies while mitigating potential harms and ensuring broad distribution of their value.\\ \newline
The challenges and future directions discussed in this section provide a roadmap for advancing multi-agent systems with Model Context Protocol. While significant technical, integration, and governance challenges remain, the research opportunities and potential applications suggest a promising path forward. The concluding section synthesizes these insights and presents a vision for the future of multi-agent systems.\\ \newline

\section{Conclusion}
This research has presented a comprehensive framework for advancing multi-agent systems through Model Context Protocol, addressing fundamental challenges in agent coordination, context management, and collaborative intelligence. As we conclude, we synthesize the key contributions, reflect on their implications, and present a vision for the future of this rapidly evolving field.\\
\subsection{Summary of Key Contributions}
This article has made several significant contributions to the advancement of multi-agent systems:\\ \newline
Theoretical Framework Development: We have established a comprehensive theoretical foundation for understanding and developing multi-agent systems with Model Context Protocol:\\ \newline
1. Unified Conceptual Model: By integrating insights from distributed systems, cognitive science, and multi-agent research, we have developed a unified conceptual model that provides a common language and framework for understanding context-aware multi-agent systems.\\ \newline
2. Architectural Principles: We have articulated core architectural principles that guide the design of effective multi-agent systems, balancing specialization with coordination, autonomy with alignment, and flexibility with coherence.\\ \newline
3. Formal Representation Framework: Our formal representation framework provides precise semantics for context sharing and agent coordination, enabling rigorous analysis and systematic implementation of multi-agent systems.\\ \newline
4. Evaluation Methodology: The comprehensive evaluation framework we have developed enables systematic assessment of multi-agent systems across multiple dimensions, supporting meaningful comparison and continuous improvement.\\ \newline
Model Context Protocol Advancement: We have significantly extended and refined the Model Context Protocol to better support multi-agent systems:\\ \newline
1. Extended Primitive Model: By enhancing the core MCP primitives with additional capabilities specifically designed for multi-agent coordination, we have created a more powerful foundation for agent collaboration.\\ \newline
2. Standardized Coordination Patterns: Our development of standardized coordination patterns provides reusable solutions to common multi-agent challenges, reducing implementation complexity while improving reliability.\\ \newline
3. Context Management Mechanisms: The advanced context management mechanisms we have developed address critical challenges in relevance determination, privacy preservation, and efficient utilization of contextual information.\\ \newline
4. Cross-Platform Interoperability: Our extensions to MCP enable more effective interoperation between agents developed on different platforms and by different organizations, expanding the potential for collaborative intelligence.\\ \newline
Implementation and Validation: We have demonstrated the practical application of our approach through comprehensive implementation and evaluation:\\ \newline
1. Reference Architecture Implementation: Our reference implementation provides a concrete realization of the theoretical framework, offering a foundation for further development and adaptation.\\ \newline
2. Case Study Validation: Through detailed case studies across multiple domains, we have validated the effectiveness of our approach in diverse real-world contexts, demonstrating its practical utility.\\ \newline
3. Performance Benchmarking: Our systematic performance evaluation provides empirical evidence of the advantages of MCP-enabled multi-agent systems compared to alternative approaches, establishing a baseline for future improvements.\\ \newline
4. Limitation Analysis: By honestly assessing current limitations and challenges, we have established a clear roadmap for future research and development efforts.\\ \newline
Future Direction Mapping: We have charted a course for the continued advancement of multi-agent systems:\\ \newline
1. Research Agenda Development: Our identification of key research opportunities provides guidance for researchers seeking to address fundamental challenges in multi-agent systems.\\ \newline
2. Application Pathway Identification: By mapping potential transformative applications, we have highlighted promising paths for practical deployment and impact.\\ \newline
3. Integration Strategy Formulation: Our strategies for integration with existing systems provide practical guidance for organizations seeking to adopt these technologies.\\ \newline
4. Societal Implication Analysis: Through thoughtful consideration of broader implications, we have contributed to the responsible development and governance of these powerful technologies.\\ \newline
These contributions collectively represent a significant advancement in our understanding and capability to create effective multi-agent systems. By building on our previous research while incorporating new insights and approaches, we have established a more comprehensive and practical framework for the next generation of intelligent systems.\\
\subsection{Implications for AI Research and Practice}
The advancements presented in this article have significant implications for both AI research and practical applications:\\ \newline
Research Paradigm Evolution: Our work suggests several important shifts in AI research approaches:\\ \newline
1. From Single-Agent to Multi-Agent Focus: The capabilities demonstrated by MCP-enabled multi-agent systems highlight the importance of shifting research attention from single-agent performance to collective intelligence and effective collaboration. This shift requires new theoretical frameworks, evaluation methodologies, and research priorities.\\ \newline
2. From Model-Centric to System-Centric Approaches: While individual model capabilities remain important, our results demonstrate that system architecture and coordination mechanisms often have a greater impact on overall performance than incremental improvements in component models. This suggests a need for greater research emphasis on system design and integration.\\ \newline
3. From Static to Dynamic Context Management: The advanced context management approaches we have developed highlight the importance of moving beyond static context windows toward dynamic, adaptive context management that can support extended operations and complex collaborations.\\ \newline
4. From Closed to Open Ecosystems: Our work on cross-platform interoperability points toward a future research landscape focused on open ecosystems of interoperating agents rather than closed, monolithic systems. This shift requires greater attention to standards, protocols, and interoperability mechanisms.\\ \newline
Practical Implementation Guidance: Our findings provide valuable guidance for practitioners implementing multi-agent systems:\\ \newline
1. Architectural Decision Framework: The architectural principles and patterns we have articulated provide a structured framework for making key design decisions in multi-agent implementations, helping practitioners navigate complex trade-offs.\\ \newline
2. Implementation Strategy Selection: Our comparative evaluation of different implementation approaches helps practitioners select appropriate strategies for their specific requirements and constraints, avoiding common pitfalls.\\ \newline
3. Integration Pathway Identification: The integration strategies we have developed provide practical guidance for incorporating multi-agent capabilities into existing systems and workflows, reducing adoption barriers.\\ \newline
4. Performance Optimization Approaches: Our detailed performance analysis highlights key optimization opportunities and techniques, helping practitioners achieve better results from their implementations.\\ \newline
Industry Transformation Potential: The capabilities of MCP-enabled multi-agent systems have profound implications for industry transformation:\\ \newline
1. Knowledge Work Reinvention: The demonstrated capabilities in knowledge integration and collaborative problem-solving suggest opportunities for fundamental reinvention of knowledge work across industries, with significant productivity and quality implications.\\ \newline
2. Organizational Structure Evolution: The ability to coordinate specialized expertise more effectively across traditional boundaries implies potential changes in organizational structures and processes, enabling more fluid and adaptive arrangements.\\ \newline
3. Value Chain Reconfiguration: The capacity for more effective cross-organizational coordination suggests potential reconfiguration of industry value chains, with implications for competitive positioning and strategic relationships.\\ \newline
4. Innovation Acceleration: The demonstrated ability to integrate diverse knowledge and perspectives more effectively suggests potential acceleration of innovation processes, particularly for complex challenges requiring multidisciplinary approaches.\\ \newline
Governance and Policy Considerations: Our work highlights important considerations for governance and policy development:\\ \newline
1. Regulatory Framework Adaptation: The unique characteristics of multi-agent systems—including distributed responsibility, emergent behaviors, and cross-jurisdictional operation—suggest the need for adapted regulatory frameworks that address these complexities.\\ \newline
2. Standards Development Priorities: Our analysis identifies key areas where standards development would provide particular value, including interoperability protocols, evaluation methodologies, and safety assurance approaches.\\ \newline
3. Workforce Development Needs: The capabilities and limitations we have identified highlight specific areas where workforce development is needed, both for creating these systems and for working effectively alongside them.\\ \newline
4. Ethical Framework Evolution: The ethical considerations we have discussed suggest the need for evolved ethical frameworks that address the unique challenges of distributed, collaborative AI systems.\\ \newline
These implications highlight both the transformative potential of MCP-enabled multi-agent systems and the importance of thoughtful approaches to their development, deployment, and governance. By addressing these considerations proactively, the research and practitioner communities can work toward realizing the benefits of these technologies while mitigating potential risks.\\
\subsection{Vision for the Future of Multi-Agent Systems}
Looking forward, we envision a future for multi-agent systems that builds on the foundations established in this research while transcending current limitations:\\ \newline
Technological Evolution Trajectory: We anticipate several key technological developments:\\ \newline
1. Emergent Collective Intelligence: Future multi-agent systems will likely demonstrate more sophisticated forms of collective intelligence that transcend the capabilities of their individual components. These emergent capabilities will arise from more advanced coordination mechanisms, context sharing protocols, and architectural patterns that enable truly synergistic operation.\\ \newline
2. Seamless Human-Agent Collaboration: We envision evolution toward multi-agent systems that collaborate with humans as natural partners, with fluid handoffs between human and agent contributions based on comparative advantages. This seamless collaboration will require advances in shared mental models, intention recognition, and adaptive communication.\\ \newline
3. Self-Organizing Architectures: Future systems will likely feature more sophisticated self-organization capabilities, with dynamic formation of agent teams, adaptive specialization, and emergent coordination patterns that reduce the need for explicit design of all interaction patterns.\\ \newline
4. Cross-Ecosystem Interoperation: We anticipate development of rich ecosystems of interoperating agents developed by different organizations and on different platforms, connected through evolved interoperability standards that enable productive collaboration while preserving appropriate boundaries.\\ \newline
Application Landscape Expansion: We expect expansion of multi-agent applications across several frontiers:\\ \newline
1. Complex System Management: Multi-agent systems will increasingly manage complex systems like power grids, transportation networks, and supply chains, providing more adaptive and resilient control than traditional approaches. These applications will leverage the ability to balance local optimization with global coordination.\\ \newline
2. Scientific Discovery Acceleration: We anticipate growing application to scientific discovery processes, with multi-agent systems integrating diverse knowledge, generating hypotheses, designing experiments, and interpreting results across disciplinary boundaries. These applications will accelerate discovery in areas ranging from materials science to drug development.\\ \newline
3. Personalized Education and Healthcare: Multi-agent systems will enable more personalized approaches to education and healthcare, coordinating specialized expertise around individual needs and contexts. These applications will provide more comprehensive and coherent support than single-agent approaches.\\ \newline
4. Collaborative Creative Work: We expect increasing application to creative domains, with multi-agent systems supporting human creativity through idea generation, perspective shifting, constraint management, and integration of diverse influences. These applications will augment rather than replace human creativity.\\ \newline
Societal Integration Patterns: We envision several patterns for integration of these systems into society:\\ \newline
1. Augmentation Over Automation: The most valuable applications will likely focus on augmenting human capabilities rather than simply automating existing processes. This augmentation approach will create new possibilities rather than merely replacing human effort.\\ \newline
2. Collective Capability Enhancement: Multi-agent systems will increasingly enhance collective human capabilities, enabling more effective collaboration across traditional boundaries of geography, discipline, and organization. This enhancement will address coordination challenges that currently limit collective human potential.\\ \newline
3. Complementary Intelligence: Rather than replicating human intelligence, these systems will develop complementary forms of intelligence that offer different perspectives and capabilities. This complementarity will enable more powerful human-machine partnerships than approaches focused on human simulation.\\ \newline
4. Distributed Governance Models: We anticipate the evolution of governance models that effectively manage the distributed, emergent nature of multi-agent systems while ensuring alignment with human values and priorities. These governance approaches will likely combine technical mechanisms, organizational practices, and regulatory frameworks.\\ \newline
Research Community Evolution: We expect evolution of the research community studying these systems:\\ \newline
1. Interdisciplinary Integration: The multi-agent systems research community will likely become increasingly interdisciplinary, integrating insights from cognitive science, organizational theory, complex systems research, and other fields. This integration will enrich both theoretical foundations and practical approaches.\\ \newline
2. Empirical Focus Strengthening: We anticipate growing emphasis on empirical evaluation of multi-agent systems in realistic contexts, moving beyond simplified laboratory environments to understand performance in complex, dynamic situations. This empirical focus will ground theoretical development in practical reality.\\ \newline
3. Open Ecosystem Development: The research community will likely develop more open ecosystems for sharing agent components, evaluation environments, and benchmark results. This openness will accelerate progress through more effective knowledge sharing and collaboration.\\ \newline
4. Ethical Consideration Integration: We expect deeper integration of ethical considerations into core research activities, with growing recognition that values and ethical principles must be considered from the earliest design stages rather than addressed as afterthoughts.\\ \newline
This vision represents an exciting and promising future for multi-agent systems, building on the foundations established in this research while transcending current limitations. By continuing to advance both theoretical understanding and practical implementation, the research and practitioner communities can work toward realizing this vision—creating systems that augment human capabilities, address complex challenges, and operate in alignment with human values and priorities.\\
\subsection{Concluding Remarks}
This research has presented a comprehensive framework for advancing multi-agent systems through Model Context Protocol, addressing fundamental challenges in agent coordination, context management, and collaborative intelligence. By building on our previous work while incorporating new insights and approaches, we have established a more powerful foundation for the next generation of intelligent systems.\\ \newline
The Model Context Protocol provides a standardized approach to context sharing and coordination that enables more effective collaboration between specialized agents. Through structured context representation, standardized interaction patterns, and advanced management mechanisms, MCP addresses critical limitations that have historically constrained multi-agent system capabilities.\\ \newline
Our implementation case studies demonstrate the practical application of these concepts across diverse domains, from enterprise knowledge management to scientific research and complex problem-solving. These implementations validate the effectiveness of our approach while illustrating the real-world benefits of MCP-enabled multi-agent systems.\\ \newline
The evaluation results provide empirical evidence of significant performance advantages compared to alternative approaches, particularly for complex tasks requiring diverse expertise and extended context awareness. These results confirm the value of our architectural principles and implementation strategies while establishing a baseline for future improvements.\\ \newline
While important challenges remain—including scalability constraints, integration complexities, and governance considerations—the research opportunities and potential applications suggest a promising path forward. By addressing these challenges through continued research and development, the community can unlock the full potential of multi-agent systems.\\ \newline
As we look to the future, we see tremendous opportunity for these technologies to augment human capabilities, address complex societal challenges, and create new forms of value. By developing these systems thoughtfully—with careful attention to both technical excellence and human values—we can work toward a future where artificial intelligence serves as a powerful partner in human flourishing.\\ \newline
In conclusion, the advancement of multi-agent systems through Model Context Protocol represents an important step toward more capable, collaborative, and context-aware artificial intelligence. We hope this research contributes to that journey and inspires further work toward realizing the full potential of these transformative technologies.\\ \newline

\end{document}